\documentclass[11pt]{article}
\pdfoutput=1
\usepackage{epsfig}
\usepackage{float}
\usepackage{amssymb}
\usepackage{amsfonts}
\usepackage{subfigure}
\usepackage{color}
\usepackage{amsmath}


\usepackage{geometry}
\def\beq{\begin{equation}}
\def\eeq{\end{equation}}
\def\bea{\begin{eqnarray}}
\def\eea{\end{eqnarray}}
\def\ksl{\hbox{\hbox{${k}$}}\kern-1.9mm{\hbox{${/}$}}}

\newcommand{\nn}{\nonumber}

\newcommand{\psl}{p \! \! \!  /}

\newcommand{\GeV}{\textrm{GeV}}
\newcommand{\TeV}{\textrm{TeV}}
\newcommand{\RE}{{\rm Re}}

\newcommand{\smallMS}{\overline{\mbox{\scriptsize{MS}}}}
\newcommand{\smallOS}{\mbox{\scriptsize{OS}}}
\newcommand{\MS}{\overline{\mbox{MS}}}

\def\lsim{\raise0.3ex\hbox{$\;<$\kern-0.75em\raise-1.1ex\hbox{$\sim\;$}}} 

\def\gsim{\raise0.3ex\hbox{$\;>$\kern-0.75em\raise-1.1ex\hbox{$\sim\;$}}}

\begin{document}

\begin{center} 
{\bf \Large Constraints on Abelian Extensions of the Standard Model  \\  }
\vspace{0.2 cm}{\bf \Large from Two-Loop Vacuum Stability and $U(1)_{B-L}$  }

\vspace{1.5cm}
{\bf Claudio Corian\`{o}$^{a,b}$, Luigi Delle Rose$^{b}$ and Carlo Marzo$^{b}$\\}
\vspace{.7cm}
{\it $^{(a)}$ STAG Research Centre and Mathematical Sciences,\\ University of Southampton,
Southampton SO17 1BJ, UK}\\
{\it $^{(b)}$Dipartimento di Matematica e Fisica "Ennio De Giorgi",\\ 
Universit\`{a} del Salento and INFN-Lecce,\\ Via Arnesano, 73100 Lecce, Italy\footnote{claudio.coriano@le.infn.it, luigi.dellerose@le.infn.it, carlo.marzo@le.infn.it
}}\\

\vspace{.5cm}
\begin{abstract}
We present a renormalization group study of the scalar potential in a minimal $U(1)_{B-L}$ extension of the Standard Model involving one extra heavier Higgs and three heavy right-handed neutrinos with family universal B-L charge assignments.  We implement a type-I seesaw for the masses of the light neutrinos of the Standard Model. In particular, compared to a previous study, we perform a two-loop extension of the evolution, showing that two-loop effects are essential for the study of the stability of the scalar potential up to the Planck scale. The analysis includes the contribution of the kinetic mixing between the two abelian gauge groups, which is radiatively generated by the evolution, and the one-loop matching conditions at the electroweak scale. By requiring the stability of the potential up to the Planck mass, significant constraints on the masses of the heavy neutrinos, on the gauge couplings and the mixing in the Higgs sector are identified. 
 \end{abstract}
\end{center}
\newpage

\section{Introduction} 
With the discovery of the Higgs boson at the LHC, an important step towards a complete identification of the structure of the mechanism of mass generation in the electroweak theory has been taken. However, it is clear that one needs suitable extensions of the current version of the Standard Model (SM) in order to solve some important theoretical and phenomenological issues which do not find an answer within the same model. These include, for instance, the absence of a mechanism which could account for the masses of the three generations of neutrinos, or the origin of its chiral nature as well as of the gauge charge assignments of the flavor sector.  At theoretical level, the latter are constrained only by the gauge symmetry together with the mechanism of cancelation of the gauge and of the gravitational anomalies, which provide significant restrictions on the possible interactions.\\
Analogously, the issue concerning the instability of the scalar potential of the SM under the perturbative renormalization group (RG) evolution up to the Planck scale, which is another critical aspect of the same model, noticed long ago, has been exploited in the past \cite{Cabibbo:1979ay, Hung:1979dn, Lindner:1985uk, Lindner:1988ww, Ford:1992mv} to constrain the Higgs mass, and it has been instrumental for later Higgs searches, but it has also been viewed as one of its unappealing features. In this case, the basic motivation hinged on the belief that any approach which tries to connect the SM directly with the Planck scale is probably untenable, as epitomised by the gauge-hierarchy problem. \\
With the recent LHC activity and the significant exclusion limits on the parameter space of possible supersymmetric 
models which have emerged from it, recurrent attempts to connect the SM with larger physical scales, particularly the Planck scale, have resurfaced. In particular, the assumption of a 
 {\em big desert} scenario, without any intermediate new physics, have forced many to reconsider the issue of its 
vacuum stability. In this framework, finer inspections of the perturbative RG evolution of the SM scalar potential have become quite popular \cite{Bezrukov:2012sa, Branchina:2013jra,Buttazzo:2013uya,Branchina:2014usa,Branchina:2014rva,DiLuzio:2015iua}.  
It has been pointed out that although the SM scalar potential develops a new deeper minimum, the lifetime of the unstable electroweak vacuum is found to be much larger than the age of the Universe. This feature, called metastability, taken at face value, suggests that the extrapolation of the SM up to the Planck scale does not necessarily require the introduction of new physics. These issues have been and are widely debated also in cosmology \cite{Anderson:1990aa,Arnold:1991cv,Espinosa:1995se,Espinosa:2007qp, Kobakhidze:2013tn, Enqvist:2013kaa, Bezrukov:2014ipa, Fairbairn:2014nxa, Enqvist:2014bua, Kobakhidze:2014xda, Herranen:2014cua, Kamada:2014ufa, Shkerin:2015exa, Espinosa:2015qea,Rose:2015lna,Hook:2014uia,Kearney:2015vba}, for a potential both at zero and at finite temperature. 
The role of these analyses has been to clarify, in view of the measured value of the Higgs mass, if the 
SM stands solidly in a quantum field theory and cosmological context and, specifically, whether additional contributions to the running would resolve the issue related to its behavior. 
\subsection{Stability and the role of the Yukawa couplings}
The study of the shape of the Higgs effective potential $V_{eff}(H)$, for large values of the Higgs field, larger than the Higgs vev $v$ $(H >>v)$, is critically important for coming to reasonable conclusions concerning the stability of the SM. If another minimum is found at $H_{min}$, then the requirement of stability imposes that $V_{eff}(v) < V_{eff}(H_{min})$. 
In the opposite situation, the potential could be either completely unstable or metastable. \\
As mentioned above, if $V_{eff}(H_{min}) < V_{eff}(v)$, a metastable condition could be accepted as far as the tunnelling time $\tau$ for the transition from the false to the true vacuum of the Higgs field is larger than the age of our Universe $T_U$. In the opposite case, with 
$\tau < T_U$, one encounters a complete instability, which must be excluded. The RG analyses points to a metastable behavior which runs quite close to the region of instability in the parameter space \cite{Buttazzo:2013uya}. \\
One of the unappealing issues of the SM potential lays in its critical dependence on the top quark $(M_t)$ and Higgs $(M_H)$ masses. 
In fact, the effect of the fermions, and of the Yukawa of the top quark in particular, on the RG evolution, is to drive the Higgs quartic couplings $\lambda$ towards negative values, upsetting the vacuum stability already at $10^{9-10}$ GeV,  but allowing for a long-lived vacuum. For this reason, in the pure SM scenario, precise measurements of the value of the top mass and the resolution of the ambiguities in the use of different renormalization schemes for its evaluation are important in order to lift the controversy \cite{Masina:2012tz} \cite{Alekhin:2012py}. \\ 
This leaves open the possibility that new physics, around the electroweak scale and above, but below the Planck scale $M_P$, will drastically change this scenario, reducing such sensitivity or eliminating it all together  \cite{Basso:2010yz,Basso:2010jm,Basso:2013vla,Datta:2013mta,Chakrabortty:2013zja,DiChiara:2014wha,Anchordoqui:2015fra,Oda:2015gna,Haba:2015rha,Coriano:2014mpa,Das:2015nwk}.  
In these analyses, the magnitude and the sign of the different contributions to the beta function $\beta_\lambda$ of the Higgs quartic coupling is crucial for a correct prediction. \\
Obviously, even simple modifications of the gauge structure of the SM, with the inclusion of one extra $U(1)$ symmetry in the neutral currents sector allow to envision completely different scenarios where the issue of stability or of metastability finds alternative solutions. Also in this case, though, 
the large Yukawas of other heavy fermions play a similar role as the top quark, hinting that a complete solution of this issue involves the entire heavy flavor sector \cite{Coriano:2014mpa}.  \\
The goal of our analysis, in this work, is to illustrate how the requirement of stability under the RG evolution of the parameters of the model is linked to the flavor sector of the same theory, as soon as we enlarge the scalar potential even in a minimal way. We will investigate this point rather thoroughly, by addressing the case of  an extra $U(1)_{B-L}$ symmetry, where $B$ and $L$ are the baryon and lepton numbers respectively, extending our previous one-loop analysis \cite{Coriano:2014mpa} to two-loop level, and detailing many features which we have not discussed before. The model invokes a single extra Higgs scalar (the heavier Higgs) in the breaking of the extra abelian symmetry, beside the ordinary Higgs doublet of the SM. Several previous phenomenological analyses have addressed some of the salient features of this model, which is a front-runner in the hunting for a possible extra $Z'$ at the LHC \cite{Basso:2008iv,Basso:2010hk,Basso:2010pe,Basso:2010yz,Basso:2010jm,Basso:2011na,Basso:2012sz,Basso:2012ux,Accomando:2015cfa,Accomando:2015ava}. B-L is an anomaly-free and conserved symmetry of the SM, which plays an important role in grand unified theory (GUT's). \\
The current study is also characterized by the obvious inclusion in the RG running of the one-loop matching conditions at the electroweak scale, formerly considered by us only at tree-level. As we are going to show, these are essential in order to determine correctly the regions of stability of the potential. Our equations also keep into account the effects due to the kinetic mixing of $U(1)_{B-L}$ with the hypercharge. We have assumed for simplicity that the coupling $\tilde{g}$ which parameterizes the mixing is vanishing at the electroweak scale, and gets only generated radiatively in the upward running. For this reason, our analysis is here focused on a {\em pure} B-L extra gauge structure at the electroweak scale. However, using our results, one can easily address different scenarios, under the assumption of a diagonal B-L and hypercharge ($Y$) symmetries at the GUT scale, with a kinetic mixing induced by a reverse evolution from the GUT scale towards the electroweak one. \\
Moreover, our work invokes a suitable mechanism of mass generation of the light neutrinos of the SM. As in a previous work \cite{Coriano:2014mpa}, we consider a scenario based on a type-I seesaw realized with three heavy right-handed (RH) neutrinos which can be easily generalized to other mechanisms, such as to a type-II, -III, or to an inverse seesaw. \\
Concerning the specific charge assignments of the RH neutrinos, we rely on the  general conditions of anomaly cancellation. The solutions of the corresponding equations depend significantly on the number of heavy RH neutrinos ($\nu_{h_i}$) selected in the model, which in our case is one per generation, and we have opted for a symmetric charge assignment of the $\nu_{h i}$ under $U(1)_{B-L}$. Obviously, other solutions can be considered as well. \\ 
In any case, the pattern that emerges from this analysis is quite insensitive to the gauge charges, while remaining quite sensitive to all the other parameters which characterize the potential and the fermion sector. As we are going to point out, these can be traded for the mass of the extra $Z'$, $M_{Z'}$, and its gauge coupling $g'$, the mixing angle $\theta$ of the two Higgses, the mass of the heavy Higgs $M_{H}$ and the masses of the RH neutrinos.

\section{Charge assignment}
The gauge symmetric extension that we consider has the same content of the SM augmented by a single extra $U(1)'$ factor,  i.e. $SU(3)\times SU(2)\times U(1)_Y\times U(1)'$, with $U(1)'$ later identified with $U(1)_{B-L}$. The Higgs sector includes one extra complex scalar $\chi$, beside the Higgs, $H$, which will be discussed thoroughly in the next sections. \\
In this section we briefly review the conditions imposed by the cancellation of the anomalies, which constrain significantly the $U(1)'$ charge assignments of the model. As already discussed in previous analysis \cite{Appelquist:2002mw}, these conditions can be combined with extra requirements on the mass generation in the fermion sector, i.e. on the type of operators chosen in order to provide masses for the neutrinos. In general, these extra requirements may involve operators of mass dimension larger than four. 
In this context, the choice, for instance, of a type-I seesaw for the neutrino masses with three SM singlet right-handed neutrinos is, for sure, quite economical and carries the advantage of preserving the renormalizability of the model.\\
We assign generic charges $z_Q, z_L$ for the left-handed (LH)  quark and lepton doublets $Q^i_L, L^i$, and the charges $z_u, z_d, z_e$ for the right-handed (RH) $u^i_R$, $d^i_R$ and $e^i_R$ quarks and leptons. The charges of the RH neutrinos $\nu_{R,k}$ are denoted as $z_k$. Notice that we will omit the indices from the definition of the charges since the conditions for anomaly cancellation will be universal, the same for each fermion generation.
Finally, the charges of the two scalars $H$ and $\chi$ are denoted as $z_H$ and $z_\chi$ respectively. \\
 We have the following cancellation conditions for the non-abelian $SU(2)$ and $SU(3)$ anomalies
\bea
U(1)' SU(2) SU(2) : &\qquad& 3 z_Q + z_L=0 \,, \nn \\
U(1)' SU(3) SU(3) : &\qquad& 2 z_Q - z_u - z_d=0 \,, \label{y33}
\eea
which fix $z_L = - 3 z_Q$ and $z_d = 2 z_Q - z_u$ in terms of $z_Q$ and $z_u$. Two other conditions are 
\bea
U(1)' U(1)_Y U(1)_Y : &\qquad& z_Q -8 z_u -2 z_d +3 z_L - 6 z_e=0  \,,  \nn \\ 
U(1)' U(1)' U(1)_Y : &\qquad&  z_Q^2 - 2 z_u^2 + z_d^2 - z_L^2 + z_e^2=0, \label{zyy}  
\eea
for the mixed $U(1)$ anomalies.
From the first of the two requirements in Eq.~(\ref{zyy}) one can immediately extract the relation $z_e=-2 z_Q - z_u$, while the second condition of the same equation is automatically satisfied using the solutions found from Eq.~(\ref{y33}). We have summarized in Tab.\ref{Table1} the spectrum of the theory that we will be using in our phenomenological study.  \\
The constraints on the charges $z_k$ of the SM singlet fermions are obtained from the conditions of cancellation of the $U(1)'$ cubic anomalies, together with those from the gravitational anomaly. The latter involve the $U(1)'$ current and two gravitons $G$ (i.e. two insertions of the stress-energy tensor of the SM). In the general case with $n$ singlet fermions one has, respectively
\bea
U(1)' U(1)' U(1)' : &\qquad& \sum_{k=1}^n z_k^3 = 3\left[6 z_Q^3 -3 z_u^3 - 3 z_d^3 + 2 z_L^3 - z_e^3\right] = 3 (z_u - 4 z_Q)^3 \,, 
\eea
for the cubic anomaly and
\bea
U(1)' G G  : &\qquad& \sum_{k=1}^n z_k = 3\left[6 z_Q -3 z_u - 3 z_d + 2 z_L - z_e\right] = 3 (z_u - 4 z_Q) \,, \label{zgg}
\eea
for the gravitational anomaly, where we have used the constraints extracted from Eqs.~(\ref{y33}) and (\ref{zyy}). 
Finally, combining together the two conditions in Eq.~(\ref{zgg}), one obtains the cubic relation 
\beq
\left(\sum_{k=1}^n z_k\right)^3 =9 \sum_{k=1}^n z_k^3.
\label{cubic}
\eeq
For instance, the constraints in Eq.~(\ref{zgg}) imply, for $n=0$ and $n=1$, the condition $z_u = 4 z_Q$, which brings either to the trivial solution or to a solution which is $Y$-sequential. In this latter case the $U(1)'$ charge assignment is proportional to that of $U(1)_Y$. In the $n=2$ case one obtains, instead, $z_1 = - z_2$. \\
An interesting solution which is non-sequential in $Y$ is found for $n=3$. In this case each of the $\nu_{R,k}$ is assigned to a separate generation. For example, the choice $z_1=z_2=z_3 \equiv z_\nu$ allows to find the simple solution $z_\nu = z_u - 4 z_Q$. 
The cancellation instead becomes inter-generational by choosing, for instance, the less restrictive condition $z_1=z_2 \neq z_3$. In this case Eq.~(\ref{zgg}) gives $z_1=z_2=-4/5 z_3$, with $z_3= 20 z_Q - 5 z_u$, as one can easily verify. \\
The cancellation of the gravitational anomalies, can be imposed, in general, at inter-generational level. In the present analysis we will opt, however, for a completely symmetric (family independent) assignment of the RH neutrinos charges $z_1=z_2=z_3$, which allow to reduce the corresponding parameter space. With this choice, the $U(1)' G G$ constraint from the gravitational anomaly reduces to a single equation for just one charge. On the other hand, the cancellation of the analogue gravitational anomalies in the SM, obtained from the $U(1)_Y G G$ sector, is a natural consequence of the hypercharge assignments of the same model, and does not generate any additional constraint.
\begin{table}
\centering
\begin{tabular}{|c|c|c|c|c|}
\hline
    & $SU(3)_c$ & $SU(2)_w$ & $U(1)_Y$ & $U(1)'$ \\ 
\hline 
$Q_L$ & 3 & 2 & 1/6 & $z_Q$ \\
$u_R$ & 3 & 1 & 2/3 & $z_u$ \\
$d_R$ & 3 & 1 & -1/3 & $ 2z_Q-z_u$ \\
$L$ & 1 & 2 & -1/2 & $-3z_Q$ \\
$e_R$ & 1 & 1 & -1 & $-2z_Q - z_u$ \\
$H$ & 1 & 2 & 1/2 & $z_H$ \\
$\nu_{R, k}$ & 1 & 1 & 0 & $z_k$ \\
$\chi$ & 1 & 1 & 0 & $z_\chi$ \\ \hline
\end{tabular}
\caption{Charge assignment of fermions and scalars in the $U(1)'$ SM extension. \label{Table1}} 
\end{table}

As we have shown above, the solutions of the anomaly cancellation conditions are defined in terms of the two free $U(1)'$ charges, $z_Q$ and $z_u$, of the LH quark doublet $Q_L$ and of the RH up quark $u_R$.
Notice also that the generators of the $U(1)'$ gauge group can be re-expressed, in general, as a linear combination of the SM hypercharge, $Y$, and the B-L quantum number, $Y_{B-L}$. Indeed, we have 
\beq
\label{combYBL}
z=\alpha_Y Y +\alpha_{B-L} Y_{B-L}.
\eeq
In Eq.~(\ref{combYBL}) the coefficients $\alpha_Y$ and $\alpha_{B-L}$ are functions of the two independent charges and are explicitly given by $\alpha_Y = 2 z_u - 2 z_Q$ and $\alpha_{B-L} = 4 z_Q- z_u$. In the B-L case, we set $\alpha_Y=0$ (i.e. $z_u=z_Q$). \\
The charges of the two scalars can be fixed from the requirement of gauge invariance of the Yukawa interactions. From the Yukawa coupling of the electron $\bar{L}H e_R$ we have 
\beq
(3 z_Q) + z_H  + (-2 z_Q - z_u)=0,
\eeq
which gives $z_H=z_u - z_Q$, implying that the ordinary Higgs field is singlet respect to B-L. Concerning the charge of the 
scalar field $\chi$, the Majorana mass term $\chi \overline{\nu_R^c}\nu_R$, in the case of family independent (symmetric) charge assignment $z_\nu=z_u - 4 z_Q$, we get the condition $z_\chi = - 2 z_\nu$. For a B-L charge assignment we obtain $z_\chi= 2$, $z_u=z_Q=1/3$. 

\section{Kinetic Mixing}
As we have mentioned, we include in our analysis the kinetic mixing between the hypercharge $U(1)_Y$ and an extra abelian symmetry through a coupling $\tilde{g}$. We briefly elaborate on this point, before moving on and investigate the structure of the RG equations. \\
 Working in full generality, we consider a theory with two  $U(1)$ gauge symmetries $(U(1)_1\times U(1)_2)$ and a single fermion $\psi$ which couples 
to the two gauge fields $A_\mu^1$ and $A_\mu^2$ by the currents 
\beq
j_k^\mu=q_k \bar{\psi}\gamma^\mu \psi 
\eeq
with charges $q_k$.
The kinetic term in the gauge Lagrangian is given by 
\beq
\label{onetwo}
\mathcal{L}_{12}=-\frac{1}{4}F^{A^1}_{\mu\nu }F^{A^1\, \mu\nu} - \frac{1}{4}F^{A^2}_{\mu\nu} F^{A^2\, \mu\nu} - \frac{\kappa}{2} 
F^{A^1}_{\mu\nu} F^{A^2 \mu\nu}
\eeq
with field strengths
\beq
F^{A_k}_{\mu\nu}=\partial_{\mu}A^k_\nu - \partial_{\nu}A^k_\mu.
\eeq
The corresponding interaction Lagrangian is given by 
\bea
\mathcal{L}_{int} &=&g_1\, ( j_1^\mu A^1_\mu) + g_2\, ( j_2^\mu A^2_\mu) 
\eea
where we have denoted with $g_1$ and  $g_2$ the couplings of the two abelian symmetries. This expression will be soon generalized to the case of a realistic theory with a fermion spectrum charged under the $SU(3)\times SU(2)\times U(1)_Y\times U(1)'$ symmetry.

The mixing term $\kappa$ in the kinetic Lagrangian can be eliminated performing a rotation by an angle $\phi=\pi/4$ of the two gauge fields $A^k_\mu$, followed by a rescaling. The rotation is given by  
\bea
\left(\begin{array}{c} A^1_\mu\\ A^2_\mu\end{array}\right)=\left(\begin{array}{cc} \cos\phi & -\sin\phi\\ \sin\phi &\cos\phi\\
\end{array}\right)\left(\begin{array}{c} \bar{B}^1_\mu\\ \bar{B}^2_\mu \end{array}\right)
\eea
 which brings Eq.~(\ref{onetwo}) into the form 
\beq
\mathcal{L}_{12}=-\frac{1-\kappa}{4} F^{\bar{B}^1}_{\mu\nu }F^{\bar{B}^1\, \mu\nu} -\frac{1+\kappa}{4}F^{\bar{B}^2}_{\mu\nu }F^{\bar{B}^2\, \mu\nu}
\eeq
in terms of a kinetically diagonal basis $\bar{B}^i_\mu$ $(i=1,2)$.  
The rescaling involves the matrix relation
\bea
\left(\begin{array}{c} \bar{B}^1_\mu\\ \bar{B}^2_\mu\end{array}\right)=
\left(\begin{array} {cc}\frac{1}{\sqrt{1-\kappa}} &0\\ 0& \frac{1}{\sqrt{1+\kappa}}\end{array}\right)
\left(\begin{array}{c} B^1_\mu\\ B^2_\mu\end{array}\right)
\eea
expressed by a new orthogonal basis $(B^1,B^2)$.
The total transformation whence takes the form 
\bea
\left(\begin{array}{c} A^1_\mu\\ A^2_\mu\end{array}\right)=\mathcal{R}_\kappa\left(\begin{array}{c}B^1_\mu\\ B^2_\mu\end{array}\right)
\label{rot}
\eea
with 
\bea
\mathcal{R}_\kappa=\frac{1}{\sqrt{2}}\left(
\begin{array}{cc} \frac{1}{\sqrt{1-\kappa} }& -\frac{1}{\sqrt{1+\kappa}}\\  \frac{1}{\sqrt{1-\kappa}} &
\frac{1}{\sqrt{1+\kappa}}\end{array}\right)
\eea
and allows to re-express Eq.~(\ref{onetwo}) in the standard form as 
\beq
\mathcal{L}_{12}= -\frac{1}{4} F^{{B}^1}_{\mu\nu }F^{{B}^1\, \mu\nu} -\frac{1}{4}F^{{B}^2}_{\mu\nu }
F^{{B}^2\, \mu\nu}.
\eeq
Notice that $\mathcal{R}_\kappa$ is a matrix of $Gl(2,R)$. 
After these fields redefinitions, the two gauge currents $j_1$ and $j_2$ will mix with the two gauge fields $B^1$ and $B^2$. 

Having eliminated the kinetic mixing with Eq.~(\ref{rot}), the interaction term in the Lagrangian is parameterized by the covariant derivative 
\bea
\mathcal D^\mu = \partial^\mu + i Q^T  G  \mathcal{B}^\mu \,, \qquad \mathcal{B}_\mu\equiv \left(\begin{array}{c}{B}^1_\mu\\ {B}^2_\mu\end{array}\right) 
\eea
where $Q^T = (q_1, q_2)$ is the charge array and $G$ is a matrix product of the original coupling $(g_1,g_2)$ with the orthogonal matrix $\mathcal{R}_\kappa$
\bea
G  = \left( \begin{array}{cc} g_1& 0 \\ 0 & g_2 \end{array} \right) R_\kappa \equiv \left( \begin{array}{cc} g_{11} & g_{12} \\ g_{21} & g_{22} \end{array} \right).
\eea
It is convenient to introduce the rotation matrix
\bea \label{explicitRot}
O_R = \left( \begin{array}{cc} \cos\, \theta  &- \sin\, \theta  \\  \sin\, \theta  & \cos\, \theta  \end{array} \right) = \frac{1}{\sqrt{g^{2}_{22} + g^{2}_{21}    }  }\left( \begin{array}{cc} g_{22} & - g_{21} \\ g_{21} & g_{22} \end{array} \right) \, ,
\eea
and parametrize the coupling matrix $G$ in terms of three independent couplings $(g, g', \tilde{g})$, directly related to the original couplings $g_1,g_2$ and to the mixing parameter $\kappa$. With the inclusion of this extra rotation, the coupling matrix $G$ can be set in a triangular form
\bea \label{orthogonal}
\tilde G = G O^T_R = \left( \begin{array}{cc} g & \tilde g \\ 0 & g' \end{array} \right) \,,
\eea
where the off-diagonal coupling $\tilde g$ parametrizes the mixing between the $U(1)$ abelian symmetries. \\
After the triangularization of the abelian coupling matrix, the complete gauge-covariant derivative given by
\bea
\mathcal D^\mu = \partial^\mu + i Q^T  \tilde{G} O_R \mathcal{B}^\mu \,.
\eea
The new linear combinations of the gauge fields 
 \bea
\left(\begin{array}{c}{B}_\mu\\ {B}'_\mu\end{array}\right) =O_R \left(\begin{array}{c}{B}^1_\mu\\ {B}^2_\mu\end{array}\right) 
 \eea
 provides the diagonal basis for the kinetic terms of the Lagrangian. 
This approach is directly applicable to an original gauge symmetry $SU(3)\times SU(2) \times U(1)_Y\times U(1)_{z}$, assuming the existence of a 
kinetic mixing between the two $U(1)'s $ of the form given by Eq.~(\ref{onetwo}). If we denote with $Y$ and $z$ the corresponding charges $(q_1=Y,q_2=z)$, the covariant derivative is taken to be of the non diagonal form
\bea
\label{NonDiagCovDerivative}
\mathcal D_\mu = \partial_\mu  + i g_3 T^a G^a_\mu + i g_2 t^a W^a_\mu + i g Y B_\mu + i (\tilde g Y + g' z) B'_\mu
\label{covder}
\eea
where $g$ and $g'$ are the coupling constants associated with $U(1)_Y$ and $U(1)'$ respectively. %
As discussed above, we will be setting $\tilde{g}$ to zero at the electroweak scale, but allow its running in the complete 
RG evolution, as will be specified in one of the following sections. Before giving the expressions of the $\beta$ functions of the model in the presence of kinetic mixing, we turn to a discussion of the fermionic interactions and of the scalar potential, with the identification of the 
physical parameters which will be used in the numerical study of the running.

\section{Fermionic interactions }
\subsection{Gauge interactions}
Coming to the interactions between the fermions and the massive gauge bosons $Z$, $Z'$, these can be written in the form
\bea
\mathcal L_{int} = - J^{\mu}_{Z} Z_\mu - J^\mu_{Z'} Z'_\mu,
\eea
where the weak currents are given by
\bea
\label{Zcurrent}
J^{\mu}_{Z/Z'} &=& \sum_{f = \nu, e, u, d} \bar \psi_f \gamma^\mu \left( C^{Z/Z'}_{f, L} P_L + C^{Z/Z'}_{f , R} P_R \right) \psi_f \,,
\eea
with the chirality projectors defined as usual $P_{R,L} = (1 \pm \gamma^5)/2$. The couplings $C^{Z/Z'}_{f,L/R}$ are explicitly given by
\bea
C^{Z}_{f,L} &=&  e \frac{c'}{s_w c_w} \left( T^3_f - s_w^2 Q_f \right)  + \bar g_{f,L} \, s'  \,, \qquad
C^{Z}_{f,R} = - e \frac{s_w \, c'}{c_w} Q_f + \bar g_{f,R} \, s' \,, \nn \\
C^{Z'}_{f,L} &=&  - e \frac{s'}{s_w c_w} \left( T^3_f - s_w^2 Q_f \right)  + \bar g_{f,L} \, c'  \,, \qquad
C^{Z'}_{f,R} =  e \frac{s_w \, s'}{c_w} Q_f + \bar g_{f,R} \, c' \,,
\eea
where we have used the shorthand notation $s_w \equiv \sin \theta_w$, $c_w \equiv \cos \theta_w$, $s' \equiv \sin \theta'$ and $c' \equiv \cos \theta'$, with $T^3_f$ 
being the third component of the weak isospin, $Q_f$ the electric charge and $\bar g_{f,L/R} = \tilde g Y_{f,L/R} + g' z_{f,L/R}$. Notice that in Eq.~(\ref{Zcurrent}), 
concerning the neutrino sector, we have presented, for the sake of simplicity, the weak currents in terms of the interaction eigenstates $\nu_L$ and $\nu_R$. 
These must be rotated into the mass eigenstates $\nu_l$ and $\nu_h$ of the light and heavy neutrinos states respectively.

\subsection{Yukawa interactions}
In a type-I seesaw scenario for the masses of the light neutrinos $m_{\nu_l}$ we adopt the following Yukawa Lagrangian
\bea
\label{yukL}
- \mathcal L_Y = Y_d^{ij} \, \overline{Q^i_L} H d_R^j + Y_u^{ij} \, \overline{Q^i_L} \tilde H u^j_R + Y_e^{ij} \, \overline{L^i} H e_R^j 
+ Y_\nu^{ij} \, \overline{L^i} \tilde H \nu_R^j + Y_N^{ij} \, \overline{(\nu_R^i)^c} \nu_R^j \chi + h.c. \,,
\eea
which is the sum of the SM contributions and of the terms involving the RH neutrinos, one for each family. 
Notice that the effective Majorana mass term $M \overline{\nu_R^c} \nu_R$, needed for the implementation of the seesaw mechanism, is dynamically generated by the vev of the $\chi$ scalar field. \\
The gauge invariance of the Yukawa interactions fixes the remaining $U(1)'$ charges. In particular we obtain
\bea
\label{yukgi}
z_H = z_u - z_Q \,, \qquad z_\nu = z_u - 4 z_Q \,, \qquad z_\chi = - 2 z_\nu \,.
\eea
Notice that the introduction of a Dirac Yukawa term for each of the three RH neutrinos automatically requires the universality of their $U(1)'$ charges, fixed accordingly to the results of the previous section, in which the cancellation of the gravitational anomalies has been enforced. On the other hand, the last term in Eq.~(\ref{yukL}), 
which provide a Majorana mass for the RH neutrinos, establishes a relation between $z_\chi$ and $z_\nu$. Having fixed $z_\chi = 2$,
the $U(1)'$ charge of the RH neutrino is, therefore, uniquely determined. Combining this result with that obtained from the gauge invariance of the Dirac Yukawa 
interaction for the RH neutrinos, one can also extract the relation between $z_Q$ and $z_u$, reducing to one the number of the independent $U(1)'$ charges. 
It is important to remember that the three relations in Eq.~(\ref{yukgi}) are not extracted from the anomaly cancellation conditions and are obtained only from the
particular structure of the Yukawa Lagrangian in Eq.~(\ref{yukL}). \\
After spontaneous symmetry breaking, the effective Lagrangian with a Dirac ($M_d$) and Majorana ($M_m$) mass terms for the LH and RH neutrinos will be of the form 
\bea \label{TypeISeesawEffective}
- \mathcal L_{Y}^\nu =   M_{d}^{ij} \, (\nu_L^i)^c  \nu_R^j + \frac{1}{2}M_{m}^{ij} \, \nu_R^i \, \nu_R^j + h.c. \, ,
\eea
where 
\beq
\label{Yuk}
M_{d} =  Y_\nu \, \frac{v}{\sqrt{2}} \,  \qquad   M_{m} = Y_N \, v' \sqrt{2} 
\eeq
inherit the flavor index structure from the corresponding Yukawa matrices. \\
To infer the order of the Yukawa couplings that this mechanism introduces, we set the values of the three light neutrinos $(\nu_i)$ masses $m_{\nu_i}$ down to the eV scale and consider the effects of having chosen $M_m$ around the TeV scale. This choice is justified, for instance, within supersymmetric scenarios \cite{Khalil:2007dr}, but finds one of its most interesting feature for being accessible at a typical LHC energy.
At the same time the value of $M_m$ results from the interplay of the Yukawa coupling $Y_{N}$ with the unknown value of the $U(1)'$ symmetry breaking scale $v'$ and, therefore, remains completely unset. We have reviewed in appendix B 
the derivation of the relation involving the light and heavy neutrino states in terms of $M_d$ and $M_m$ which is given by 
\bea \label{TypeISeesawMasses1}
M_{\nu_l} \simeq  - M_d^T M_m^{-1} M_d\, , \qquad M_{\nu_h} \simeq   M_m \, .
\eea
From the formula for the light neutrino mass in Eq.~(\ref{TypeISeesawMasses1}) it is instead possible to extract the size of the Yukawa coupling $Y_\nu$. Choosing the scale of $M_m$ around TeV, and considering $m_{\nu_l} \sim$ eV, the Yukawa $Y_\nu$ must be $\lesssim 10^{-6}$, a value which is too small to affect the evolution of the RG evolution. On the other hand, $Y_N$ could be even of $O(1)$ and, henceforth, it can play a significant role. \\
For the sake of simplicity, 
we consider diagonal Yukawa couplings which lead to diagonal neutrino mass matrices $M_{\nu_l}$ and $M_{\nu_h}$.
The neutrino spectrum generated by the type-I seesaw mechanism is now at hand presenting, for each generation, two kind of different states 
\bea
\label{nurotation}
\left( \begin{array}{c} \nu_{l_i} \\ \nu_{h_i}^c \end{array} \right) = 
\left( \begin{array}{cc} \cos \alpha_i & \sin \alpha_i \\ - \sin \alpha_i & \cos \alpha_i \end{array} \right)
\left( \begin{array}{c} \nu_{L_i} \\ \nu_{R_i}^c \end{array} \right) \qquad \mbox{with} \quad \tan 2\alpha_i = - 2 \frac{M_{d_i}}{M_{m_i}} \,.
\eea
We find that a light physical neutrino with mass in $m_{\nu_l}$ is a mixing of SM LH neutrino with a highly damped RH sterile part. In the opposite, the heavy counterpart is a linear combination of the SM sterile RH neutrino with a weak LH component. The damping being generated by the ratio between the Dirac mass $M_d$ and the big scale of the Majorana mass term $M_m$.

\section{The Higgs sector and spontaneous symmetry breaking}
We now turn to a description of the parameters of the extended Higgs sector.
This sector is characterized by the usual $SU(2)$ doublet $H$ and a SM singlet complex scalar $\chi$. With this field content the most general renormalizable scalar potential is given by
\bea
\label{pot}
V(H,\chi) = m_1^2 H^\dag H + m_2^2 \chi^\dag \chi + \lambda_1 (H^\dag H)^2 + \lambda_2 (\chi^\dag \chi)^2 + \lambda_3 (H^\dag H)(\chi^\dag \chi) \,
\eea
as a function of 5 parameters, two masses $m_1^2, m_2^2$ and of three quartic coupling $\lambda_i$.  
The stability of the potential is achieved by the following conditions
\bea
\label{stabilitycond}
\lambda_1 > 0\,, \quad \lambda_2 >0 \,, \quad 4 \lambda_1 \lambda_2 - \lambda_3^2 > 0 \,,
\eea
which are obtained by requiring the corresponding Hessian matrix to be positive definite at large field values. 
The spontaneous symmetry breaking pattern is obtained when the two scalar fields acquire vacuum expectation values (vev)
\bea
< H > = \frac{1}{\sqrt{2}} \left( \begin{tabular}{c} 0 \\ $v$ \end{tabular} \right) \,, \qquad <\chi> = \frac{v'}{\sqrt{2}}\,,
\eea
whose expressions, determined by the minimization conditions, are
\bea
v^2 = \frac{ m_2^2 \, \lambda_3/2 - m_1^2 \, \lambda_2 }{\lambda_1 \lambda_2 - \lambda_3^2 /4} \,, \qquad
v'^2 = \frac{ m_1^2 \, \lambda_3/2 - m_2^2 \, \lambda_1 }{\lambda_1 \lambda_2 - \lambda_3^2 /4} \,.
\label{vevs}
\eea
After spontaneous symmetry breaking, the mixing between the two scalar fields can be removed by an orthogonal transformation which rotates $H$ and $\chi$ into the two mass eigenstates $h_1$ and $h_2$
\bea
\left( \begin{array}{c} h_1 \\ h_2 \end{array} \right) = \left( \begin{array}{cc} \cos \theta & - \sin \theta \\  \sin \theta & \cos \theta \end{array} \right)  \left( \begin{array}{c} H  \\ \chi \end{array} \right)
\eea 
with $- \pi/2 < \theta < \pi/2$. The masses of the scalar eigenstates are
\bea
m_{h_{1,2}}^2 = \lambda_1 v^2 + \lambda_2 v'^2 \mp \sqrt{\left( \lambda_1 v^2 - \lambda_2 v'^2\right)^2 + \left( \lambda_3 v v' \right)^2} \,,
\eea
and one can easily derive the relations
\bea
\label{lambdas}
\lambda_1 &=& \frac{m_{h_1}^2}{4v^2}(1+\cos 2\theta) + \frac{m_{h_2}^2}{4 v^2}(1-\cos 2 \theta) \,, \nn \\
\lambda_2 &=& \frac{m_{h_1}^2}{4v'^2}(1-\cos 2\theta) + \frac{m_{h_2}^2}{4 v'^2}(1+\cos 2 \theta) \,, \nn \\
\lambda_3 &=& \sin 2 \theta \left( \frac{m_{h_2}^2 - m_{h_1}^2}{2 v v'}\right) \,,
\eea
which can be used to set the initial conditions on the scalar couplings through the physical masses $m_{h_{1,2}}$, the two vevs $v, v'$ and the mixing angle $\theta$. From Eq.~(\ref{lambdas}) one can immediately derive the relation
\bea
\tan 2 \theta = \frac{\lambda_3 v v'}{\lambda_1 v^2 - \lambda_2 v'^2} \,,
\label{theta}
\eea
which allows to express the mixing of the two scalars in terms of the three independent couplings of the potential and the two vevs $v, v'$. Notice that while the couplings $\lambda_i$ in Eq.~(\ref{lambdas}) can be taken as the independent parameters of the potential after imposing the vacuum conditions Eq.~(\ref{vevs}), Eq.~(\ref{theta}) is a trivial identity, derived from the former. 

After spontaneous symmetry breaking, one obtains for the $W^\pm$ mass the usual SM expression, $M_W = g_2 v/2$, while the masses of the neutral gauge bosons are obtained from the Lagrangian 
\bea
\frac{v^2}{8}(g_2 W^3_\mu - g B_\mu - \bar g B'_\mu)^2 + \frac{v'^2}{2} (g' z_\chi B'_\mu)^2,
\eea
where we have assumed a nonzero kinetic mixing of the two $B, B'$ gauge fields, as expressed in Eq.~(\ref{covder}).
We have also defined the new coupling $\bar g = \tilde g + 2 g' z_H$, which involves the kinetic mixing $\tilde{g}$ and the 
$U(1)'$ gauge coupling $g'$. $z_H$ and $z_\chi$ are the $U(1)'$ charges of the two scalar fields. In the case of 
a B-L symmetry, since $z_H=0$, then the new coupling  $\bar{g}$ coincides with the coupling of the kinetic mixing, $\bar{g}=\tilde{g}$. \\ 
The diagonalization of the mass matrix extracted from the previous expression gives the relations between mass and interaction eigenstates
\bea
\left( \begin{array}{c} B^\mu \\ W_3^\mu \\ B'^\mu \end{array} \right) = \left( \begin{array}{ccc} 
\cos \theta_w & - \sin \theta_w \cos \theta' & \sin \theta_w \sin \theta' \\
\sin \theta_w & \cos \theta_w \cos \theta' & - \cos \theta_w \sin \theta' \\
0 & \sin \theta'  & \cos \theta'  
\end{array} \right)
\left( \begin{array}{c} A^\mu \\ Z^\mu \\ Z'^\mu \end{array} \right)
\eea
where $\theta_w$ is the usual Weinberg angle of the SM, and $\theta'$ is a new mixing angle, with $-\pi/4 \le \theta' \le \pi/4$ defined as
\bea \label{ThetaPrime}
\tan 2 \theta' = \frac{2 \bar g \sqrt{g_2^2 + g^2}}{\bar g^2 + ( 2 z_\chi g' \, v'/v )^2 - g_2^2 - g^2} \,.
\eea
The relation above can be re-expressed in terms of the masses of the two gauge bosons.  
The masses of the $Z$ and $Z'$ gauge bosons are given by
\bea \label{ZZpMasses}
M_{Z, Z'} = \sqrt{g_2^2 + g^2} \frac{v}{2} \left[ \frac{1}{2} \left( \frac{\bar g^2 +  ( 2 z_\chi g' \, v'/v )^2 }{g_2^2 + g^2} + 1\right)  \mp \frac{\bar g}{\sin 2 \theta' \sqrt{g_2^2 + g^2}} \right]^{\frac{1}{2}} \,.
\eea
A bound on the mixing angle $\theta'$ has been obtained from the LEP experiment \cite{Abreu:1994ria} which constrains $\theta'$ to small values, namely $|\theta'| \lesssim 10^{-3}$. In this case the expressions for the gauge boson masses simplify considerably to
\bea \label{ZZpMassesSmallTh}
M_Z \simeq \frac{v}{2} \sqrt{g_2^2 + g^2}\,, \qquad M_{Z'} \simeq \frac{v}{2} \sqrt{\bar g^2 + (2 z_\chi g' \, v'/v)^2} \,,
\eea
and
\bea \label{ThetaPrimeMasses}
\theta' \simeq \bar g \frac{ M_{Z}^2 }{M_{Z'}^2-M_{Z}^2} \,
\eea
can be used in order to quickly grasp the dependence of the mixing angle $\theta'$ in terms of $M_{Z,Z'}$. \\
Notice that the kinetic mixing between the two $U(1)$ gauge fields can be taken to vanish, without loss of generality, at any particular scale, but it can be reintroduced by the RG evolution \cite{delAguila:1988jz,delAguila:1995rb,Chankowski:2006jk}. As already mentioned above, in the present analysis we have chosen a vanishing $\tilde g$ at the electroweak scale, which is one of the boundary conditions of our RG study. Also notice that, for $\tilde g = 0$, the tree-level mixing between the $Z'$ and the SM gauge bosons is absent only for $z_H = 0$, which occurs when the $U(1)'$ charges are those of $U(1)_{ B-L}$.

\subsection{The parameter choice}
We conclude this section with few comments about the physical parameters of the model, which we will be using in our 
numerical study in the sections below. The 
original 5 parameters of the potential $(m_1^2,m_2^2,\lambda_i)$, after imposing the vacuum conditions Eq.~(\ref{vevs}), can be replaced with the two physical masses of the Higgses, the two vevs $v, v'$, and 
the scalar mixing angle $\theta$. After eliminating $v$ and $m_{h_1}$  (respectively identified with the SM vev and Higgs mass, $v \simeq 246$ GeV, $m_{h_1} \simeq 125$ GeV) using the current LHC data, we are left with 
$m_{h_2}$, the mass of the extra Higgs, $\theta$ and $v'$ as extra parameters. We can re-express $v'$ in terms of the physical mass of the $Z'$ using Eq.~(\ref{ZZpMassesSmallTh}), which involves as relevant unknown parameter the gauge coupling $g'$ and the kinetic mixing $\tilde{g}$. Henceforth, the potential can be conveniently investigated in terms of the physical masses $M_{Z'}$ and $m_{h_2}$, and of the mixing angle $\theta$. For a given mass of the extra $Z'$, one can use the two couplings $(g',\tilde{g})$ as additional parameters, whose values are taken as boundary conditions at the electroweak scale. A typical study of the stability of the potential would then involve $(M_{Z'}, m_{h_2}, \theta)$, together with $(g',\tilde{g})$. The inclusion of the RH neutrinos renders this study more complex, since the evolution will couple the previous parameters to the mass of the RH neutrino $m_{\nu_h}$. As we are going to see, the requirement of stability of the potential through the running of its RG equations, will in general select specific regions of the $(m_{\nu_h}, m_{h_2})$ plane, for assigned values of the couplings, $M_{Z'}$ and $\theta$. \\
As we will discuss in the sections below, the variables $M_{Z'}/g'$, $\theta$ and $m_{h_2}$ will have to satisfy some  bounds from LEP (Eqs.~(\ref{STU1}) and (\ref{STUbound}) respectively) while $(g',\tilde{g})$ will be constrained by the requirements of perturbativity along the entire RG flow (Eq.~(\ref{pertconds})).\\
One final comment concerns the mass of the heavy RH neutrino, which is also directly related to the vev $v'$ as for 
$M_{Z'}$, as shown in Eqs.~(\ref{Yuk},\ref{TypeISeesawMasses1}). Therefore, an independent variation of $m_{\nu_h}$ respect to $M_{Z'}$, causes $Y_N$, the Yukawa of the RH neutrino, to vary. For a given mass of the extra $Z'$, with fixed couplings $(g',\tilde{g})$ at the electroweak scale, the variation of 
$m_{\nu_h}$ is entirely accounted for by a variation of $Y_N$.

\section{One-loop $\beta$-functions for a general charge assignment}
We are now in the position to
start a renormalization group analysis using a basis of diagonal kinetic terms and with Eq.~(\ref{NonDiagCovDerivative}) as the minimal form of the covariant derivative. 
Consequently, the contribution coming from the abelian sector to the couplings include  $g$, $g'$ and $\tilde{g}$ as given by Eq.~(\ref{orthogonal}), the latter accounting for
the mixing. We will simplify the structure of the SM Yukawa sector by retaining only the contribution coming from the top quark 
($Y_u^{3,3} = Y_t$), thereby omitting the Yukawas of the 3 light neutrinos. On the other hand, we have taken the Yukawas of the heavy neutrinos to be universal (generation independent) $Y_N^{i,j}  = Y_N \delta^{i,j}$. \\
We present, in this section, the one-loop $\beta$-functions for the dimensionless parameters for a generic charge assignment. The corresponding two-loop corrections can be found in Appendix \ref{betafuncs}.\\
We start from the rotated gauge coupling sector which, as already said, contain no trace of $g_{21}$, if the $\beta$-functions undergo the same rotation given in Eq.~(\ref{explicitRot}).
At one-loop order, the evolution of the gauge couplings is defined by the $\beta$ functions 
\bea \label{abelianBeta}
\beta^{(1)}_{{g}} &=& \frac{41 {g}^3}{6}\,,\quad
\beta^{(1)}_{{g_2}} = -\frac{19 {g_2}^3}{6}\,,\quad
\beta^{(1)}_{{g_3}} = -7 {g_3}^3 \,,\quad \nn \\
\beta^{(1)}_{{g'}} &=& {g'}^2 \left(\frac{46 \tilde{g} z_Q}{3}+\frac{50 \tilde{g} z_u}{3}\right)+\frac{41 {g'} \tilde{g}^2}{6}+{g'}^3
   \left(-44 z_Q z_u+134 z_Q^2+18 z_u^2\right)     ,
\eea
coupled with the $\beta$-function accounting for the mixing between the two $U(1)$ factors
\bea \label{mixBeta}
\beta^{(1)}_{{\tilde{g}}} &=& \tilde{g} \left(\frac{41 g^2}{3}-44 {g'}^2 z_Q z_u+134 {g'}^2 z_Q^2+18 {g'}^2 z_u^2\right)+\tilde{g}^2
   \left(\frac{46 {g'} z_Q}{3}+\frac{50 {g'} z_u}{3}\right)\nn \\   &+&\frac{41 \tilde{g}^3}{6}+\frac{46}{3} g^2 {g'}
   z_Q+\frac{50}{3} g^2 {g'} z_u\, .\eea  
It is clear from Eq.~(\ref{mixBeta}) that even imposing  a zero starting value for $\tilde{g}$, an abelian mixing can be radiatively generated by the evolution. \\ 
The Yukawa sector is described by the evolution of just two real terms 
\bea
\beta^{(1)}_{Y_t} &=& Y_t \left(-{g'} \tilde{g} z_Q-4 {g'} \tilde{g} z_u-\frac{17 \tilde{g}^2}{12}-\frac{17 g^2}{12}-\frac{9 g_2^2}{4}-8
   g_3^2-3 {g'}^2 z_Q^2-3 {g'}^2 z_u^2\right)+\frac{9 Y_t^3}{2} \nn \\
   \beta^{(1)}_{Y_N} &=& Y_N \left(48 {g'}^2 z_Q z_u-96 {g'}^2 z_Q^2-6 {g'}^2 z_u^2\right)+10 Y_N^3  \,,   
\eea
and together with the evolution of the scalar quartic parameters 
\bea \label{betal1}
\beta^{(1)}_{{\lambda_{1}}} &=& \lambda _1 \left(12 {g'} \tilde{g} z_Q-12 {g'} \tilde{g} z_u-3 \tilde{g}^2-3 g^2-9 g_2^2+24 {g'}^2 z_Q z_u-12
   {g'}^2 z_Q^2-12 {g'}^2 z_u^2+12 Y_t^2\right) \nn \\ && -3 g^2 {g'} \tilde{g} z_Q+3 g^2 {g'} \tilde{g}
   z_u+\frac{3}{4} g^2 \tilde{g}^2-36 {g'}^3 \tilde{g} z_Q z_u^2+36 {g'}^3 \tilde{g} z_Q^2 z_u-12 {g'}^3
   \tilde{g} z_Q^3+12 {g'}^3 \tilde{g} z_u^3   \nn \\
   &&  -18 {g'}^2 \tilde{g}^2 z_Q z_u+9 {g'}^2 \tilde{g}^2 z_Q^2+9
   {g'}^2 \tilde{g}^2 z_u^2-3 {g'} \tilde{g}^3 z_Q-3 g_2^2 {g'} \tilde{g} z_Q+3 {g'} \tilde{g}^3 z_u+3 g_2^2
   {g'} \tilde{g} z_u \nn \\ 
   &&+\frac{3 \tilde{g}^4}{8}+\frac{3}{4} g_2^2 \tilde{g}^2+\frac{3 g^4}{8}-6 g^2 {g'}^2 z_Q z_u+3 g^2
   {g'}^2 z_Q^2+3 g^2 {g'}^2 z_u^2+\frac{3}{4} g_2^2 g^2-6 g_2^2 {g'}^2 z_Q z_u \nn \\ 
   && +3 g_2^2 {g'}^2 z_Q^2  +3 g_2^2
   {g'}^2 z_u^2+\frac{9 g_2^4}{8}-24 {g'}^4 z_Q z_u^3+36 {g'}^4 z_Q^2 z_u^2-24 {g'}^4 z_Q^3 z_u+6
   {g'}^4 z_Q^4\nn \\ 
   &&+6 {g'}^4 z_u^4+24 \lambda _1^2+\lambda _3^2-6 Y_t^4\,,
\eea
\bea \label{betal2}
\beta^{(1)}_{{\lambda_{2}}} &&= -1536 {g'}^4 z_Q z_u^3+9216 {g'}^4 z_Q^2 z_u^2-24576 {g'}^4 z_Q^3 z_u+24576 {g'}^4 z_Q^4+96 {g'}^4
   z_u^4 \nn \\ &&+\lambda _2 \left(384 {g'}^2 z_Q z_u-768 {g'}^2 z_Q^2-48 {g'}^2 z_u^2+24 Y_N^2\right)+20 \lambda _2^2+2
   \lambda _3^2-48 Y_N^4\,,
\eea
\bea
\beta^{(1)}_{{\lambda_{3}}} &&=\lambda _3 \left(6 {g'} \tilde{g} z_Q-6 {g'} \tilde{g} z_u-\frac{3 \tilde{g}^2}{2}-\frac{3 g^2}{2}-\frac{9
   g_2^2}{2}+204 {g'}^2 z_Q z_u-390 {g'}^2 z_Q^2-30 {g'}^2 z_u^2 \right. \nn \\ && \left. +12 \lambda _1+8 \lambda _2+12 Y_N^2+6
   Y_t^2\right)-432 {g'}^3 \tilde{g} z_Q z_u^2+1152 {g'}^3 \tilde{g} z_Q^2 z_u-768 {g'}^3 \tilde{g} z_Q^3+48
   {g'}^3 \tilde{g} z_u^3  \nn \\ &&  -96 {g'}^2 \tilde{g}^2 z_Q z_u+192 {g'}^2 \tilde{g}^2 z_Q^2+12 {g'}^2 \tilde{g}^2
   z_u^2-480 {g'}^4 z_Q z_u^3+1584 {g'}^4 z_Q^2 z_u^2 \nn \\&&-1920 {g'}^4 z_Q^3 z_u+768 {g'}^4 z_Q^4+48 {g'}^4
   z_u^4+4 \lambda _3^2 \,,
\eea
provide a closed system of differential equations to evolve the scalar potential towards any final scale. 
%%%%%%%%%%%%%%%%%%%%%

\section{The matching conditions at the electroweak scale}
\label{MatCondSection}
Before coming to a discussion of the main phenomenological implications of our analysis, we pause for a moment, discussing a point which is essential in order to secure the consistency of the evolution, which concerns the determination of the initial (matching) conditions.\\
We recall that the RGE's employed in this work have been computed in the $\MS$ renormalization framework at two-loop order.
These equations must be supplemented with suitable one-loop boundary conditions defined in the same scheme. These consist of $\MS$ renormalized couplings and masses evaluated at a given energy scale which correspond to the starting scale of the RG running.\\
In general, the initial conditions can be unknown free parameters introduced by the specific model which is under investigation, and are directly associated to some measured observables. In order to determine the latter, the $\MS$ parameters at the starting scale must be related to these physical observables. This task can be accomplished in two different ways: 1) one can adopt the $\MS$ renormalization from the very beginning and obtain the needed $\MS$ parameters directly from a set of measured observables or 2) use a scheme, as the on-shell ($\textrm{OS}$) one largely used in the electroweak theory, in which the renormalized parameters are expressed in terms of the physical quantities, the pole masses and the Fermi constant, and then translate the on-shell parameters to the corresponding $\MS$ expressions through appropriate matching conditions. In this work we adopt the second strategy which is quite common in the literature on the perturbative corrections in the SM. \\
The matching conditions are easily extracted from the obvious relation
\bea
\label{match}
\alpha_0 = \alpha_{\smallOS} + \delta \alpha_{\smallOS} = \alpha_{\smallMS}(\mu) + \delta \alpha_{\smallMS},
\eea  
where $\alpha_0$, $\alpha_{\smallOS}$ and $\alpha_{\smallMS}$ denote, respectively, the bare, the on-shell and the $\MS$ expressions of a generic parameter $\alpha$. 
From Eq.~(\ref{match}) one can extract a $\MS$ parameter in terms of its on-shell version obtaining, at one-loop order,
\bea
\label{OStoMS}
\alpha_{\smallMS} = \alpha_{\smallOS} +  \delta \alpha_{\smallOS} - \delta \alpha_{\smallMS} =  \alpha_{\smallOS} +  \delta \alpha_{\smallOS}|_{finite} 
\eea
where the last expression is simply a consequence of the definition of the $\MS$ renormalization scheme, in which the counterterms only subtract the UV singular parts. It is clear from Eq.~(\ref{OStoMS}) that the matching conditions between the OS and the $\MS$ schemes are defined from the finite part of the OS counterterm. Notice that, at tree level, the $\MS$ parameters coincide with their OS version.\\
In our analysis we have adopted a mixed renormalization procedure in which the known SM parameters are renormalized in the on-shell scheme, while the $\MS$ is used for the additional couplings and masses introduced by the extended gauge and neutrino sectors and for the vacuum expectation value $v'$ of the extra scalar. On the other hand, for the sake of simplicity, all the remaining parameters in the scalar sector, containing both the Higgs-like quartic coupling and the two new quartic interactions, are renormalized in the on-shell scheme.
Such a mixed scheme is not unusual in QFTs as it is already employed, for instance, in the SM, and in particular for the computation of the strong corrections to electroweak observables. \\
As shown in \cite{Chankowski:2006jk}, one of the interesting features of this hybrid renormalization scheme, besides its simplicity, is that the Appelquist-Carrazzone decoupling theorem is explicit manifest. Indeed, the SM limit of measurable quantities is straightforwardly obtained for $v' \rightarrow \infty$, with $v'$ defined in the $\MS$ scheme. This has been the main motivation for our renormalization setup.

\begin{figure}[H]
\centering
\subfigure[]{\includegraphics[scale=0.45]{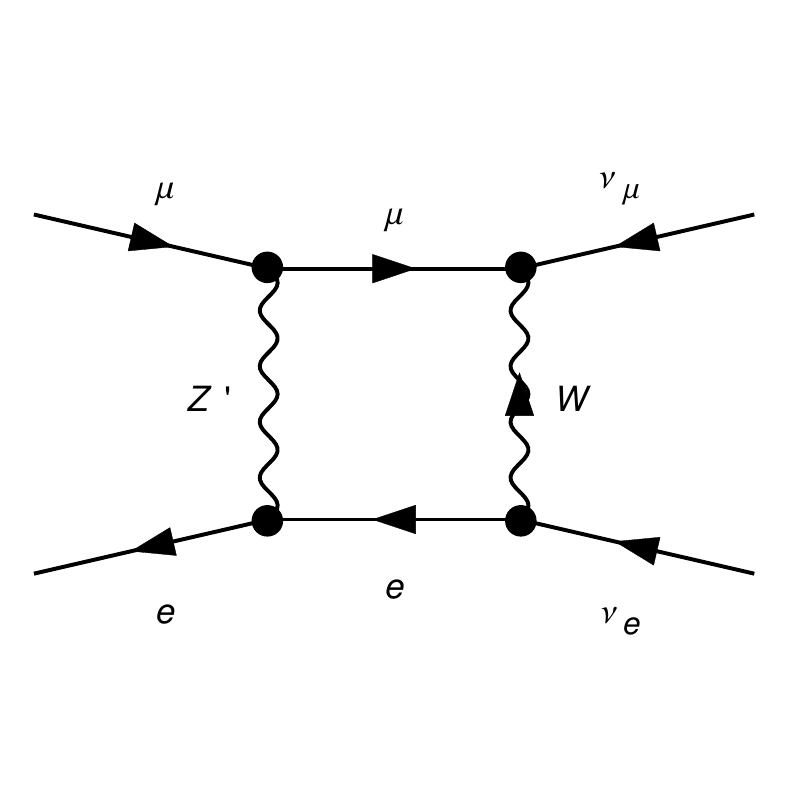}} 
\subfigure[]{\includegraphics[scale=0.45]{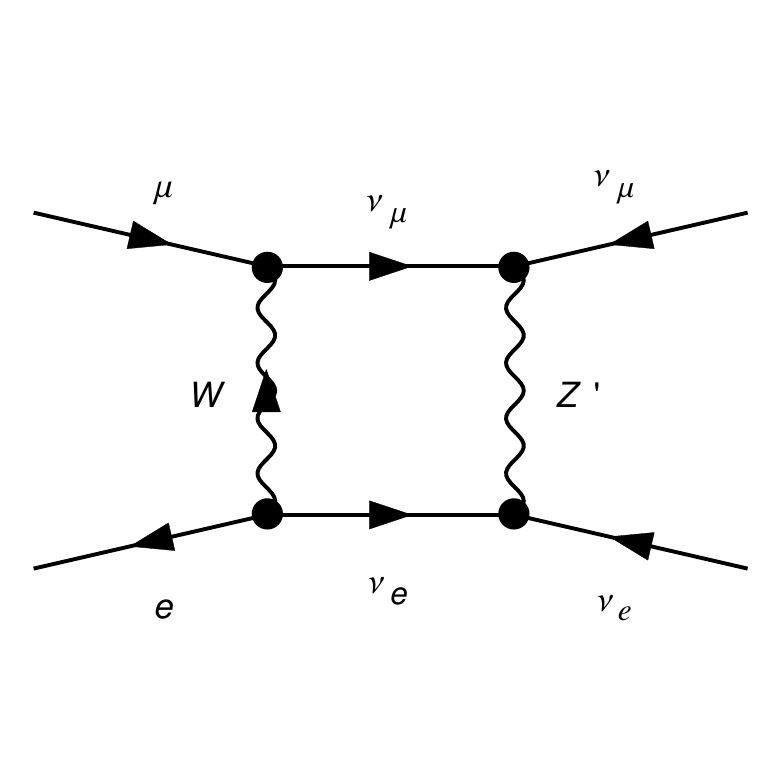}} 
\subfigure[]{\includegraphics[scale=0.45]{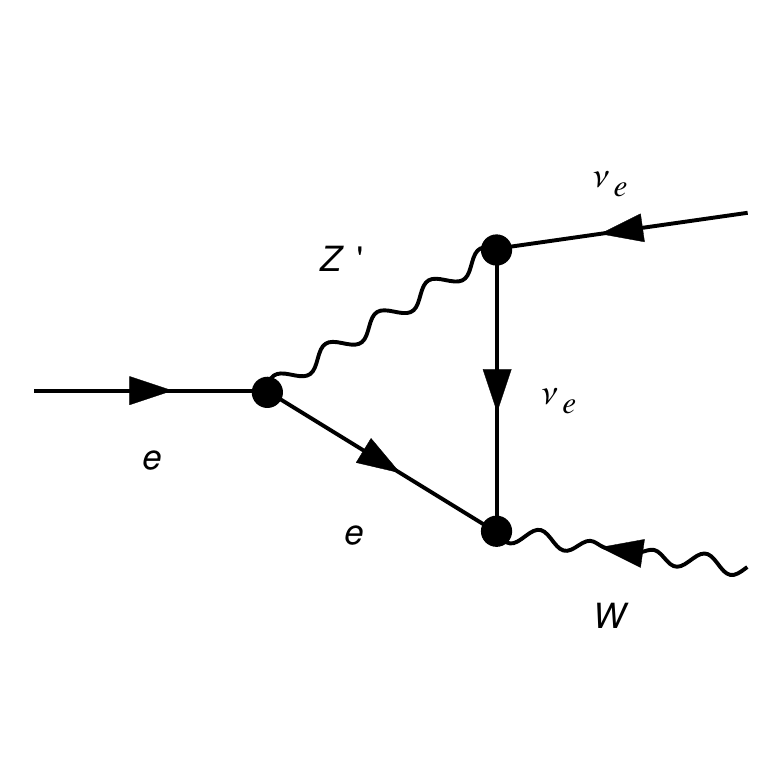}} 
\subfigure[]{\includegraphics[scale=0.45]{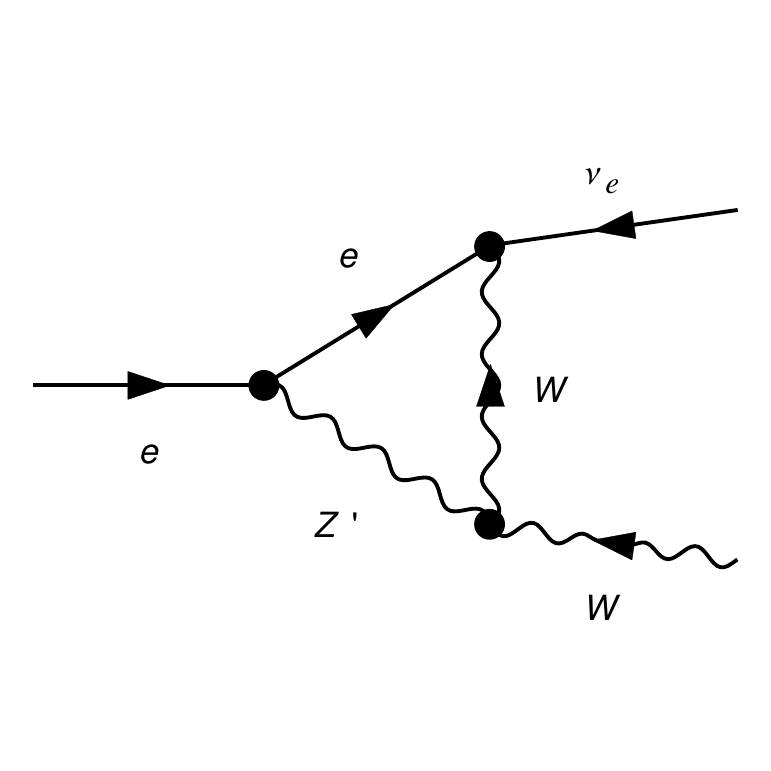}}\\
\subfigure[]{\includegraphics[scale=0.45]{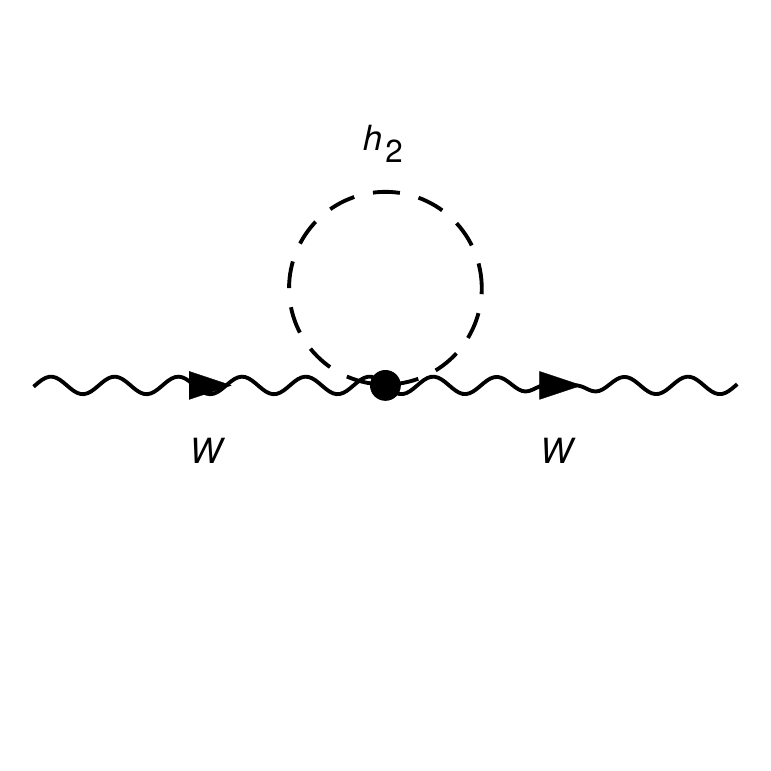}} 
\subfigure[]{\includegraphics[scale=0.45]{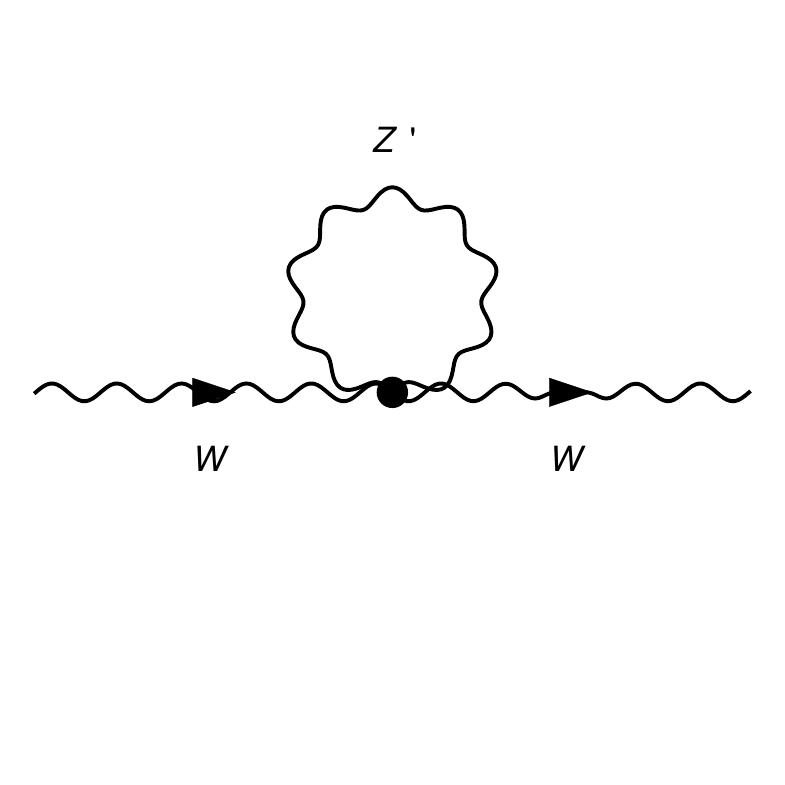}} 
\subfigure[]{\includegraphics[scale=0.45]{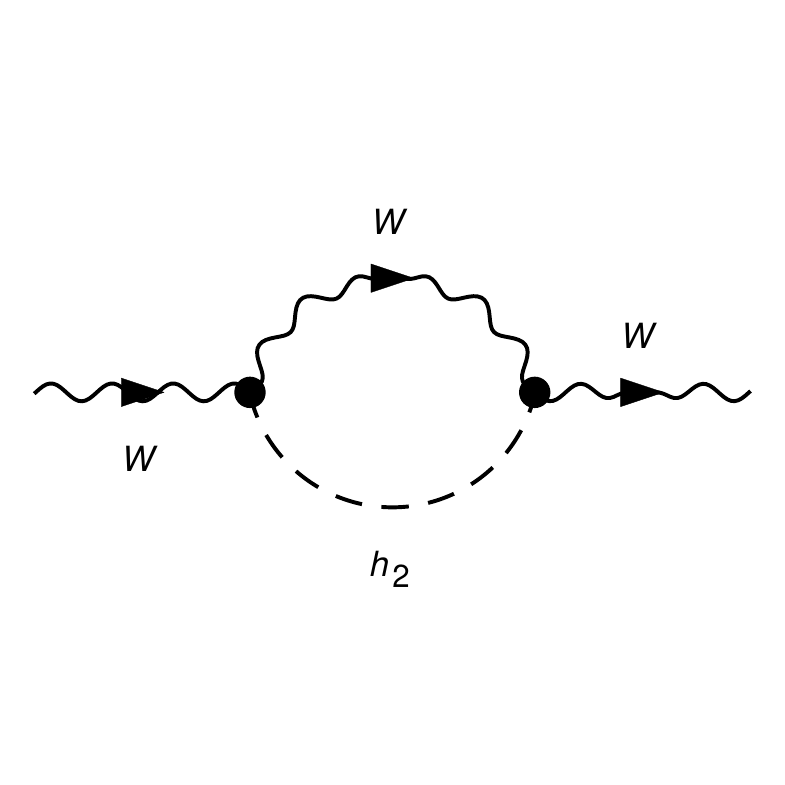}} 
\subfigure[]{\includegraphics[scale=0.45]{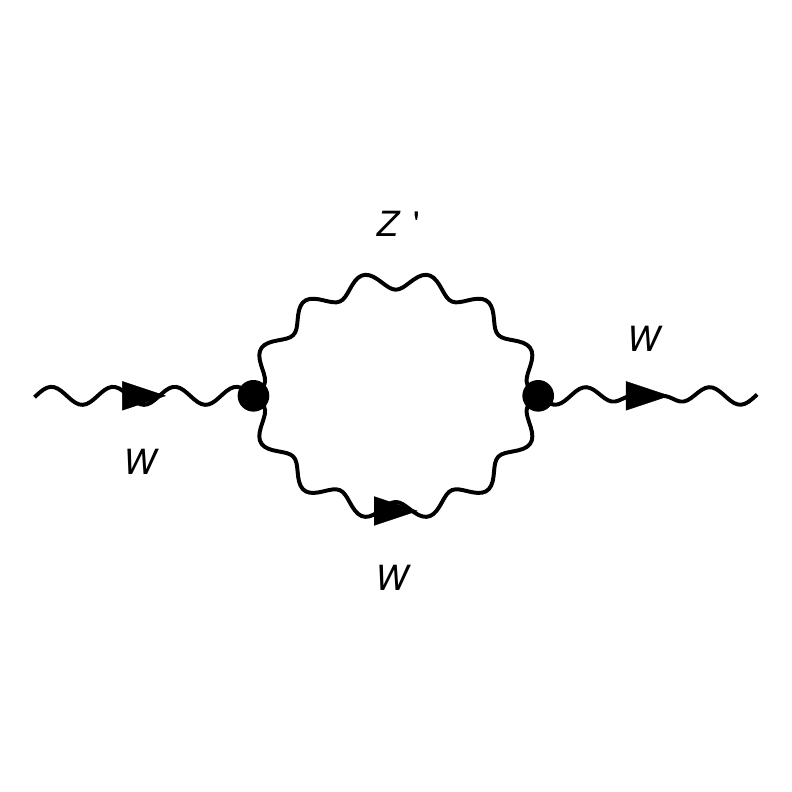}}\\
\subfigure[]{\includegraphics[scale=0.45]{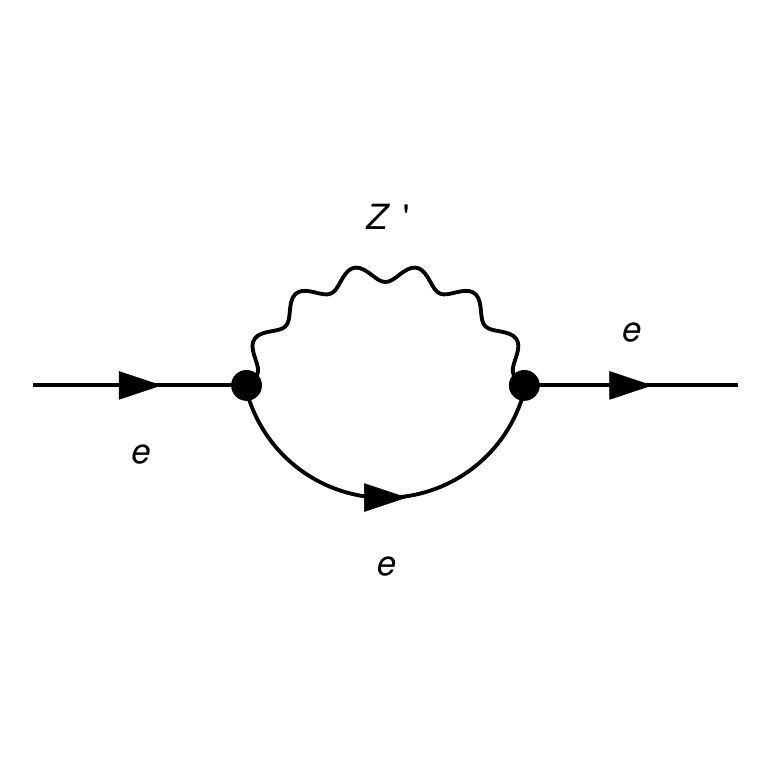}} 
\subfigure[]{\includegraphics[scale=0.45]{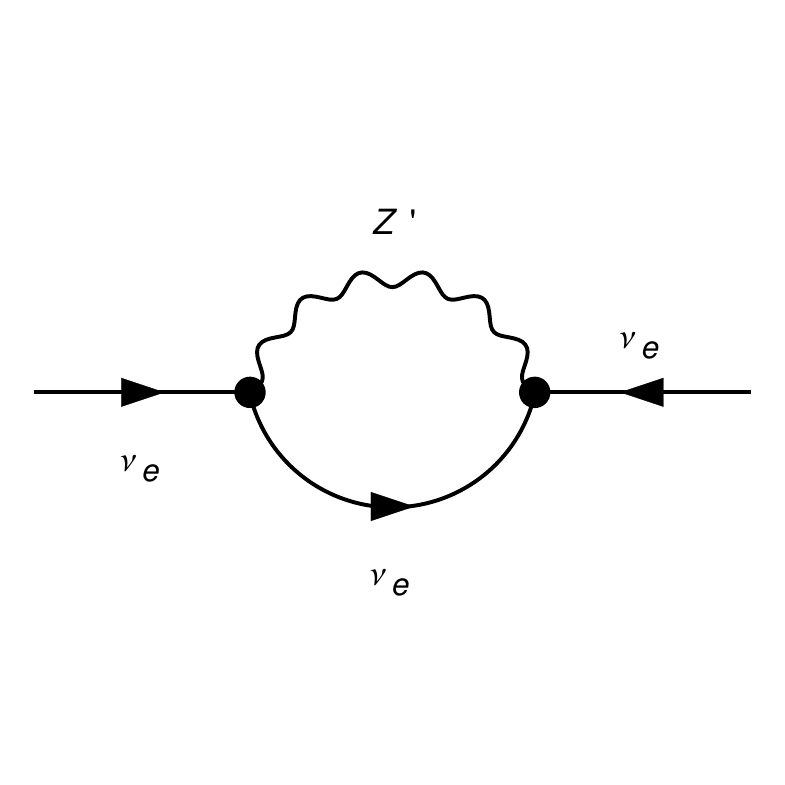}}  
\caption{ Some of the new-physics diagrams appearing in the one-loop perturbative expansion of the $\mu$ decay. These define the radiative corrections to $G_F$. In particular, the diagrams in Figs.(a),(b) enter in the computation of $B$, those in Figs.(c),(d) define $V$, Figs.(e)-(h) enter in the calculation of the $W$ self-energy and Figs.(i)-(j) in the external leg corrections $E$.  \label{Fig.PertGF}}
\end{figure}

\begin{figure}[H]
\centering
\subfigure[]{\includegraphics[scale=0.45]{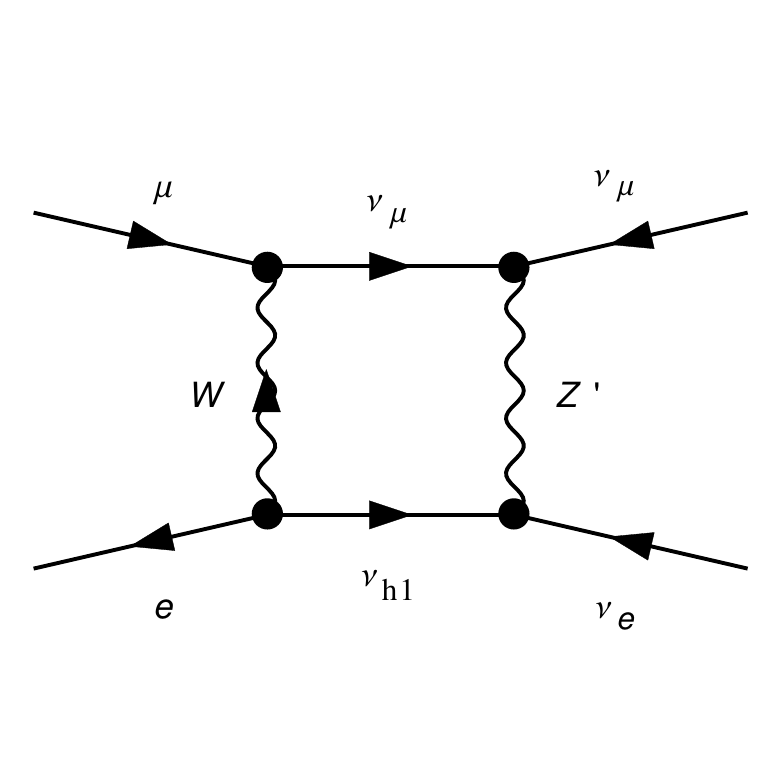}} 
\subfigure[]{\includegraphics[scale=0.45]{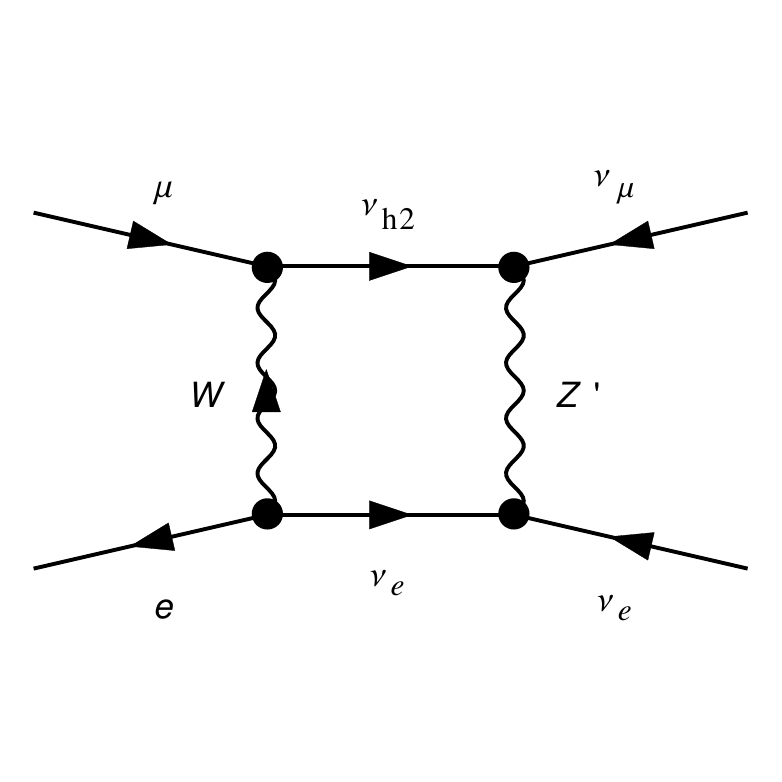}} 
\caption{Some of the one-loop box diagrams defining the $\mu$ decay proportional to the neutrino mixing angle. \label{Fig.PertGF2}}
\end{figure}

The SM-like parameters which enter into this RG study are the quartic coupling $\lambda_1$ of the $SU(2)$ doublet $H$, the Yukawa of the top quark $Y_t$ and the gauge coupling constants $g$, $g_2$ and $g_3$. These are computed in terms of the pole masses of the Higgs $m_{h_1}$, of the top $M_t$, of the weak gauge bosons $M_Z$ and $M_W$, and of the Fermi constant $G_F$. All these quantities are then translated in the $\MS$ scheme using Eq.~(\ref{OStoMS}). Notice that for the $SU(3)$ strong coupling constant $g_3$ there is no need to introduce matching conditions because it is directly extracted in the $\MS$ renormalization framework as $\alpha_3(M_Z)$.\\
On the other hand, the other unknown dimensionless parameters introduced by the $U(1)'$ extension, with the only exception of the those of the scalar potential, are directly employed in the $\MS$ scheme from the very beginning (or, equivalently, are matched to their counterparts at tree level). These are the abelian gauge coupling constants, $g'$ and $\tilde g$, and the Yukawa of the right-handed neutrinos $Y_N$. Instead, the quartic couplings $\lambda_2$ and $\lambda_3$ are matched at one-loop from the on-shell physical mass of the heavy scalar $m_{h_2}$ and from the the mixing angle $\theta$. \\
One of the most important SM parameters needed in the determination of the initial conditions of our RG study is the Fermi constant $G_F$.
Using its definition in the effective Fermi theory
\bea
\frac{G_F}{\sqrt{2}} = \frac{g_0^2}{8 M_{W, 0}^2} (1 + \Delta r_0) = \frac{1}{2 v_0^2}(1 + \Delta r_0) \,,
\eea 
we obtain the counterterm of the vev $v$ in the on-shell scheme $\delta v^2_{\smallOS} = \Delta r_0/(\sqrt{2} G_F)$. We recall that $G_F$ is extracted from the muon lifetime, computed in the Fermi theory augmented by QED corrections. As a consequence, the computation of the $\Delta r_0$ electroweak corrections to the $\mu$ decay requires the subtraction of the pure QED contributions. 
At one-loop order $\Delta r_0$ can be decomposed as 
\bea
\Delta r_0 = V - \frac{\Pi_{WW}}{M_W^2} + \frac{\sqrt{2}}{G_F} B + E \,,
\eea
where $V$ and $B$ denote vertex and box corrections, $\Pi_{WW}$ is the $W$ boson self-energy evaluated at zero momentum and $E$ corresponds to the wave-functions contributions. All of them are computed at zero external momenta and are affected by SM-like corrections as well as new-physics effects. Notice also that we have chosen a renormalization prescription in which the tadpoles are included in the perturbative expansion. This property has the advantage to provide a gauge-independent definition of the mass counterterms and of $\Delta r_0$. Nevertheless, the dimensionless parameters appearing in the Lagrangian, which are the interesting ones for our analysis, are not affected by this choice and the results are independent of the tadpole corrections. In Fig.\ref{Fig.PertGF} we show some of the one-loop perturbative contributions to the $\mu$ decay which are needed in the computation of the matching condition of $G_F$.  The corrections proportional to the neutrino mixing angle $\alpha_i$, as the ones depicted in Fig.\ref{Fig.PertGF2}, have been neglected due to the smallness of the ratio between the Dirac and the Majorana masses. 
The counterterms of the top Yukawa and of the $SU(2)$ gauge couplings, needed in the matching procedure, are obtained exploiting the relations
\bea
M_t = Y_{t \, \smallOS} \frac{v_{\smallOS}}{\sqrt{2}}\,, \qquad M_W^2 = g_{2\,\smallOS}^2 \frac{v_{\smallOS}^2}{4} \,,
\eea
which preserve their SM-like form, and read as 
\bea
\delta Y_{t \, \smallOS} = Y_{t \, \smallOS} \left( \frac{\delta M_t}{M_t} - \frac{\delta v_{\smallOS}}{v_{\smallOS}} \right)\,, \qquad
\delta g_{2 \, \smallOS} = g_{2 \, \smallOS}  \left( \frac{\delta M_W^2}{2 M_W^2} - \frac{\delta v_{\smallOS}}{v_{\smallOS}} \right) \,.
\eea
The determination of counterterms of the abelian gauge coupling $g$ deserves a special attention because its defining relation gets modified in the $U(1)'$ extension with respect to the SM case as one can see from Eq.~(\ref{ZZpMasses}). Nevertheless, the departure from the SM expression, parametrized by the mixing angle $\theta'$, is very small and Eq.~(\ref{ZZpMassesSmallTh}) can be safely employed leading to
\bea
\delta g_{\smallOS} = g_{\smallOS} \left( \frac{1}{2} \frac{\delta M_Z^2 - \delta M_W^2}{M_Z^2- M_W^2} - \frac{\delta v_{\smallOS}}{v_{\smallOS}} \right) \,.
\eea
In the previous equation $\delta v_{\smallOS}$ is obtained from $\Delta r_0$ as explained above, while the top, the $W$ and the $Z$ boson mass counterterms in the on-shell renormalization scheme are computed from the corresponding self-energies
\bea
\delta M_t = \RE \, \Sigma_{t}(\psl = M_t)\,, \quad \delta M_W^2 = \RE \, \Pi_{WW}(p^2 = M_W^2)\,, \quad \delta M_Z^2 = \RE \, \Pi_{ZZ}(p^2 = M_Z^2) \,.
\eea
The remaining two abelian coupling constants, $g'$ and $\tilde g$, could be related, in principle, to the pole mass of the $Z'$ and to the OS expression of the mixing angle $\theta'$ through Eq.~(\ref{ThetaPrime}) and (\ref{ZZpMasses}). Nevertheless, being these two free parameters of the $U(1)'$ extension, the use of the matching conditions in this case is not mandatory and we can work directly with their $\MS$ expression. Therefore, in all of the following analyses, $g'$ and $\tilde g$ should always be understood as computed within the $\MS$ . \\   
The major differences in the matching relations, with respect to the SM case, are found in the scalar sector due to the presence of new quartic interactions. Using the defining equations in Eq.~(\ref{lambdas}) we obtain 
\bea
\delta \lambda_{1 \, \smallOS} &=& \frac{m_{h_1}^2}{4 v_{\smallOS}^2}(1+ \cos 2 \theta_{\smallOS}) \left( \frac{\delta m_{h_1}^2}{m_{h_1}^2} - 2 \frac{\delta v_{\smallOS}}{v_{\smallOS}} - 2 \delta \theta_{\smallOS} \tan \theta_{\smallOS} \right) \nn \\
&+&
\frac{m_{h_2}^2}{4 v_{\smallOS}^2}(1- \cos 2 \theta_{\smallOS}) \left( \frac{\delta m_{h_2}^2}{m_{h_2}^2} - 2 \frac{\delta v_{\smallOS}}{v_{\smallOS}} + 2 \delta \theta_{\smallOS} \cot \theta_{\smallOS} \right) \,, \nn \\
\delta \lambda_{2 \, \smallOS} &=& \frac{m_{h_1}^2}{4 v_{\smallOS}'^2}(1- \cos 2 \theta_{\smallOS}) \left( \frac{\delta m_{h_1}^2}{m_{h_1}^2} 
+ 2 \delta \theta_{\smallOS} \cot \theta_{\smallOS} \right) \nn \\
&+&
\frac{m_{h_2}^2}{4 v_{\smallOS}'^2}(1 + \cos 2 \theta_{\smallOS}) \left( \frac{\delta m_{h_2}^2}{m_{h_2}^2} 
- 2 \delta \theta_{\smallOS} \tan \theta_{\smallOS} \right) \,, \nn \\
\delta \lambda_{3 \, \smallOS} &=& \lambda_{3 \, \smallOS} \left( \frac{\delta m_{h_2}^2 - \delta m_{h_1}^2}{m_{h_2}^2 - m_{h_1}^2} - \frac{\delta v_{\smallOS}}{v_{\smallOS}} 
+ 2 \delta \theta_{\smallOS} \cot 2 \theta_{\smallOS}\right) \,,
\eea
where 
\bea
\delta m_{h_i}^2 = \RE \, \Pi_{h_i h_i}(p^2 = m_{h_i}^2) \,, \qquad \delta \theta_{\smallOS} = \frac{\RE \, \Pi_{h_1 h_2}(p^2 = m_{h_1}^2)}{m_{h_1}^2 - m_{h_2}^2} \,.
\eea
Notice that, in the last of the previous equations, the mixed scalar self-energy $\Pi_{h_1 h_2}$ has been evaluated at $p^2 = m_{h_1}^2$. This choice enforces the absence of mixing between the two tree-level mass eigenstates $h_1, h_2$ also at one-loop level and at a particular scale, given by $m_{h_1}$.

\section{Two-loop numerical analysis}

The allowed parameter space of the model can be strongly restricted if we require the vacuum stability and the perturbativity 
up to a given scale $ Q $. This may be chosen to coincide with the GUT, Planck or any other scale where new physics scenarios are expected to appear. Obviously, these conditions are often not sufficient to select a bounded region of the same space, and new assumptions must be added, for instance in the form of suitable benchmark points where to sharpen the physics predictions.\\
Significant bounds are derived 
mainly from previous LEP-II analysis \cite{Abreu:1994ria}, which limit the gauge mixing angle $|\theta'| \lesssim 10^{-3}$, and force us to consider only the
sector for which the condition  
\bea
M_{Z'}/g' \geq 7\, \, \TeV \,, 
\label{STU1}
\eea
holds \cite{Cacciapaglia:2006pk}. Besides, the addition of a new heavy scalar in the spectrum modifies the coupling of the light Higgs to the SM particles by a factor of $\cos\theta$. Therefore
the electroweak precision measurements, through the $S,T,U$ parameters \cite{Dawson:2009yx}, may be used to constrain the scalar mixing angle together with 
the heavy Higgs mass leading to the bound
\bea \label{STUbound}
\theta  \lesssim 0.44\,\, , \, m_{h_2} \ge 500 \, \GeV  
\eea
for $m_{h_1} = 125$ GeV.

\subsection{Weak coupling evolution for a general $U(1)'$ charge assignment}
%
%
%.   
%
\begin{figure}[H]
\centering
\subfigure[]{\includegraphics[scale=0.407]{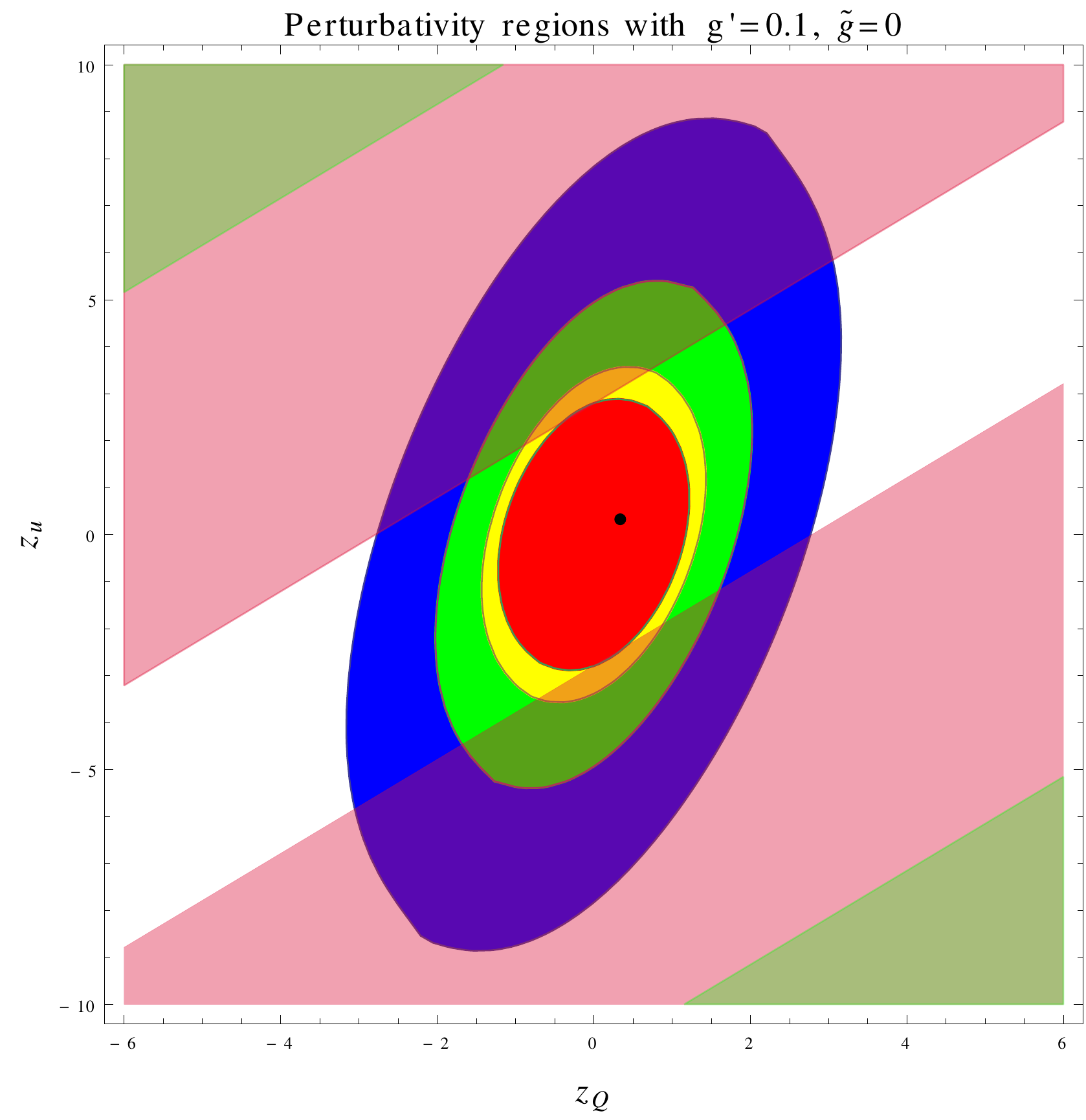}}
\subfigure[]{\includegraphics[scale=0.401]{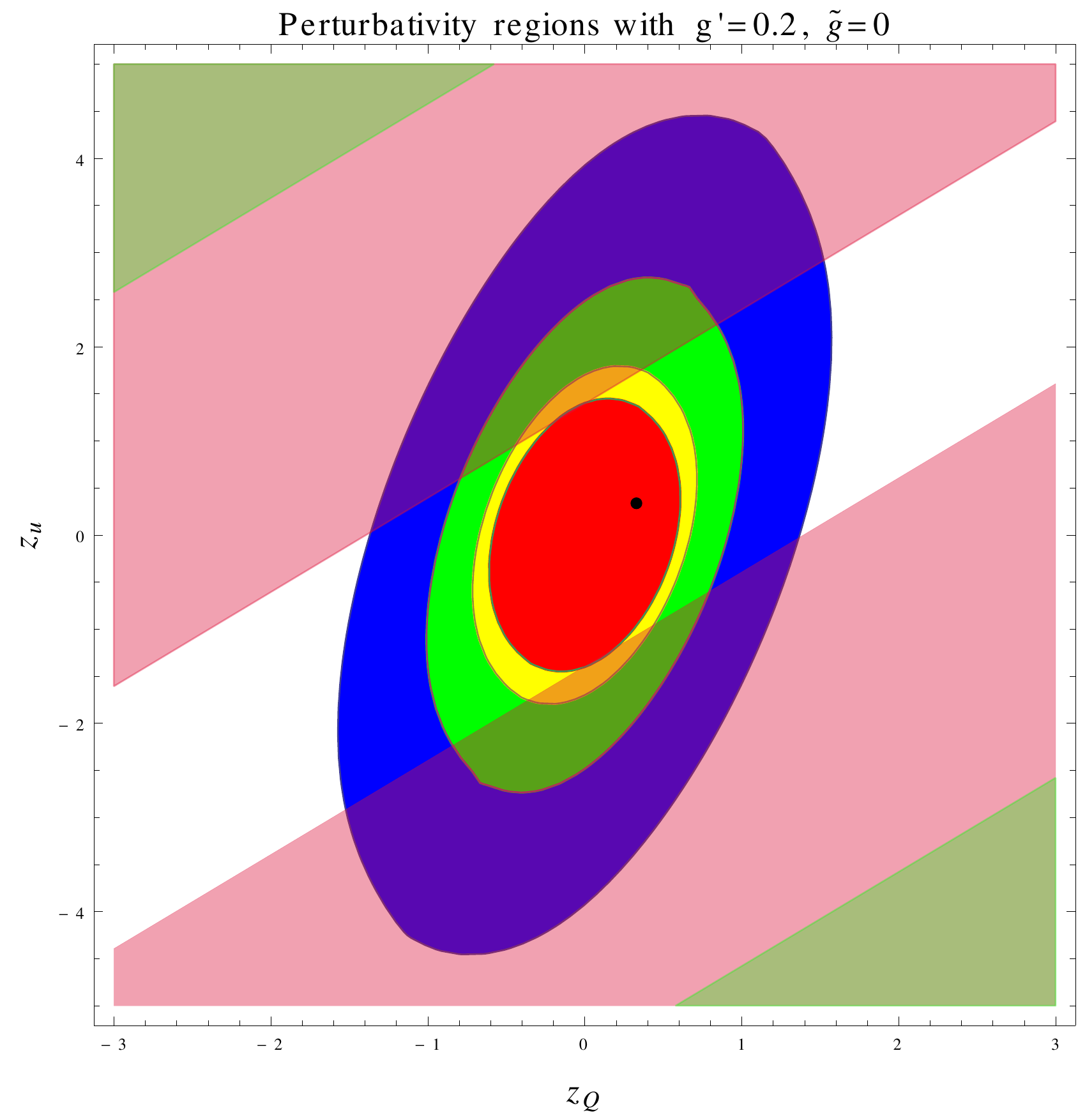}}
\subfigure[]{\includegraphics[scale=0.4]{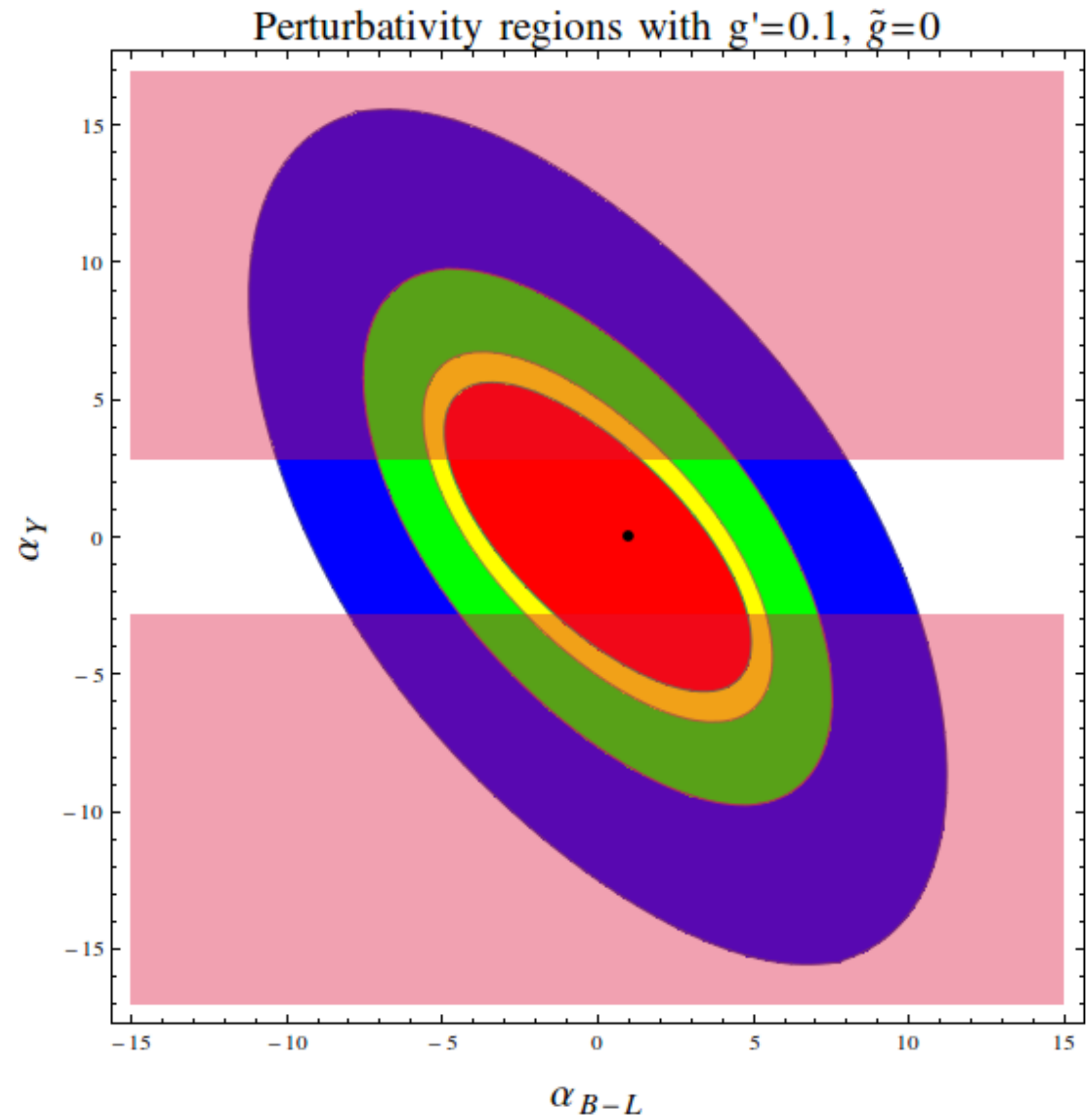}}
\caption{Values of the free $U(1)'$ charges $z_Q$ and $z_u$ for which the perturbativity constraint is satisfied up to $10^5$ GeV
(blue region), $10^9$ GeV (green region), $10^{15}$ GeV (yellow region) and $10^{19}$ GeV (red region) for $g' = 0.1$ (a) and $g' = 0.2$ (b). In (c) it is shown the 
same study for the charges $\alpha_Y \,=\,2\, z_u - 2\, z_Q$ and $\alpha_{B-L} \,=\,4\, z_Q - z_u$ and $g = 0.1$. The shadowed areas are excluded regions corresponding to $M_{Z'}=2.5$ TeV (pink shadows) and $M_{Z'}=5$ TeV (green shadows) respectively. The black thick dot indicates the B-L charge assignment.
\label{Fig.PertCoupGeneric} }
\end{figure}
\begin{figure}[H]
\centering
\includegraphics[scale=0.5]{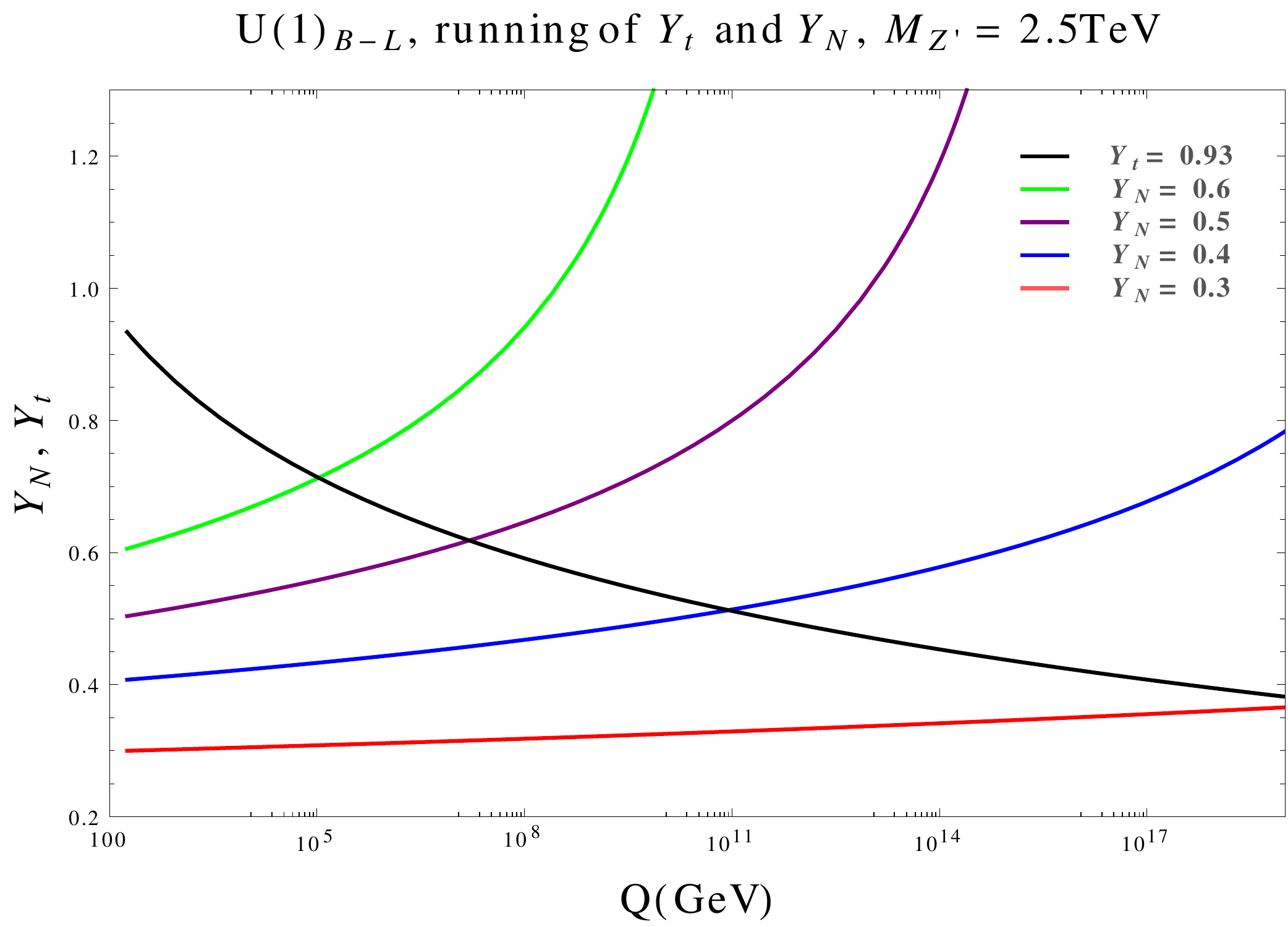}
\caption{Evolutions of the Yukawa coupling of the RH neutrinos for different initial values at the electroweak scale. The running of the top quark $Y_t$ is shown for comparison. 
\label{Fig.YNYT} }
\end{figure}
\begin{figure}[H]
\centering
\subfigure[]{\includegraphics[scale=0.325]{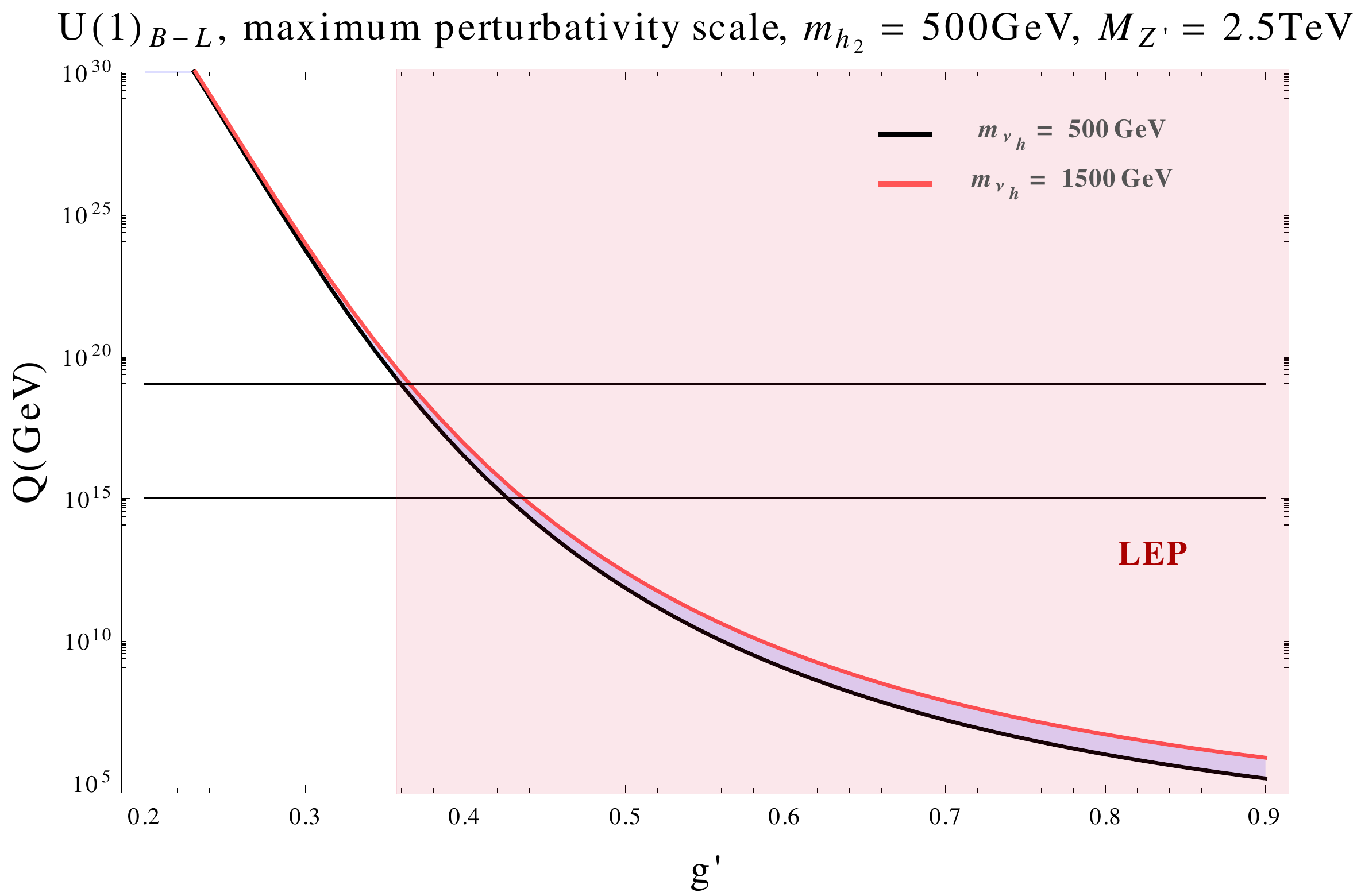}}
\subfigure[]{\includegraphics[scale=0.32]{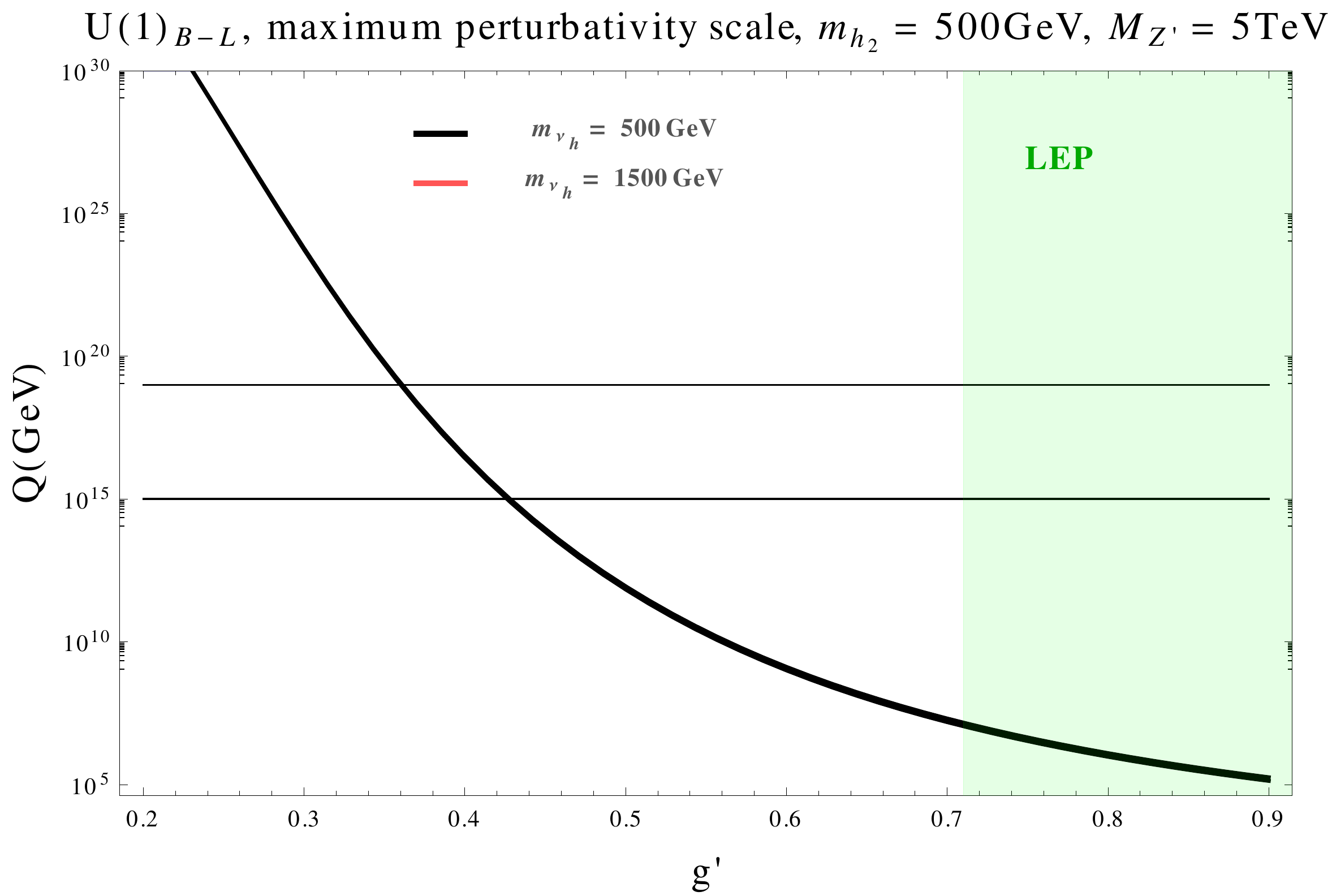}}
\subfigure[]{\includegraphics[scale=0.32]{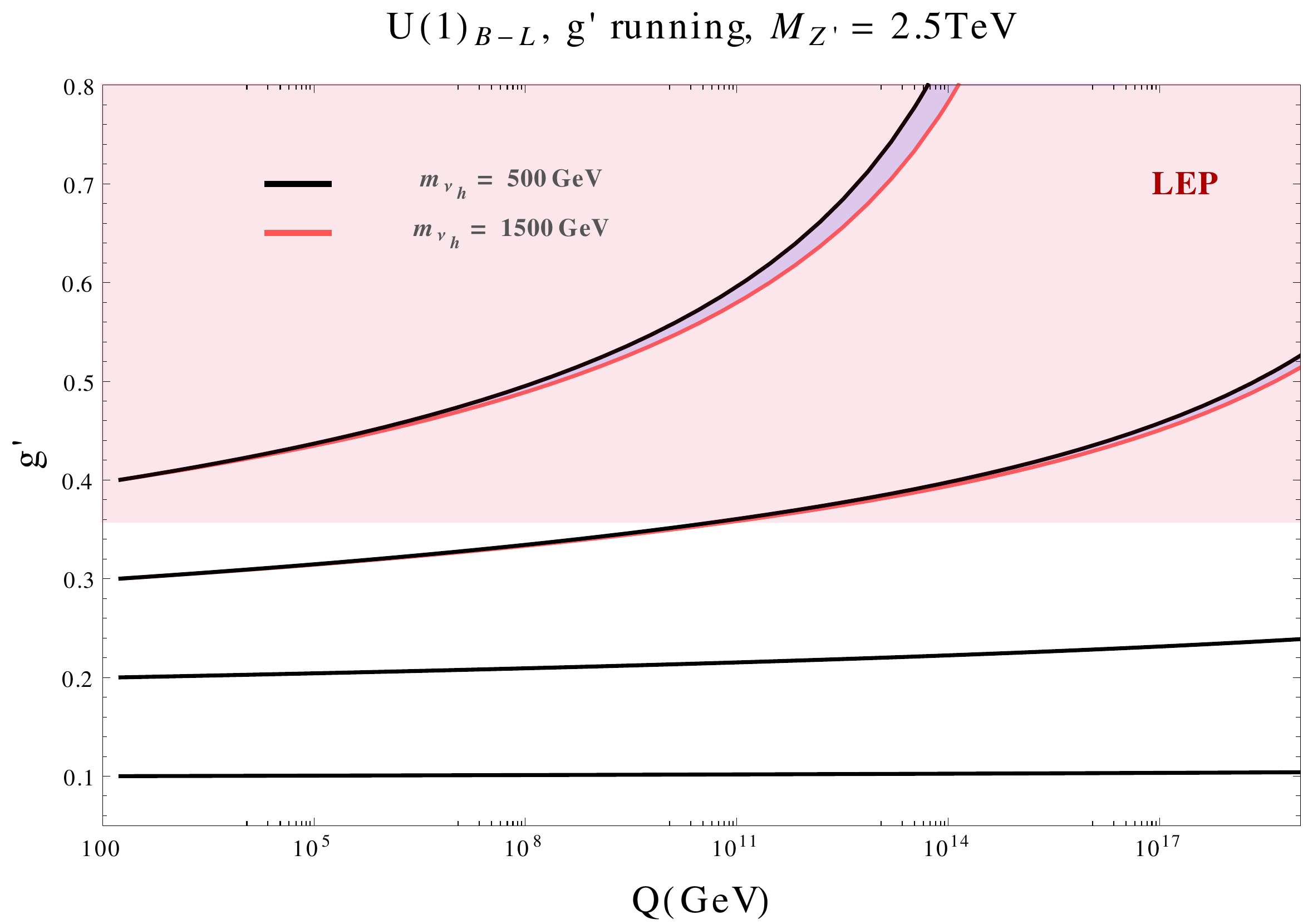}}
\subfigure[]{\includegraphics[scale=0.32]{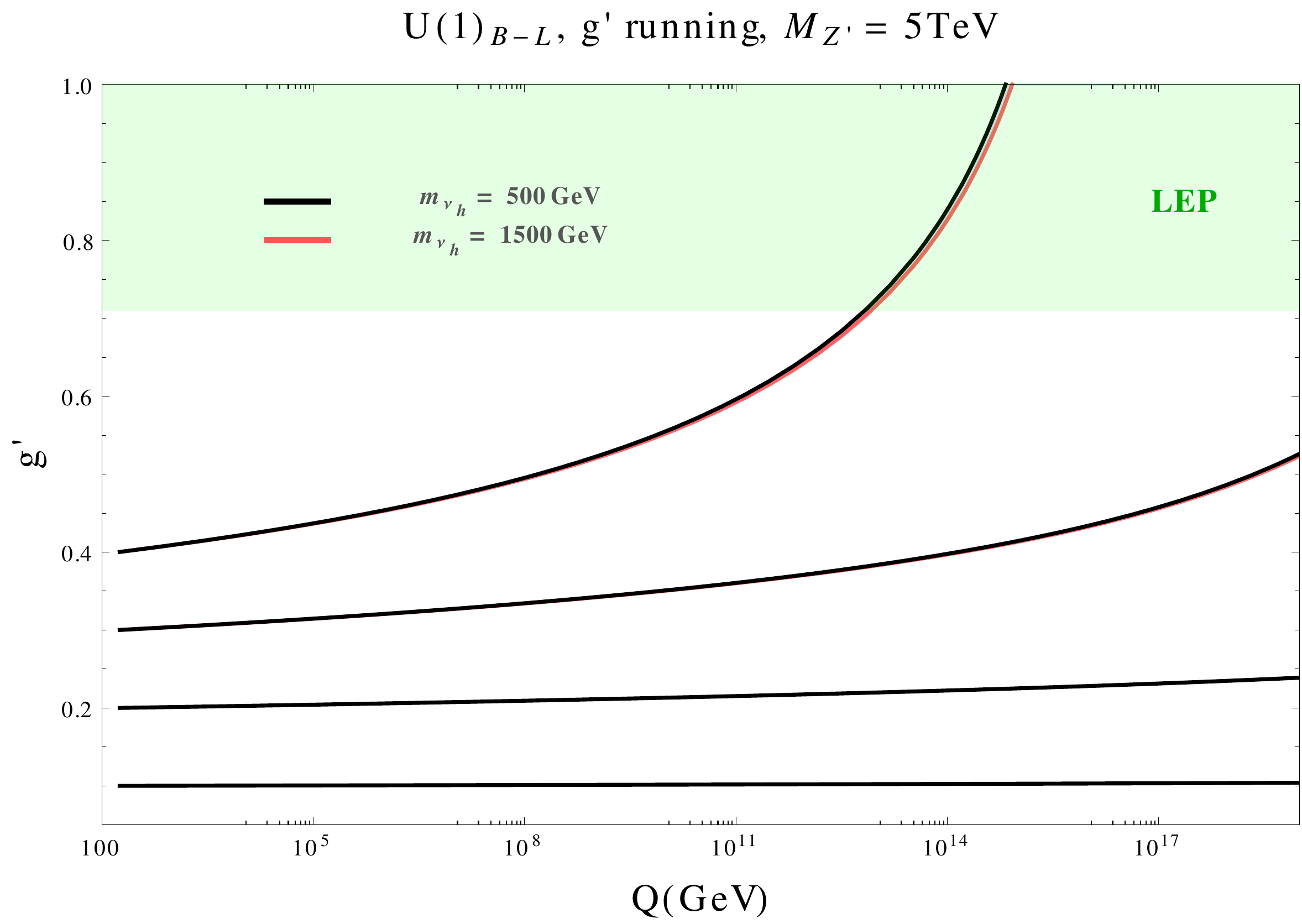}}
\subfigure[]{\includegraphics[scale=0.32]{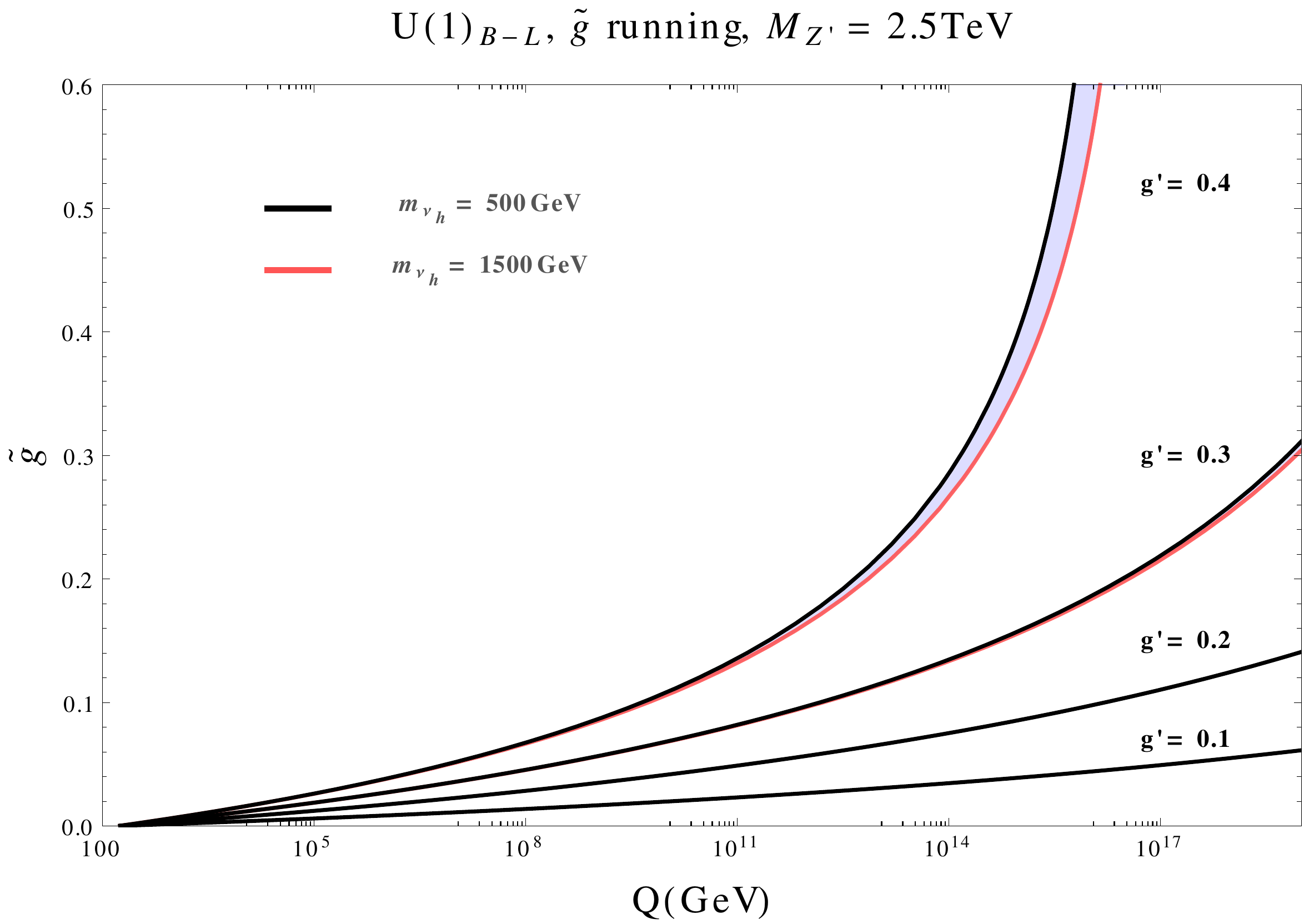}}
\caption{(a) and (b): Maximum evolution scale $Q$ where the abelian gauge sector remains perturbative, as a function of the initial condition on $g'$, fixed at the electroweak scale. Results are shown for $M_{Z'}$ = 2.5 TeV (panel (a)) and 5 TeV (panel (b)). The shadowed regions are excluded by LEP. 
(c), (d) and (e): Running of the $g'$ and $\tilde{g}$ couplings with initial conditions $\tilde{g} = 0$ and $g'=0.1$ (lower curve), $g'=0.2,0.3$ and $g'=0.4$ (upper curve) at the electroweak scale, for $m_{\nu_h}$ ranging between 500 GeV (black thick line) and 1.5 TeV (red thick line). 
\label{Fig.PertCoup}}
\end{figure}
As we have discussed in the previous sections, the requirement of cancellation of the gauge and of the gravitational anomalies has allowed us to reduce the $U(1)'$ charges to just two parameters
$z_Q$ and $z_u$.  Together with the constraint from the LEP-II  results, reported in Eq.~(\ref{STU1}), we also require the perturbativity of the evolution of the coupling of the new abelian sector which gives the conditions
\bea
\label{pertconds}
g'(Q') < \sqrt{4 \pi}\,, \quad \tilde{g}(Q') < \sqrt{4 \pi} \qquad Q' \le Q,
\eea
up to the given final evolution scale $Q$.
Here, the free parameters are the values $g'(Q_{ew}) = g'$ and $\tilde{g}(Q_{ew}) = \tilde{g}$ of the couplings at the electroweak scale, identified 
with the top quark mass. In addition, to process any allowed point in parameter space and investigate the perturbativity of the couplings and the stability of the scalar potential under the two-loop RG evolution we need to implement the appropriate matching conditions as illustrated in section \ref{MatCondSection}. 

We show in Fig.~\ref{Fig.PertCoupGeneric} the region of parameter space in which the perturbativity is maintained 
up to $10^5$ GeV (blue region), $10^9$ GeV (green region), $10^{15}$ GeV (yellow region) and $10^{19}$ GeV (red region), as a function of the free charges $z_Q$ and $z_u$, or $\alpha_Y$ and $\alpha_{B-L}$. The shadowed areas in pink and green are the excluded 
regions according to Eq.~(\ref{STU1}). They correspond to two different choices of $M_{Z'}$ equal to $2.5$ and $5 $ TeV respectively. The plots shows that the validity of perturbation theory up to $M_P$ requires the charge values $|z_Q| \lesssim 1.5$ and $|z_u| \lesssim 3$, and that this bound becomes more stringent as $g'$ grows.
The thick point in each of the three plots corresponds to the charge assignment of a  $U(1)_{B-L}$ abelian symmetry. We have set to zero the parameter of the kinetic mixing ($\tilde{g}$) and varied only $g'$, 
the coupling of the extra $Z'$, set to 0.1 and 0.2 in panel (a) and (b) respectively. Clearly, in both cases, the $U(1)_{B-L}$ charge assignment 
guarantees a perturbative evolution up to the Planck scale. In panel (c) we have plotted the same regions of perturbativity and the corresponding 
exclusion areas of (a) using the $(\alpha_Y, \alpha_{B-L})$ parameterization for a generic extra $U(1)'$ symmetry.

As we have already discussed in the previous sections, the cancellation of gauge and gravitational anomalies naturally requires the introduction of SM singlet fermions which can dynamically acquire a Majorana mass through their Yukawa coupling to the extra scalar $\chi$ controlled by $Y_N$ in a type-I seesaw. In this scenario the mass of the heavy neutrinos can be of the same order of magnitude of the vev of the extra Higgs, which can be easily chosen in the TeV range, with a Yukawa coupling $\sim O(1)$. Values of $Y_N \sim O(1)$ require a detailed RG analysis in order to identify the perturbativity and stability regions in the parameter space of these models. As shown below, the Yukawa coupling $Y_N$, and, equivalently, the mass of the heavy neutrinos, has a fundamental role in this study. \\
In Fig. \ref{Fig.YNYT} we show the evolution of $Y_N$ for different initial values at the electroweak scale. Differently from the Yukawa of the top quark, $Y_N$ grows at a higher rate as we increase the starting value of $Y_N$. The analysis shows that for $g'=0.1$ and $M_{Z'} = 2.5$ TeV, a value of $Y_N \sim 0.45$ at the electroweak scale spoils the perturbativity of the model at $Q = M_P$. The decreasing behavior of $Y_t$ is due to the negative contributions of the QCD corrections to the corresponding $\beta$ function which are obviously absent in the evolution of the Yukawas of the RH neutrinos.

The study of the perturbativity of the abelian sector, investigated by the evolution of the coupling of the extra $Z'$ and the size of the kinetic mixing $\tilde{g}$, provides significant information on the possible scenarios available in the B-L case. This analysis is presented in Fig. \ref{Fig.PertCoup} where we show five plots related to the evolution of the abelian couplings. In particular, we try to identify the regions in the parameter space of this model where the evolution is perturbative along the RG running up to a given final scale $Q$.  
We show in panel (a) the maximum value of such evolution scale ($Q$) allowed for a given initial value of 
the coupling $g'$, with $\tilde g=0$ (pure B-L case) and $M_{Z'}=2.5$ TeV, $m_{h_2}= 500$ GeV. For instance, for $g'\sim 0.45$ at the electroweak scale, the RG evolution remain 
perturbative only up to the GUT scale ($10^{15}$ GeV), but it violates the limits from LEP given in Eq.~(\ref{STU1}). For smaller values of the same coupling, such as $g' \lesssim 0.35$, the model is weakly coupled for any value of the final scale $Q$ up to $M_P$ and satisfies Eq.~(\ref{STU1}). 
We have varied the mass of the heavy neutrinos $m_{\nu_h}$ continuously from 500 up to 1500 GeV, thereby identifying a colored area bounded by the two black and red curves. The continuous boundary curves refer to the choices of $m_{\nu_h}=500$ GeV and 1.5 TeV respectively. The two horizontal black lines in the same plot identify the values of $Q$ corresponding to the GUT ($10^{15}$ GeV) and Planck scales 
($10^{19}$ GeV). \\
The analysis is repeated in the case of $M_{Z'}=5$ TeV (panel (b)), where now the experimentally allowed initial values of  $g'$ are for $g' \lesssim 0.7$, with the shadowed region being therefore excluded by LEP data.  Notice that in this case the two boundary curves describing the variation with the heavy neutrino mass appear to be superimposed and undistinguishable.  As it can be inferred from Eq.~(\ref{abelianBeta}), the Yukawa of the heavy neutrino affects the evolution of the abelian coupling only at two-loop level, thus explaining such mild dependence on it.
Moreover, for a heavier $Z'$ the effect of the RH neutrino mass is more suppressed, since for a given $m_{\nu_h}$, a larger $M_{Z'}$ (thus a bigger $v'$) corresponds to a smaller $Y_N$ . This explains the different bandwidths in the two panels for the same span of $m_{\nu_h}$.   \\
Notice that in both panels the bounds from the requirement of perturbativity are implicitly taken into account in the identification of the maximum stability scale, while the LEP bounds are shown an shadowed regions.
 Incidentally, both for $M_{Z'} = 2.5$ and 5 TeV and for the chosen initial parameters, we find that the gauge coupling has to satisfy the condition $g' \lesssim 0.35$ at the electroweak scale in order to maintain perturbativity up to Planck scale and to fulfill the LEP bound Eq.~(\ref{STU1}). In panel (a) the LEP and the perturbativity bounds tend to coincide, while in 
 (b) the latter is far more significant, with the shadowed region excluding values of the gauge coupling $g' > 0.35$. \\
In panels (c), (d) and (e) we have investigated the evolution of the $g'$ and of the $\tilde{g}$ couplings, the latter describing the impact of the kinetic mixing, which are responsible for the behaviors discussed above. 
In panels (c) and (d), we plot the running of $g'$ as a function of the evolution scale $Q$, for a varying heavy neutrino mass. 
The shadowed areas are the regions excluded by the bound Eq.~(\ref{STU1}). 
Notice that the growth of $g'$ is very similar in both panels, and in the case of a heavier extra $Z'$ it is quite insensitive to the change of the heavy neutrino mass, for the same reasons illustrated for the previous plots. 
The growth of this coupling is quite moderate for initial conditions $g' \lesssim 0.2$ but its RG trajectory undergoes a rapid steepening for larger initial values. \\
A similar analysis (panel (e)) is presented for the coupling which parameterizes the kinetic mixing of $U(1)_Y$ and $U(1)_{B-L}$.
The evolution of the kinetic mixing $\tilde{g}$ is strongly correlated with the size of the initial condition on $g'$. The RG trajectories, in this case, steepen very fast as $g'$ grows. We have chosen four values of $g'$ at the electroweak scale equally spaced in the interval (0.1-0.4). We start with a vanishing kinetic mixing at the electroweak scale and let the coupling be radiatively generated by the evolution. By selecting initial values of $g' \lesssim 0.3$ the evolution of the mixing satisfies the perturbativity bound up to the Planck scale. For larger initial values, 
such as $g' \sim 0.4$, both couplings $(g',\tilde{g})$ will violate the perturbativity below the Planck scale. In this plot, the shadowed regions are bounded by the two trajectories generated by masses of the heavy RH neutrinos $m_{\nu_h}=500$ GeV (black curve) and 1.5 TeV (red curve)  which appear to be separated only for $g' \gsim 0.3$ and $Q \gsim 10^{13}$ GeV. 

%.
%
 \subsection{Two-loop effects and the vacuum stability bounds}
\begin{figure}[t]
\centering
\subfigure[]{\includegraphics[scale=0.33]{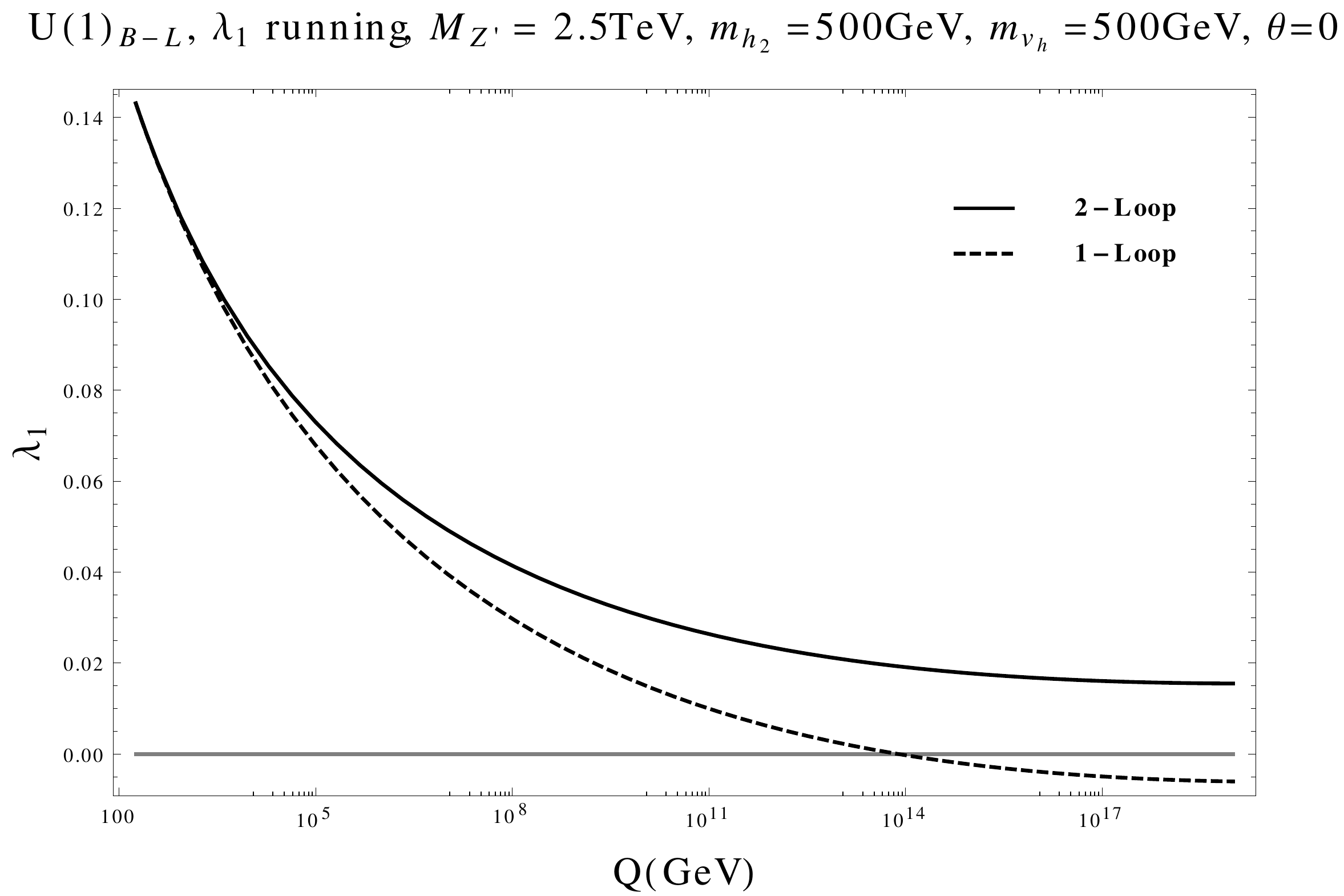}} 
\subfigure[]{\includegraphics[scale=0.31]{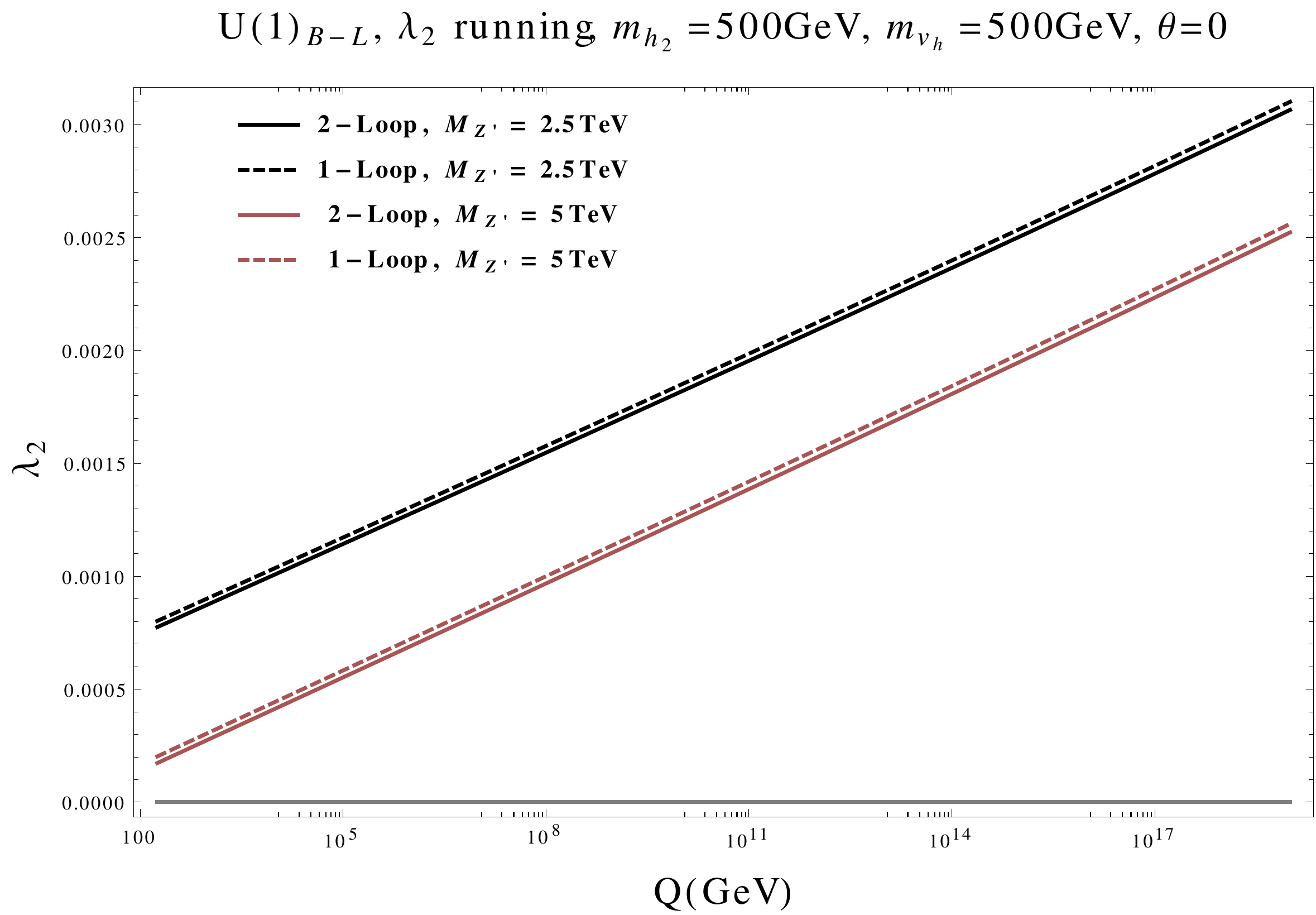}}
\subfigure[]{\includegraphics[scale=0.355]{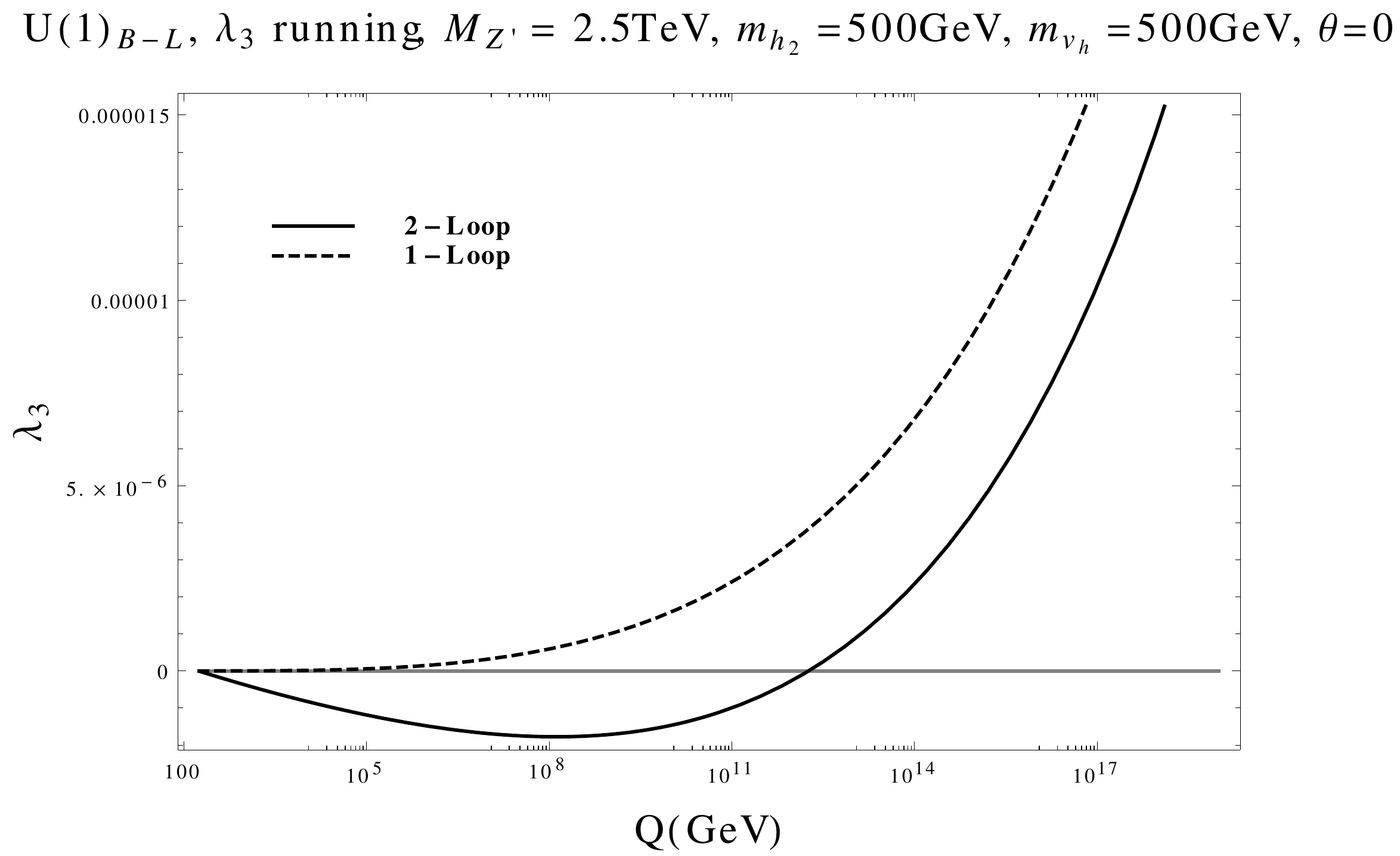}}
\subfigure[]{\includegraphics[scale=0.31]{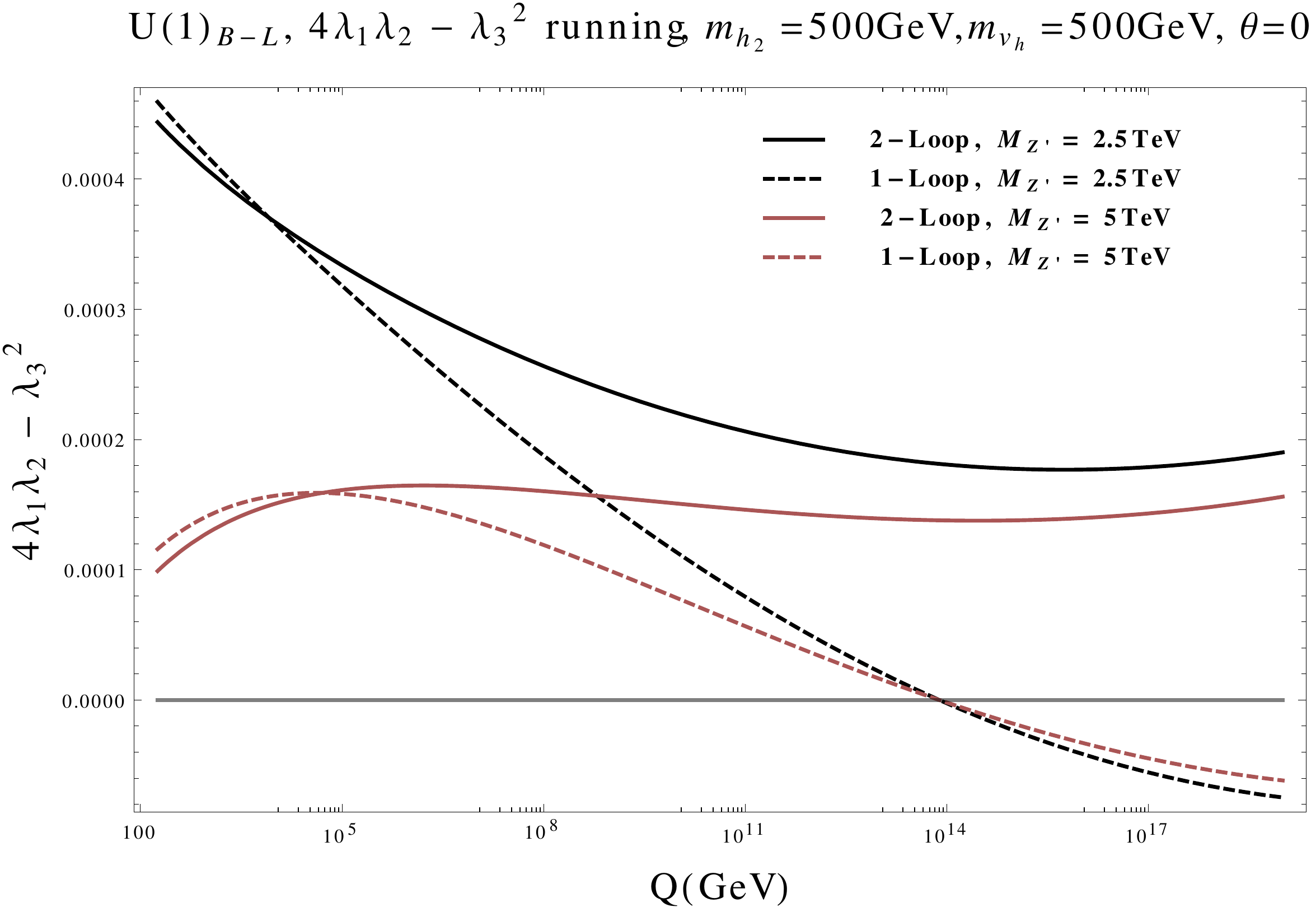}}
\caption{  RG running of the quartic couplings in the scalar potential: (a) evolution of $\lambda_1$, (b) evolution of $\lambda_2$, (c) evolution of $\lambda_3$, (d) evolution of $\sigma = 4 \lambda_1 \lambda_2 - \lambda_3^2$. One- and two-loop effects are shown.  \label{Fig.CoupEv2}}
\end{figure}

\begin{figure}[t]
\centering
\includegraphics[scale=0.4]{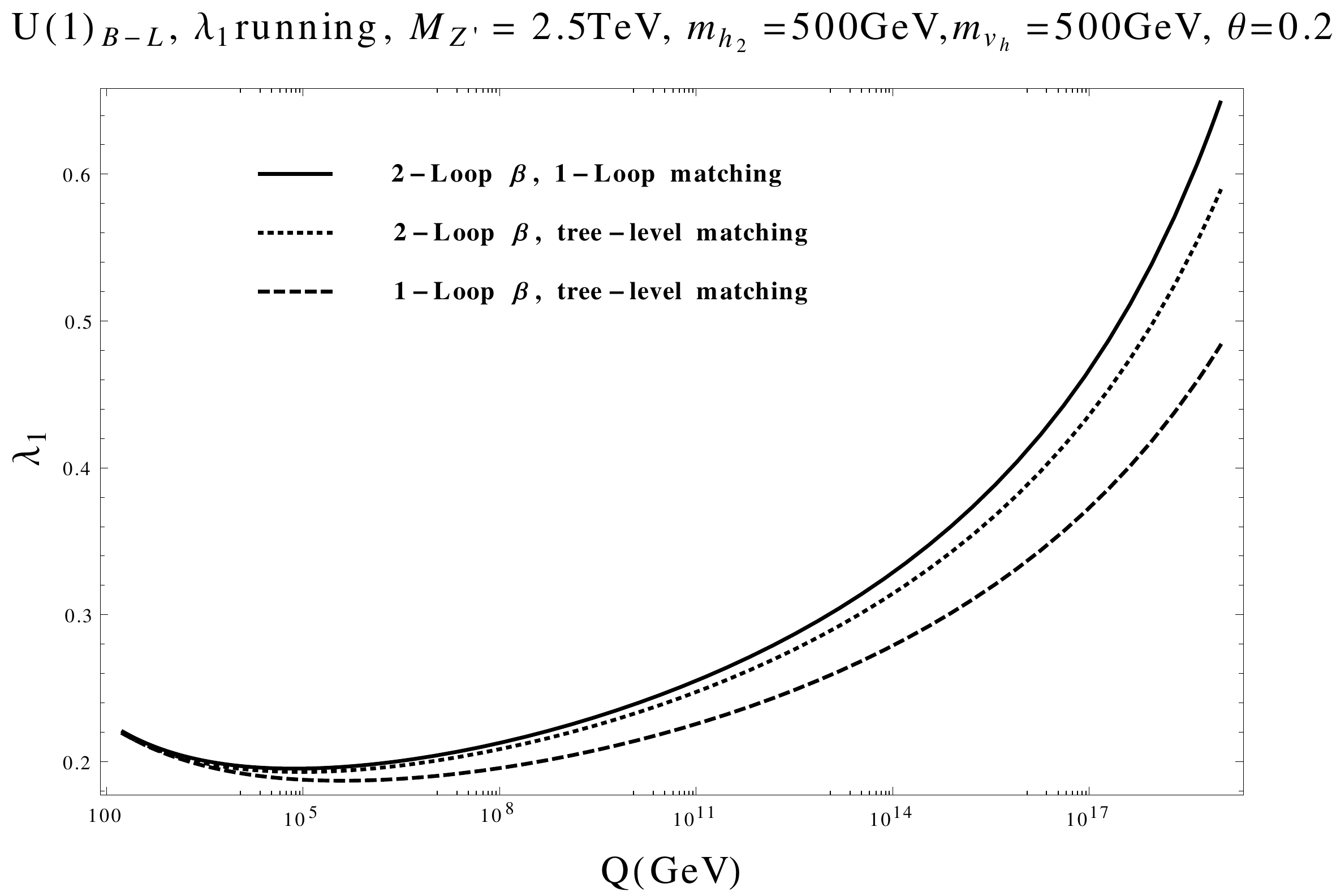}
\caption{The evolution of $\lambda_1$. Two-loop evolution with the one-loop matching (continuous line);  two-loop evolution without matching (dotted line); one-loop evolution without matching (dashed line). 
}
\label{c3}
\end{figure}

\begin{figure}[t]
\centering
\subfigure[]{\includegraphics[scale=0.4]{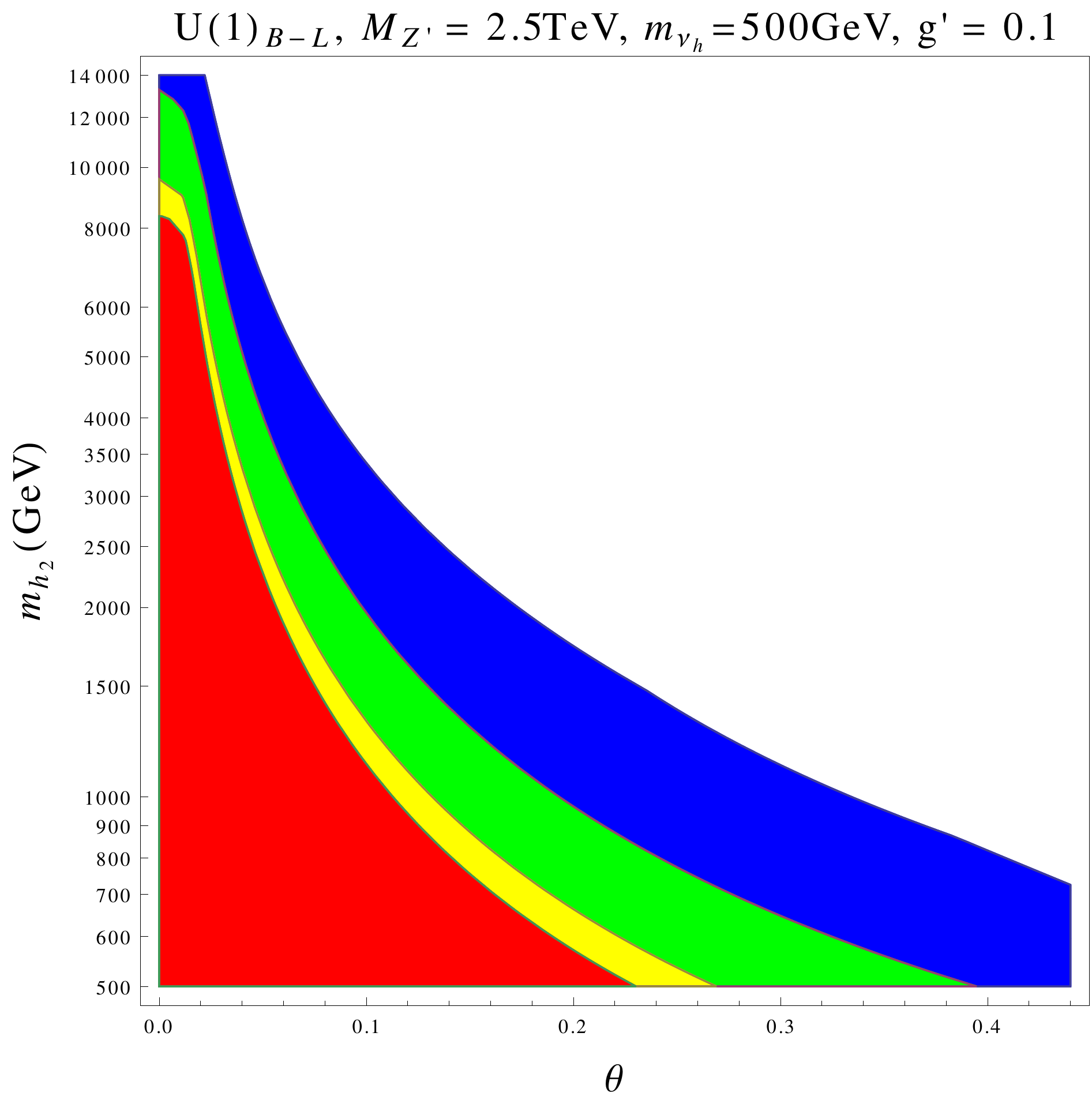}}
\subfigure[]{\includegraphics[scale=0.406]{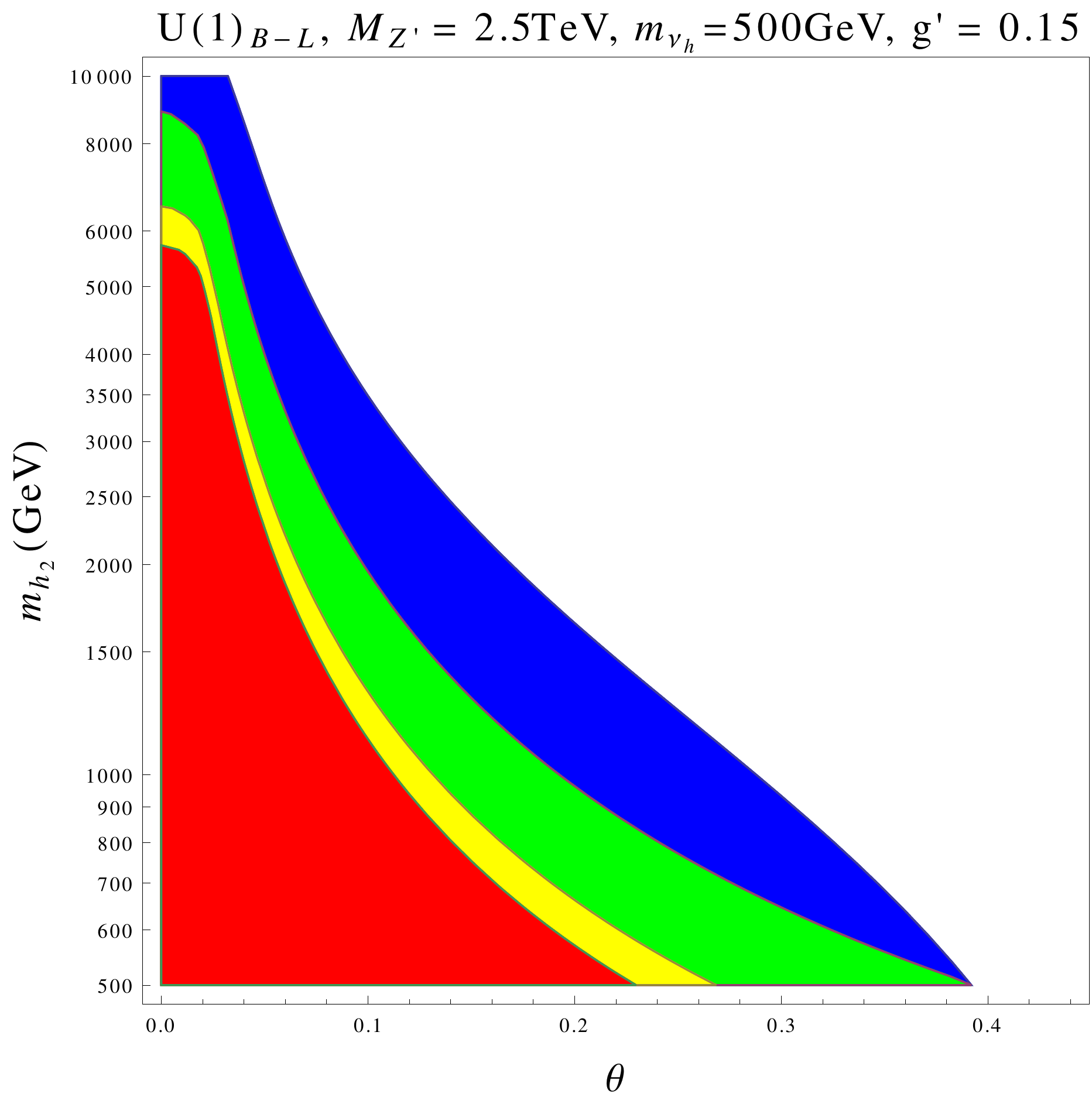}}
\caption{(a) Maximum stability and perturbativity scale in the ($m_{h_2},\theta$) parameter space for $M_{Z'} =$ 2.5 TeV and $g' = 0.1$, 
(b) Maximum stability and perturbativity scale in the ($m_{h_2},\theta$) parameter space for $M_{Z'} =$ 2.5 TeV and $g' = 0.15 $ .
\label{mtheta}}
\end{figure}

\begin{figure}[t]
\centering
\subfigure[]{\includegraphics[scale=0.32]{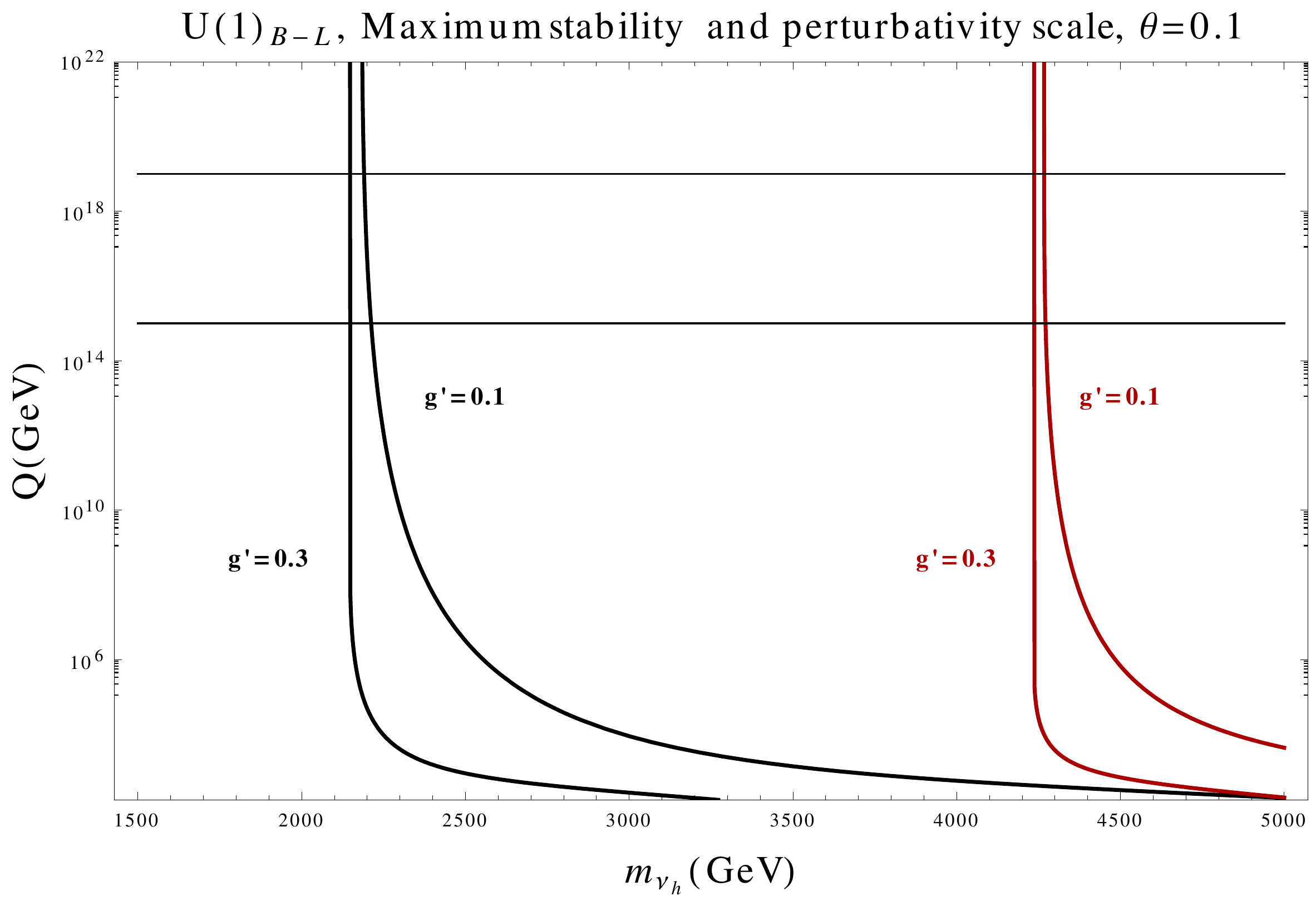}}
\subfigure[]{\includegraphics[scale=0.33]{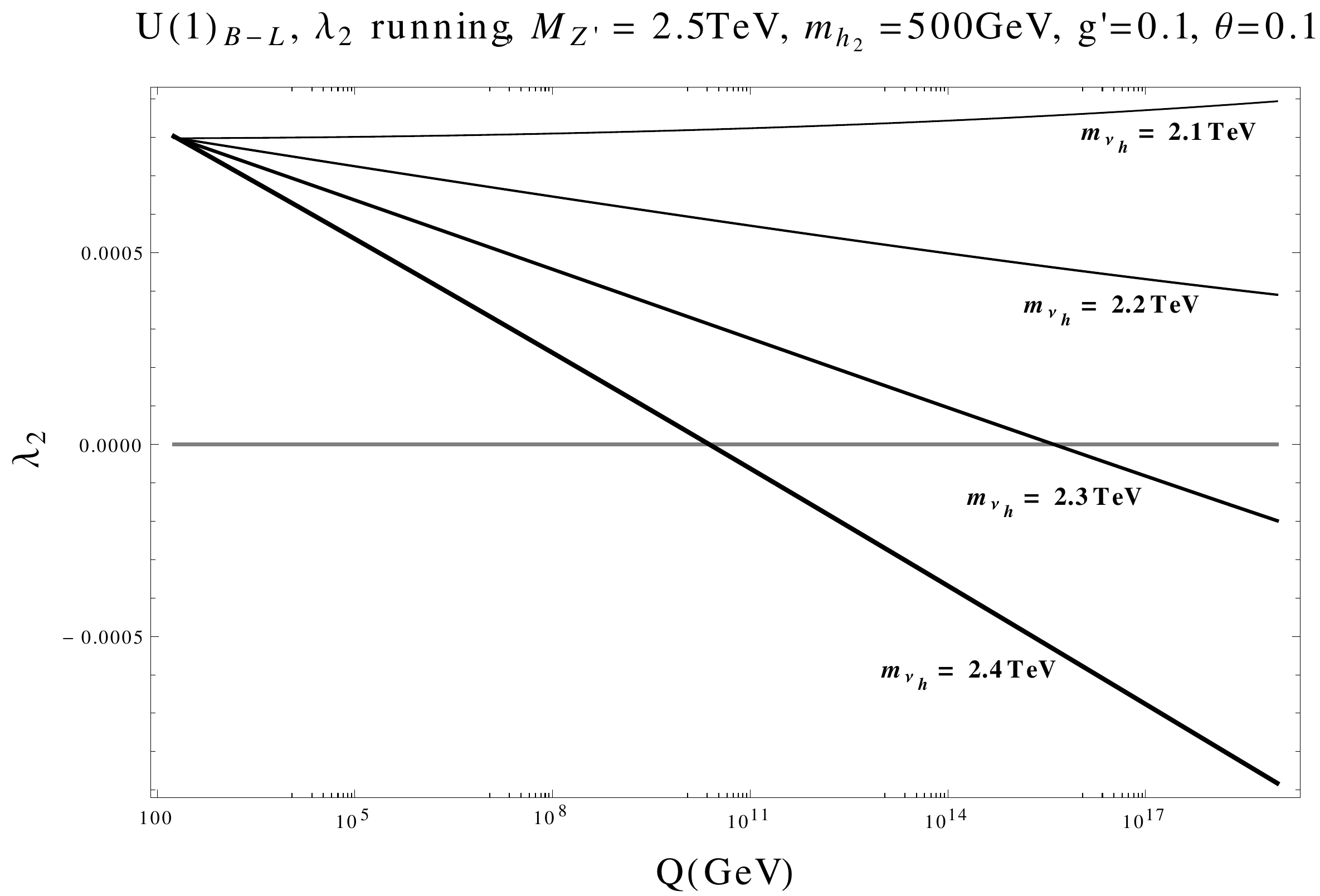}}
\subfigure[]{\includegraphics[scale=0.305]{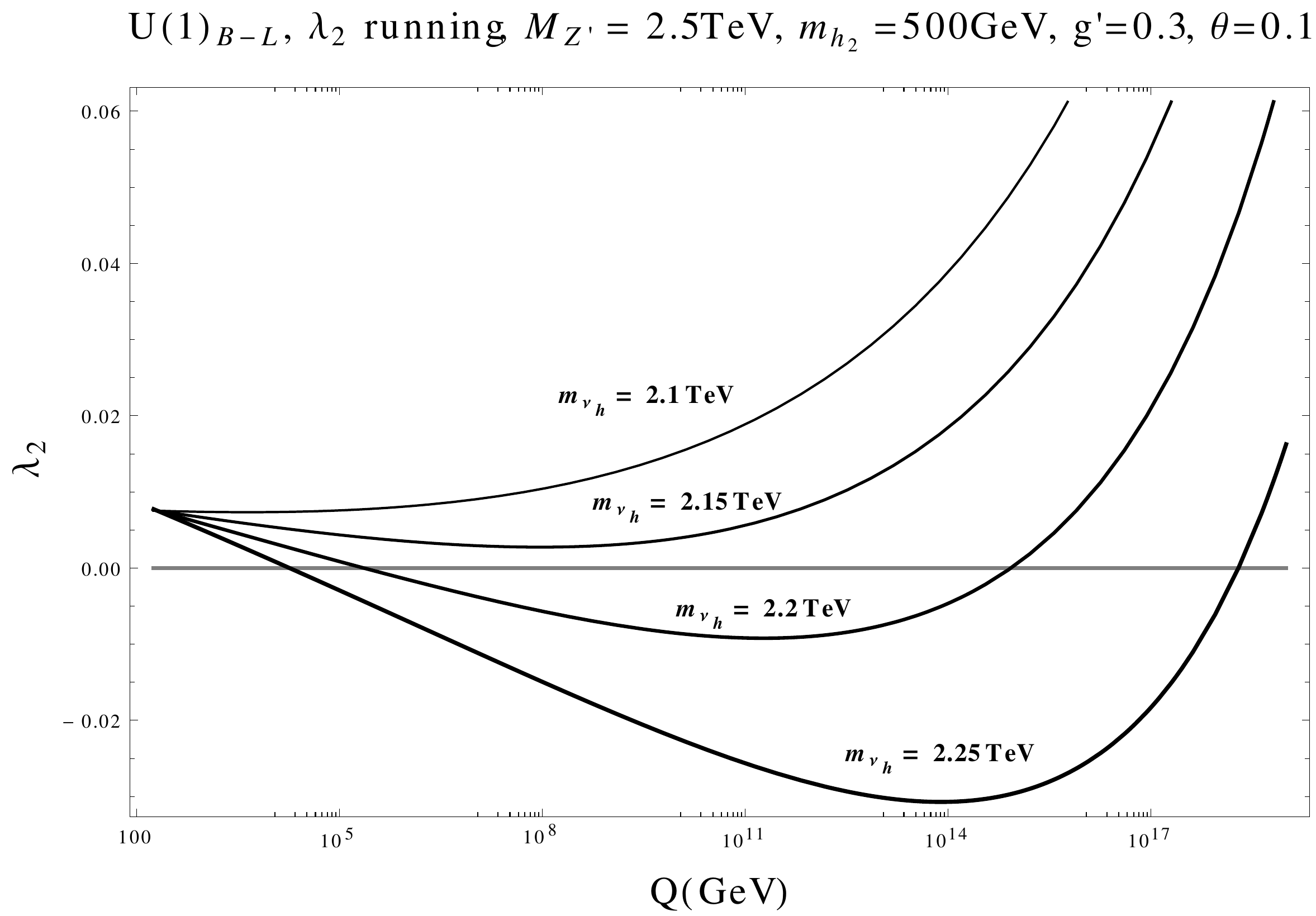}}
\caption{(a) Maximum stability and perturbativity scale as a function of $m_{\nu_h}$ for  $M_{Z'} = 2.5$ TeV (black lines) and $M_{Z'}$ = 5
TeV (red lines) each with two different values of $g'$. (b)  The evolution of $\lambda_2$ with $M_{Z'} = 2.5$TeV and $g' = 0.1$ for values of $m_{\nu_h}$ between $2.1$ TeV and $2.4$ TeV. (c)  The evolution of $\lambda_2$ with $M_{Z'} = 2.5$ TeV and $g' =0.3$ for values of $m_{\nu_h}$ between $2.1$ TeV and $2.25$ TeV.
\label{shadows}}
\end{figure}
\subsubsection{Dependence of the stability region on higher-order effects }

The information about the vacuum stability is enclosed in the structure of the potential for large fields value, that is, in the quartic scalar interaction terms, stated in the three positivity conditions of Eq.~(\ref{stabilitycond}).
In the SM, where only one term of this kind is present ($\lambda$), the stability condition coincides with the positivity of this coupling. As we are going to show in this section, the stability bound is significantly affected also by the order of the evolution, which is crucial in order to reach definitive conclusions regarding the dynamics of the $U(1)_{B-L}$ model. The result of this analysis is shown in Fig. \ref{Fig.CoupEv2}. In this study we want to emphasize the role of the two-loop corrections on the RG running and of the one-loop matchings on the initial conditions. For the sake of simplicity we choose fixed values for the mass of the heavy Higgs and of the right-handed neutrinos both of 500 GeV, and $\theta = 0$. The mass of the $Z'$ is reported in each plot. \\
We show in panel (a) the RG running of $\lambda_1$ at one- and two-loop levels. Notice that at one-loop level, for the chosen initial conditions, the maximum stability scale $Q$ in the RG trajectory of this coupling is below the GUT scale, at $10^{14}$ GeV. Two loop effects are clearly essential in order to verify its positivity up to the Planck scale. \\
A similar investigation is carried out for the quartic coupling of the heavy Higgs field $\lambda_2$. In panel (b)  it is shown that both the one-loop and the two-loop evolutions indicate that this coupling evolves rather slowly and stays positive up to the Planck scale for the chosen $m_{h_2}$ and $m_{\nu_h}$. We have selected two different initial conditions at the electroweak scale corresponding to two different values of the mass of the $Z'$, $M_{Z'} = 2.5$ TeV (black curve), $M_{Z'} = 5$ TeV (red curve). In this case the two-loop effects are quite small, independently of the chosen initial conditions. \\
In panel (c) we extend our analysis to the mixed $\chi-H$ quartic coupling, which enters into the stability condition by the quantity $\sigma\equiv 4 \lambda_1 \lambda_2 - \lambda_3^2$, the latter plotted in panel (d). 
  Notice that, in this case, the flip in sign of $\lambda_3$, as shown in (c), is not problematic for the stability bound if the two-loop RG equations are employed, since $\sigma$ stays anyhow positive over the entire range of the evolution, as also shown in (d). Also in this case, it is clear that two-loop effects are essential in order to revert the trend of the evolution, which otherwise causes $\sigma$ to turn negative 
below the GUT scale, thereby generating a vacuum instability. In (d) we show the trajectories of $\sigma$ for two values of $M_{Z'}$ reported in the same figure. In both cases an instability is generated at $10^{14}$ GeV by the one-loop running, which is completely eliminated by the two loop effects. Notice that two loop effects, in general, tend to change the concavity of the $\sigma$ evolution. This interesting behavior can be understood from a cursory look at all the panel (a), (b) and (c), due to the different slopes of the corresponding curves over the entire evolution region.
We conclude that the inclusion of two-loop effects in the RG evolution is mandatory for a consistent analysis of the stability of the scalar potential and for the identification of the available parameter space of the model.

In order to show the impact of the matching conditions we study the behavior of $\lambda_1$ in Fig. (\ref{c3}) where the matched and the unmatched evolutions are shown.  We have indicated with a continuous line the two-loop RG running of the same coupling with one-loop matching and with a dotted line the unmatched case. The dashed line refers to the one-loop result, with the tree-level matching. It is clear that, as for the two-loop $\beta$ functions, the one-loop matching relations on the initial condition improve the stability of the potential and are necessary for the consistency of the perturbative approach. Moreover, we have explicitly verified that these contributions become more sizeable as soon as we switch on the scalar mixing parameters $\theta$ of the two Higgses.

\subsubsection{Dependence of the stability region on $\theta, m_{h_2}$, and $m_{\nu_h}$ }
Having shown the importance of the two-loop corrections in the analysis of the stability of $U(1)_{B-L}$, we are now going to investigate this issue more closely by focusing our attention on its parametric dependence on the scalar mixing 
angle $\theta$, on the mass of the heavy Higgs and of the RH neutrinos.\\
The bounds from the electroweak precision measurements reported in Eq.~(\ref{STUbound}) can then be used to reduce the space of parameters in the $(m_{h_2},\theta)$ plane. In Fig. \ref{mtheta} we plot in different colors the regions of stability and perturbativity for different final evolution scales $Q$. The $(m_{h_2},\theta)$ region shown in all the panels is the one allowed by the bounds in Eq.~(\ref{STUbound}), with the areas marked in blue, green, yellow and red identifying stable evolutions with final scales $Q$ equal to $10^5$, $10^9$, $10^{15}$ and $10^{19}$ GeV respectively. We have selected $M_{Z'}=2.5$ TeV and a mass of the RH neutrinos $m_{\nu_h}$= 500 GeV.  The choices of the values of the couplings, $g'=0.1$ in panel (a) and $g'=0.15$ in panel (b) show the sensitivity of these regions on such small changes of $g'$. Notice that in panel (b) the requirement of having a stable evolution up to the Planck scale requires that $m_{h_2} \lesssim 6$ TeV. In general, as clear from the plots, such a requirement reduces significantly the bound given in Eq.~(\ref{STUbound}), suggesting that the smallness of the scalar mixing can be inferred by the condition of RG stability. In particular, 
$\theta$ has to be smaller that $0.3$ in order to have a stable evolution of the scalar potential at least up to the GUT scale.

A study of the dependence of the stability region on the mass of the RH neutrino is presented in Fig. \ref{shadows}, where (panel (a)) the curves in black and red refer to values of the mass of the extra $Z'$ of 2.5 and 5 TeV respectively, each with a more stable ($g' =$ 0.1) and a less stable ($g' =$ 0.3) coupling assignment.  Notice that both branches are characterized by a drastic change of the $Q$ value of maximum stability, around $m_{\nu_h} \sim $ 2.2 and 4.3 TeV respectively. 
As the mass of the heavy neutrino grows, the effects of its Yukawa coupling $Y_N$ dominate on the evolution of $\lambda_2$ driving it towards negative values and, therefore, destabilizing the vacuum of the scalar potential. This  behavior is caused by the large and negative $Y_N^4$ contribution to the $\beta$ function of $\lambda_2$, as one can see already at one-loop level from Eq.~(\ref{betal2}) \cite{Lyonnet:2013dna,Basso:2013vla,Coriano:2014mpa}. Clearly, 
the role of this coupling is similar to that of $Y_t$, the Yukawa of the top, which is responsible for driving the SM vacuum towards an instability. Once more we want to stress  that the destabilizing role of a large $Y_N$ is twofold: it can spoil both the stability of the scalar potential and the perturbativity, as shown in Fig. \ref{Fig.YNYT}. Moreover, it is clear from the same figure that $Y_N$ grows significantly for large final evolution scales, thus increasing its destabilizing effect on vacuum.  \\
More insight into this behavior of the potential 
can be deduced also from panel (b) and (c) of the same figure. Here we show a plot of the $\lambda_2$ coupling versus the evolution scale $Q$. The region of parameter space explored in panel (b) corresponds to the case $M_{Z'}$=2.5 TeV and $g' = 0.1$, with the mass of the RH neutrinos between 2.1 and 2.4 TeV. As shown an instability emerges for $m_{\nu_h}$ around 2.2 TeV  turning  $\lambda_2$ negative and destabilizing the potential. A similar analysis is illustrated in panel (c) where $g'=0.3$ has been chosen. 
It is clear, from this analysis, that a stable evolution up to a significantly large scale, such as the GUT scale or larger, is favored by values of $m_{\nu_h} \lesssim $ 2 TeV and is worsened by an increase in the size of the coupling $g'$.

\subsubsection{Dependence of the stability region in the $(m_{\nu_h}, m_{h_2})$ plane} 
\begin{figure}[H]
\centering
\subfigure[]{\includegraphics[scale=0.4]{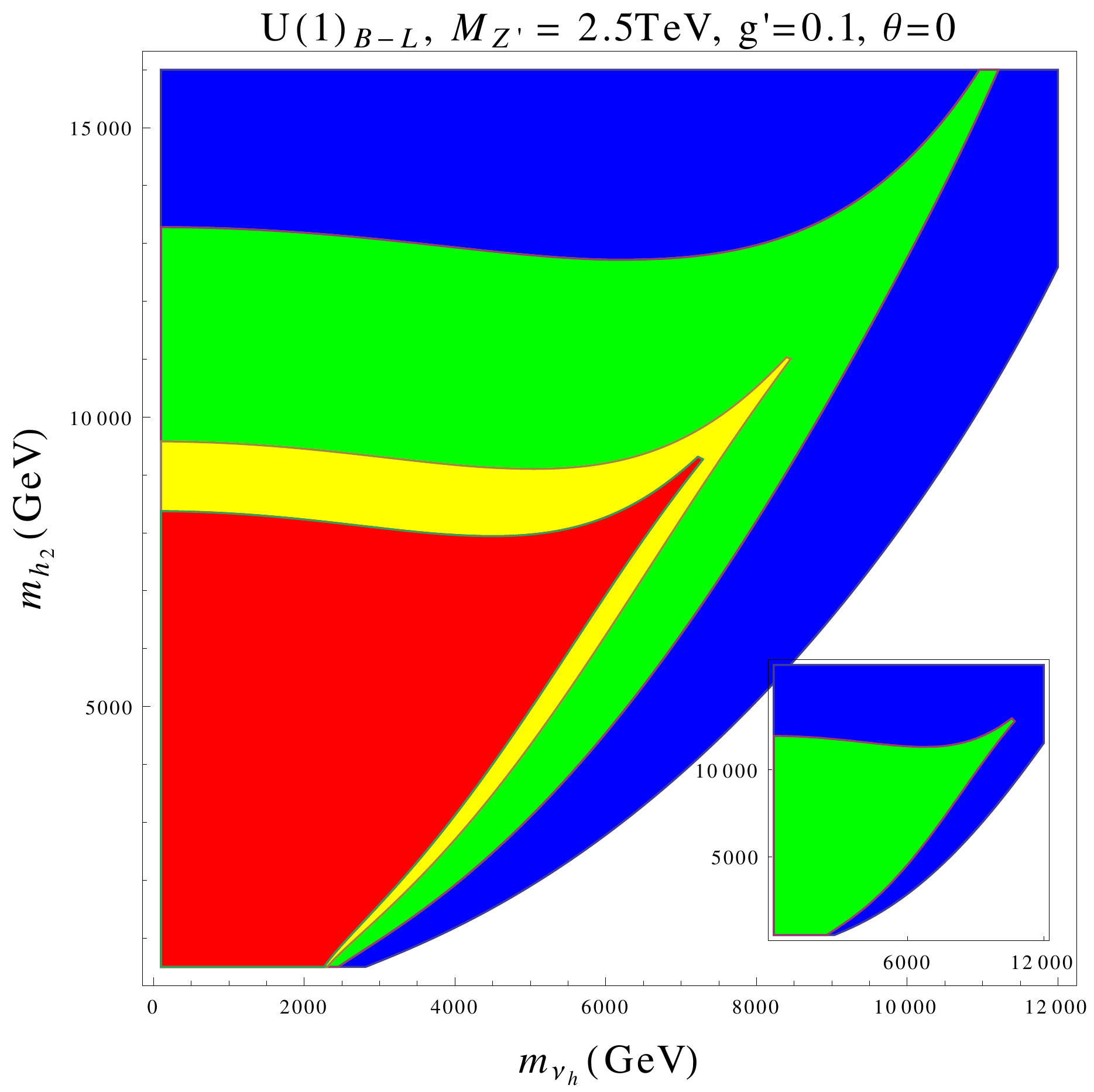}}
\subfigure[]{\includegraphics[scale=0.37]{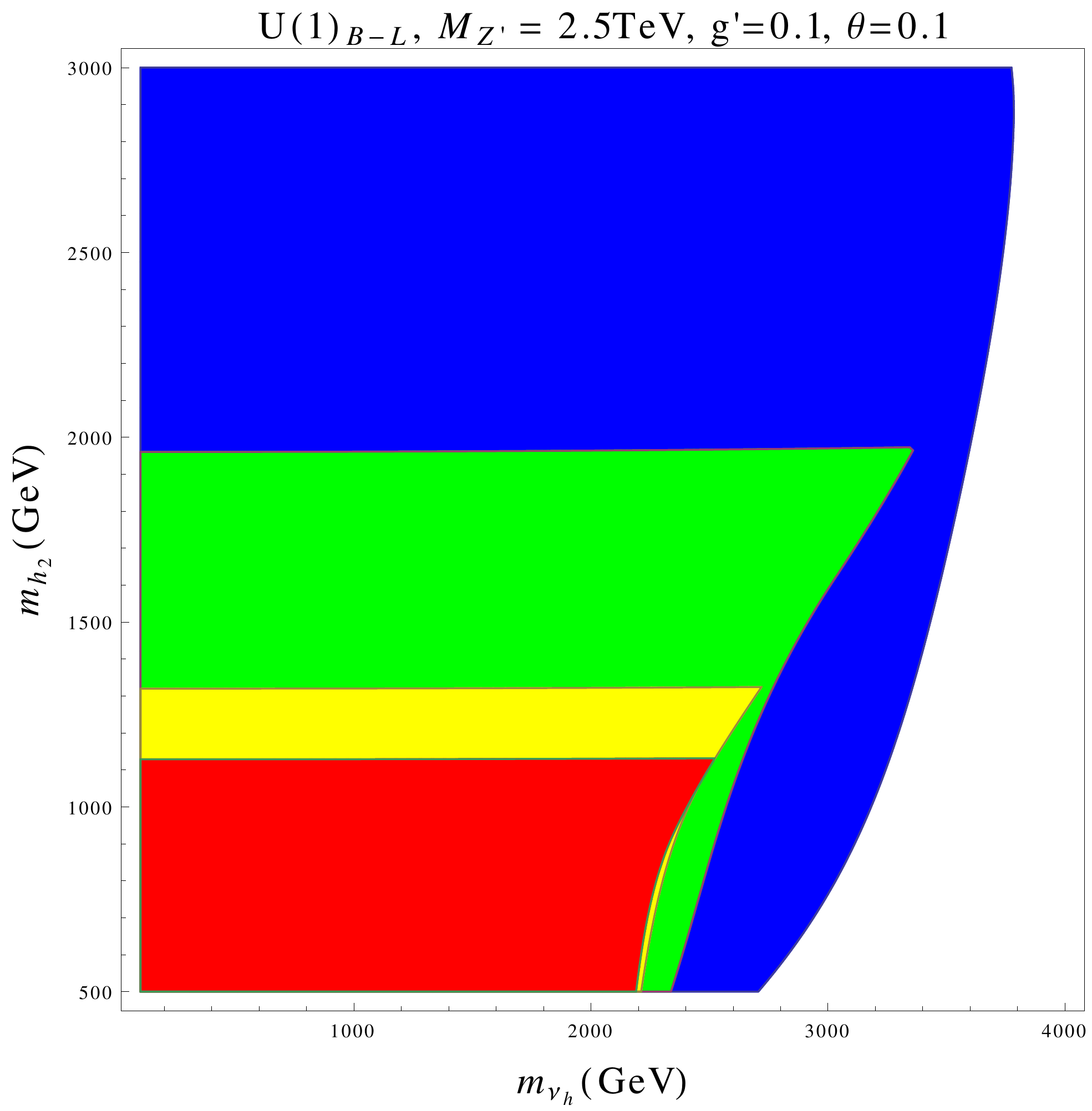}}
\subfigure[]{\includegraphics[scale=0.4]{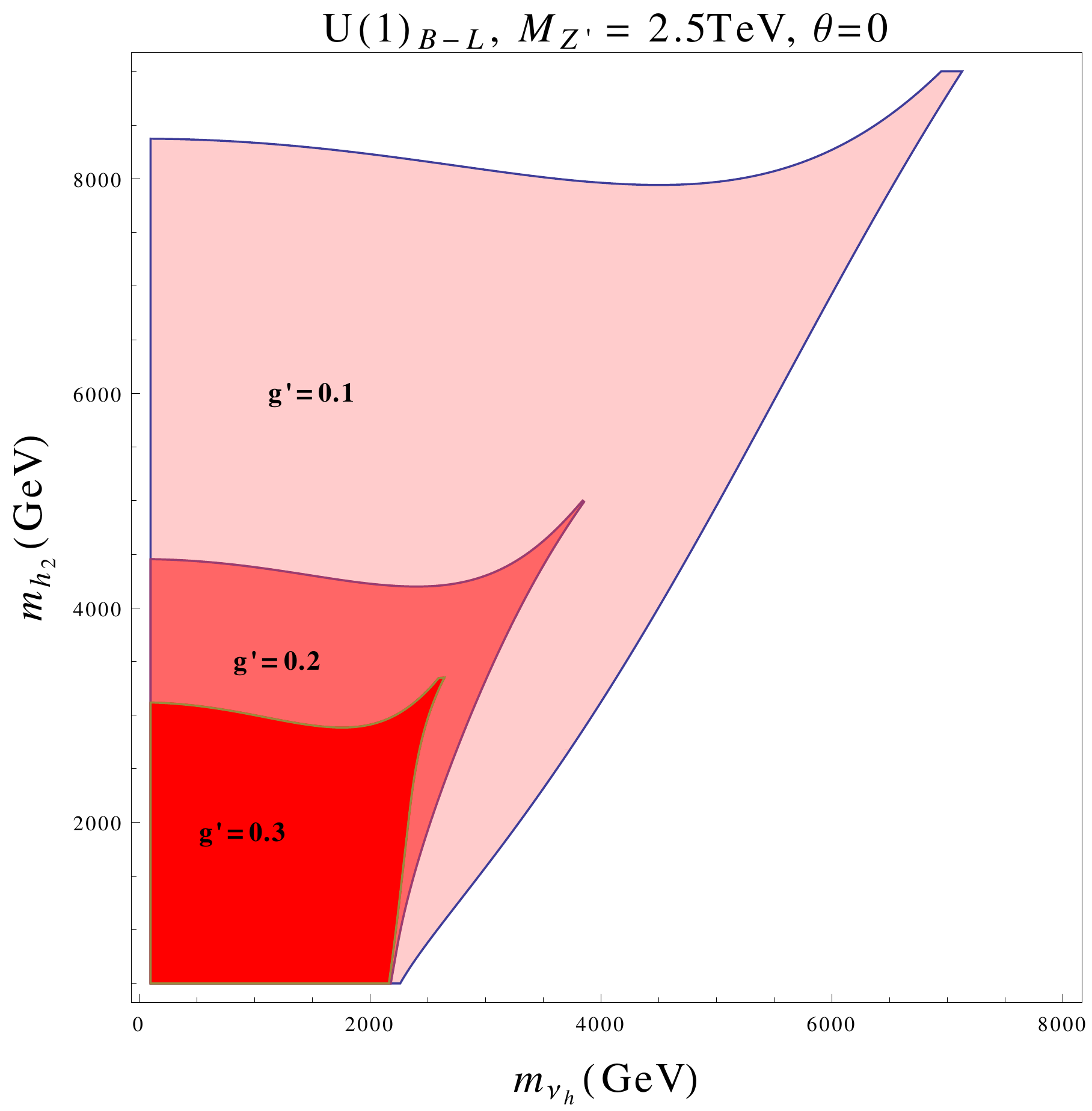}}
\subfigure[]{\includegraphics[scale=0.42]{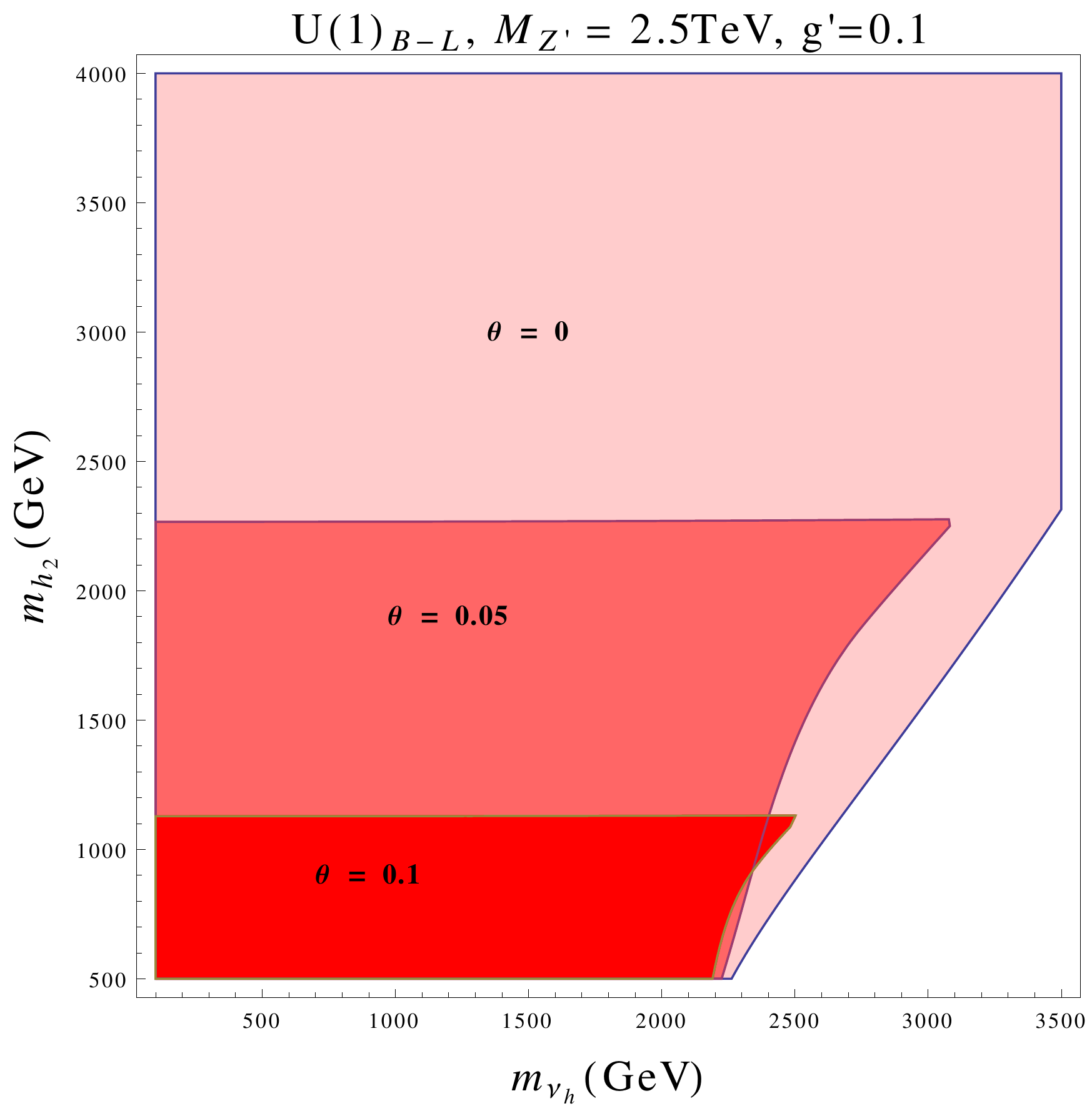}}
\caption{ (a): Stability regions at two-loops (large panel) and at one-loop (small subpanel) for $\theta=0$ in the 
$(m_{h_2},m_{\nu_h})$ plane. (b) Stability regions at two-loops for $\theta=0.1$.
(c) Regions of stability up to Planck scale for different values of $g'$ and  $\theta=0$. 
(d) Regions of stability up to Planck scale for different values of $\theta$ and  $g'=0.1$}
\label{c4c}
\end{figure}

In Fig. \ref{c4c} we show the stability regions up to $10^5$ GeV (blue region), $10^9$ GeV (green region), $10^{15}$ GeV (yellow region) and $10^{19}$ GeV (red region) in the $(m_{\nu_h}, m_{h_2})$ plane. In panel (a) we compare their shapes in the case of a two-loop versus a one-loop implementation of the running for a vanishing scalar mixing $\theta$. The two-loop RG result 
is pictured in the larger panel, while the smaller subpanel on the bottom right reproduces the same plot but uniformly rescaled and at one-loop level. As we have already observed, the two plots show clearly that the inclusion of the 
two-loop corrections is essential in order to improve the stability of the potential up to the Planck scale. Notice that in the small subpanel, the red and the yellow regions, which in the larger plot characterize a stable potential up to the Planck and GUT scale respectively, are replaced by a green area, 
corresponding to a stability which can reach only $10^{9}$ GeV. \\
In panel (b) we have varied the scalar mixing angle to the value of $\theta=0.1$. Also in this case the effects of a nonzero $\theta$ to the shape of the stability regions are quite significant. 
In general, the upper bound on $m_{h_2}$ in the $(m_{\nu_h},m_{h_2})$ plane is far stronger compared to panel (a), showing that, for a RH neutrino of 2 TeV, no values of the heavy Higgs mass can accomodate the requirement of stability above $10^5$ GeV. Notice also that the upper bound on the heavy Higgs mass $m_{h_2}$ is barely influenced by the heavy neutrino mass but it is strongly controlled by its vacuum expectation value $v'$ and the scalar mixing angle $\theta$, see Fig.~\ref{mtheta}.\\
In panels (c) and (d) the region of Planck scale stability is investigated in terms of its dependence on the initial conditions  at the electroweak scale of the gauge coupling $g'$ (panel (c)) and of $\theta$ (panel (d)). We find that the region of stability grows larger as the gauge coupling and the scalar mixing decrease.

\section{Comments and conclusions} 
The identification of possible approaches which could set significant exclusion limits on the parameter space of new physics models at the LHC is, for sure, 
of direct theoretical and experimental significance. Among these, the requirement of stability of the scalar potential up to a very large scale, such as the GUT scale
or even the Planck scale, has a clear physical appealing, although metastable vacuum states, as in the case of the SM, could also be acceptable. 
The case of simple abelian extensions of the gauge structure, which require a minimally enlarged Higgs sector, with just one extra scalar,  define a minimal framework
in which to explore the requirements of stability. 
\\
Our results clearly indicate that the sensitivity of the evolution on the large Yukawa coupling of the top quark, present in the SM, is probably part of a more general 
scenario which includes other similar couplings sharing a similar behaviour. This picture emerges as soon as we allow a more complete description of the light neutrino 
masses. With these motivations, we have investigated in some detail the case of a B-L extension of the SM, accompanied by a type-I seesaw for the generation of the neutrino
masses, realized by RH neutrinos. In such $U(1)'$ extensions, the gauge charges of the RH neutrinos are naturally constrained by the condition of cancellation of the gauge 
and gravitational anomalies. In our analysis we have specifically chosen a family universal charge assignment for these neutrinos, one per generation.
The choice of a type-I seesaw, obviously, is model dependent, but the approach can be extended to other mechanisms as well. The phenomenological advantage of a mechanism
of this type, besides its simplicity, is to involve a vev of the extra Higgs which can be easily accommodated in the TeV range, at the reach, in principle, of LHC searches. In this regard the model has been largely investigated starting from early LEP searches \cite{Abreu:1994ria}, 
to the most recent analyses of its LHC visibility. Such minimal yet rich extension of the SM has found a lot of attention along the years, covering the entire spectrum of the model, in search of specific experimental signatures at the LHC. 
 From the extra scalar sector \cite{Basso:2013vla} to its gauge 
\cite{Basso:2008iv,Basso:2010pe,Basso:2012sz,Basso:2012ux,Accomando:2015cfa,Accomando:2015ava} and neutrino sector \cite{Basso:2008iv}, the TeV B-L scenario has shown to provide
a promising window of experimental visibility, if realized in nature, for the present and forthcoming experiments at the LHC.
Moreover the aforementioned analysis have been carried out in the minimal $U(1)_{B-L}$ extension in which the mixing between the $Z$ and the $Z'$ gauge bosons 
has been neglected. 
A more complete analysis of the kinetic mixing of this model, although quite constrained by electroweak precision data \cite{Salvioni:2009jp}, 
and the possibility of exploring different charge assignments for the new abelian gauge symmetry are left to future studies. 
\\
The RG analysis has been performed employing two-loop $\beta$ functions and one-loop matching relations on the initial conditions at the electroweak scale. 
We have shown that such higher-order corrections affect considerably the running of the couplings and, probably, should always be kept into account in any 
significant phenomenological analysis. They are essential for a consistent identification of the allowed regions in the parameter space of a given model, 
under the requirements of perturbativity and of vacuum stability. \\
Following this reasoning, one finds that a charge assignment of B-L type, as shown in Figs. \ref{Fig.PertCoupGeneric} and \ref{Fig.PertCoup}, can be characterized 
by a perturbative behaviour up to the Planck scale. \\
We have seen that the requirements of perturbativity and stability of the scalar potential constrain the masses of the heavy Higgs and of the RH neutrinos quite 
significantly. The boundaries of the allowed region, for a given mass of the $Z'$, depend on the values of the scalar mixing angle $\theta$ and of the abelian 
coupling $g'$. 
We have also verified that the transition from a region of stability to an unstable one can be quite abrupt, and may depend quite significantly on the mass of 
the RH neutrino, as illustrated in Fig. \ref{shadows}. This feature can be ascribed to the evolution of the scalar coupling $\lambda_2$, and, in particular, 
to the Yukawa ($Y_N$) of the RH neutrinos which affects their masses. As already mentioned above, the roles of $Y_N$ and of $Y_t$, from this perspective, 
appear to be quite similar. Finally, we have also investigated the impact of the kinetic mixing on the evolution of the $U(1)_Y\times U(1)_{B-L}$ symmetries. \\
There are other aspects, not touched in this analysis, and that we leave to future studies, which concern the inclusion of this model into a specific unified theory. 
It is well know that multiple $U(1)$ factors are  often encountered in the breaking of a GUT scenario, or in the construction of effective theories from string models 
\cite{Alon1,Faraggi:2014ica}
so that it is natural to ask what a vacuum stability analysis can add in predictivity using the corresponding RGE's.  From our point of view its use in this context 
can be dual. Without any prior knowledge of a particular breaking pattern, the study of the instabilities of the scalar potential gives an indication of the validity of the $SM\times U(1)_{B-L}$ gauge symmetry 
as a residual one, valid at low energy, before calling for a UV completion (either of GUT type or else). More significant theoretical constraints on the parameter space can instead 
be drawn when the model is embedded in a particular GUT scenario. In this case the stability of the vacuum between 
the threshold scales of the intermediate breakings is a necessary condition for its consistency. 
From this perspective, our study clarifies that a heavy neutrino mass may challenge specific GUT models. Indeed, we have shown that the running of the corresponding Yukawa couplings 
could spoil the perturbativity and the stability of the vacuum if a new threshold is not sufficiently near the TeV scale.
\\
Another important point concerns the role of a TeV B-L scenario respect to the constraints that could be imposed by requiring a successful leptogenesis. 
Analysis in this direction have been presented in \cite{Abbas:2007ag,Iso:2010mv,Blanchet}. 
We have verified that also in the model that we have considered, the value of the baryon abundance $n_B \sim 10^{-2}\,\mathcal{\epsilon}\,\,\eta_{_{T_{Sp}}}$,
which depends on product of the CP-asymmetry of the heavy neutrinos decays ($\epsilon$) and on the efficiency factor $(\eta_{_{T_{Sp}}})$, can account for the current estimated value of $n_B\sim  10^{-10}$. $\eta_{_{T_{Sp}}}$ is temperature dependent and is calculated at the sphaleron freeze-out temperature  $(\eta_{_{T_{Sp}}})$. It is responsible for the amount of washout of the baryon asymmetry, which can be dangerously low when the condition $M_{Z'} > 2\,M_{\nu_h}$ is realized \cite{Blanchet}. 
Always in a context of a very strong near-degeneracy of the heavy neutrinos, a pattern which can be completely encompassed by our choice of a scalar $Y_N$, we have found regions, in the 
parameter space that we have analysed, which are compatible with leptogenesis. 
For instance, for a mass of the extra $Z'$ of 2.5 TeV and a coupling $g'=0.1$, we have verified that the requirement of a maximal value of the CP-asymmetry parameter $\epsilon$ starts to
allow leptogenesis ($\eta_{_{T_{Sp}}}\sim 10^{-8}$) if $M_{\nu_h} \geq 200$ GeV. Meanwhile, less stringent values of $\epsilon$, $\epsilon \sim O(10^{-1})-O(10^{-2})$, are feasible if
$M_{\nu_h} \geq 400$ GeV and $1$ TeV respectively. We have found, as expected, that an increment of the strength of the $Z'$ coupling $g'$ strongly disfavours
the leptogenesis scenario, since it intensifies the washout of the asymmetry, for a 2.5 TeV gauge boson.  \\

\centerline{\bf Acknowledgements} 
We thank Alon Faraggi and Pasquale Di Bari for discussions. L.D.R. thanks Kostas Skenderis for hospitality at 
the University of Southampton, and Nora Brambilla and Antonio Vairo for their hospitality during a recent visit at the Technical University of Munich. L.D.R. is grateful to F. Lyonnet for pointing out a typo in the $\beta$ functions and for accurate cross-checks. The work of C.C. is supported by a {\em The Leverhulme Trust Visiting Professorship} at the University of Southampton in the STAG Research Centre and Mathematical Sciences Department, whose members he thanks for the kind hospitality.

\appendix

\section{Appendix. Two-loop $\beta$-functions in B-L }
\label{betafuncs}
We give in this appendix the complete set of the two-loop $\beta$-functions, for the charge assignment  $z_Q = \frac{1}{3}$ and $z_u = \frac{1}{3}$, required for the stability analysis. 
\subsection*{$\beta$-functions of gauge coupling constants}
\allowdisplaybreaks
\bea
\beta^{(2)}_{{g_2}} =
g_2^3 \left(4 {g'}\, \tilde{g}+\frac{3 \tilde{g}^2}{2}+\frac{3 g^2}{2}+12 g_3^2+4  {g'}^2-\frac{3 Y_t^2}{2}\right)+
\frac{35 g_2^5}{6}
\eea
\bea
\beta^{(2)}_{{g_3}} = g_3^3 \left(\frac{4 {g'} \tilde{g}}{3}+\frac{11\, \tilde{g}^2}{6}+\frac{11 g^2}{6}+\frac{9 g_2^2}{2}+\frac{4 {g'}^2}{3}-2 Y_t^2\right)
- 26 g_3^5
\eea
\bea
\beta^{(2)}_{{g}} = g^3 \left(\frac{164 {g'} \tilde{g}}{9}+\frac{199
   \tilde{g}^2}{18}+\frac{9 g_2^2}{2}+\frac{44
   g_3^2}{3}+\frac{92 {g'}^2}{9}-\frac{17
   Y_t^2}{6}\right)+\frac{199 g^5}{18}
\eea
\bea
&\beta^{(2)}_{{\tilde{g}}}&= \tilde{g}^2 \left(\frac{656 g^2 {g'}}{9}+12 g_2^2
   {g'}+\frac{32 g_3^2 {g'}}{3}+\frac{448 {g'}^3}{9}-\frac{10 {g'}  Y_t^2}{3}\right) \nonumber \\ 
   && +\tilde{g}^3 \left(\frac{199  g^2}{6}+\frac{9 g_2^2}{2}+\frac{44 g_3^2}{3}+\frac{184  {g'}^2}{3}-\frac{17 Y_t^2}{6}\right) \nonumber \\ 
   && +\tilde{g}   \left(\frac{199 g^4}{9}+\frac{644 g^2 {g'}^2}{9}-\frac{17}{3} g^2 Y_t^2+9 g_2^2  g^2+\frac{88}{3} g_3^2 g^2+12 g_2^2   {g'}^2 \right. \nonumber \\
           && \left. + \frac{32}{3} g_3^2 {g'}^2+\frac{800 {g'}^4}{9}-12 {g'}^2 Y_N^2-\frac{4}{3}   {g'}^2 Y_t^2\right) \nonumber \\
   &&+\frac{328 {g'} \tilde{g}^4}{9}+\frac{199 \tilde{g}^5}{18}+\frac{164 g^4 {g'}}{9}+\frac{224 g^2 {g'}^3}{9} \nonumber \\ 
   &&-\frac{10}{3} g^2 {g'} Y_t^2+12 g^2 g_2^2 {g'}+\frac{32}{3} g^2 g_3^2 {g'} 
\eea

\bea
&\beta^{(2)}_{g'} &=
{g'}^3 \left(\frac{184 \tilde{g}^2}{3}+\frac{92
   g^2}{9}+12 g_2^2+\frac{32 g_3^2}{3}-12 Y_N^2-\frac{4
   Y_t^2}{3}\right) \nonumber \\
   && +{g'}^2 \left(\frac{164 g^2
   \tilde{g}}{9}-\frac{10}{3} \tilde{g} Y_t^2+\frac{328
   \tilde{g}^3}{9}+12 g_2^2 \tilde{g}+\frac{32}{3} g_3^2
   \tilde{g}\right) \nonumber \\ 
   &&+{g'} \left(\frac{199}{18} g^2
   \tilde{g}^2-\frac{17}{6} \tilde{g}^2 Y_t^2+\frac{199
   \tilde{g}^4}{18}+\frac{9}{2} g_2^2
   \tilde{g}^2+\frac{44}{3} g_3^2
   \tilde{g}^2\right)\nonumber \\
   &&+\frac{448 {g'}^4
   \tilde{g}}{9}+\frac{800 {g'}^5}{9}
\eea

\subsection*{$\beta$-functions of Yukawa interactions}
\allowdisplaybreaks
\bea
&\beta^{(2)}_{{Y_{t}}} &= Y_t^3 \left(\frac{25 {g'} \tilde{g}}{4}+\frac{131
   \tilde{g}^2}{16}+\frac{131 g^2}{16}+\frac{225
   g_2^2}{16}+36 g_3^2+3 {g'}^2-12 \lambda
   _1\right) \nonumber \\ 
   &&+Y_t \left(\frac{502}{27} g^2 {g'}
   \tilde{g}+\frac{1187}{108} g^2 \tilde{g}^2+\frac{665
   {g'}^3 \tilde{g}}{27}+\frac{1085}{36} {g'}^2
   \tilde{g}^2+\frac{502 {g'}
   \tilde{g}^3}{27} \right. \nonumber \\ 
   &&\left. +\frac{9}{4} g_2^2 {g'}
   \tilde{g}-\frac{20}{9} g_3^2 {g'}
   \tilde{g}+\frac{1187 \tilde{g}^4}{216}-\frac{3}{4}
   g_2^2 \tilde{g}^2+\frac{19}{9} g_3^2
   \tilde{g}^2+\frac{1187 g^4}{216}\right. \nonumber \\ && \left.+\frac{91 g^2
   {g'}^2}{12}   -\frac{3}{4} g_2^2 g^2+\frac{19}{9}
   g_3^2 g^2+\frac{3}{4} g_2^2 {g'}^2-\frac{8}{9}
   g_3^2 {g'}^2-\frac{23 g_2^4}{4}\right. \nonumber \\ && \left.-108 g_3^4+9 g_2^2
   g_3^2+\frac{203 {g'}^4}{27}  +6 \lambda
   _1^2+\frac{\lambda _3^2}{2}\right)-12 Y_t^5
\eea
\bea
&\beta^{(2)}_{{Y_{N}}} &= Y_N \left(-\frac{32 {g'}^3
   \tilde{g}}{3}-\frac{35}{6} {g'}^2 \tilde{g}^2-127
   {g'}^4+4 \lambda _2^2+\lambda
   _3^2\right) \nonumber \\ && +\left(206 {g'}^2-32 \lambda _2\right)
   Y_N^3-44 Y_N^5
\eea
\subsection*{$\beta$-functions of quartic scalar interactions}
\allowdisplaybreaks
\bea
&\beta^{(2)}_{{\lambda_{1}}} &= -\frac{32}{3} g^4 {g'} \tilde{g}-\frac{379}{16} g^4
   \tilde{g}^2-13 g^2 {g'}^2 \tilde{g}^2-10 g^2
   {g'} \tilde{g} Y_t^2-\frac{64}{3} g^2 {g'}
   \tilde{g}^3 \nonumber \\
   &&-\frac{32}{3} g_2^2 g^2 {g'}
   \tilde{g}+\lambda _1^2 \left(36 \tilde{g}^2+36 g^2+108
   g_2^2-144 Y_t^2\right) \nonumber \\
   &&-\frac{19}{2} g^2 \tilde{g}^2
   Y_t^2-\frac{379}{16} g^2 \tilde{g}^4-\frac{559}{24}
   g_2^2 g^2 \tilde{g}^2+\lambda _1 \left(\frac{80}{3}
   g^2 {g'} \tilde{g}+\frac{629}{12} g^2
   \tilde{g}^2  
   \right. \nonumber \\ && \left.   
   + 34 {g'}^2 \tilde{g}^2+\frac{50}{3}
   {g'} \tilde{g} Y_t^2+\frac{80 {g'}
   \tilde{g}^3}{3}+\frac{85}{6} \tilde{g}^2
   Y_t^2+\frac{629 \tilde{g}^4}{24}+\frac{39}{4} g_2^2
   \tilde{g}^2\right. \nonumber \\ && \left.+\frac{629 g^4}{24}+\frac{85}{6} g^2
   Y_t^2+\frac{39}{4} g_2^2 g^2+\frac{45}{2} g_2^2
   Y_t^2+80 g_3^2 Y_t^2-\frac{73 g_2^4}{8}\right. \nonumber \\ && \left.+\frac{20}{3}
   {g'}^2 Y_t^2-10 \lambda _3^2-3 Y_t^4\right)+20
   {g'}^2 \lambda _3 \tilde{g}^2-4 {g'}^2
   \tilde{g}^2 Y_t^2-13 {g'}^2 \tilde{g}^4 \nonumber \\ && -13 g_2^2
   {g'}^2 \tilde{g}^2-\frac{20}{3} {g'}
   \tilde{g} Y_t^4-10 {g'} \tilde{g}^3 Y_t^2+6 g_2^2
   {g'} \tilde{g} Y_t^2-\frac{32 {g'}
   \tilde{g}^5}{3}\nonumber \\ && -\frac{32}{3} g_2^2 {g'}
   \tilde{g}^3-\frac{8}{3} \tilde{g}^2 Y_t^4-\frac{19}{4}
   \tilde{g}^4 Y_t^2+\frac{21}{2} g_2^2 \tilde{g}^2
   Y_t^2-\frac{379 \tilde{g}^6}{48}-\frac{289}{48} g_2^4
   \tilde{g}^2\nonumber \\ &&-\frac{559}{48} g_2^2 \tilde{g}^4-\frac{379
   g^6}{48}-\frac{19}{4} g^4 Y_t^2-\frac{559}{48} g_2^2
   g^4-\frac{8}{3} g^2 Y_t^4+\frac{21}{2} g_2^2 g^2
   Y_t^2\nonumber \\ &&-\frac{289}{48} g_2^4 g^2-32 g_3^2
   Y_t^4-\frac{9}{4} g_2^4 Y_t^2+\frac{305 g_2^6}{16}+32
   {g'}^2 \lambda _3^2-\frac{8}{3} {g'}^2
   Y_t^4\nonumber \\ &&-312 \lambda _1^3-4 \lambda _3^3-12 \lambda _3^2
   Y_N^2+30 Y_t^6
   \eea
\bea
&\beta^{(2)}_{{\lambda_{2}}} &= -\frac{8192 {g'}^5 \tilde{g}}{3}-\frac{5344}{3}
   {g'}^4 \tilde{g}^2+40 {g'}^2 \lambda _3
   \tilde{g}^2+ \nonumber \\ && \lambda _2 \left(\frac{1280 {g'}^3
   \tilde{g}}{3}+\frac{844}{3} {g'}^2
   \tilde{g}^2 + 2112 {g'}^4+120 {g'}^2 Y_N^2-20
   \lambda _3^2+48 Y_N^4\right)  \nonumber \\ && +4 \lambda _3^2
   \tilde{g}^2+4 g^2 \lambda _3^2+12 g_2^2 \lambda
   _3^2-7168 {g'}^6+768 {g'}^4 Y_N^2+\lambda
   _2^2 \left(448 {g'}^2-240 Y_N^2\right) \nonumber \\ && +192
   {g'}^2 Y_N^4-240 \lambda _2^3-8 \lambda _3^3+768
   Y_N^6-12 \lambda _3^2 Y_t^2
   \eea
\bea
&\beta^{(2)}_{{\lambda_{3}}} &= -\frac{512}{3} g^2 {g'}^3 \tilde{g}+\lambda _3^2
   \left(\tilde{g}^2+g^2+3 g_2^2+16 {g'}^2-72
   \lambda _1-48 \lambda _2-24 Y_N^2-12
   Y_t^2\right)\nonumber \\&&-\frac{713}{3} g^2 {g'}^2
   \tilde{g}^2+\lambda _3 \left(\frac{40}{3} g^2
   {g'} \tilde{g}+\frac{557}{24} g^2
   \tilde{g}^2+\frac{640 {g'}^3
   \tilde{g}}{3}+\frac{497}{3} {g'}^2
   \tilde{g}^2\right. \nonumber \\ && \left.+\frac{25}{3} {g'} \tilde{g}
   Y_t^2+\frac{40 {g'} \tilde{g}^3}{3}+24 \lambda _1
   \tilde{g}^2+\frac{85}{12} \tilde{g}^2 Y_t^2+\frac{557
   \tilde{g}^4}{48}+\frac{15}{8} g_2^2
   \tilde{g}^2+\frac{557 g^4}{48}\right. \nonumber \\ && \left.+24 g^2 \lambda
   _1+\frac{85}{12} g^2 Y_t^2+\frac{15}{8} g_2^2 g^2+72
   g_2^2 \lambda _1+\frac{45}{4} g_2^2 Y_t^2+40 g_3^2
   Y_t^2-\frac{145 g_2^4}{16}\right. \nonumber \\ && \left.+672 {g'}^4+256
   {g'}^2 \lambda _2+60 {g'}^2
   Y_N^2+\frac{10}{3} {g'}^2 Y_t^2-60 \lambda
   _1^2-40 \lambda _2^2-96 \lambda _2 Y_N^2\right. \nonumber \\ && \left.-72 Y_N^4-72
   \lambda _1 Y_t^2-\frac{27 Y_t^4}{2}\right)-656
   {g'}^4 \tilde{g}^2-160 {g'}^3 \tilde{g}
   Y_t^2-\frac{1024}{3} {g'}^3 \tilde{g}^3\nonumber \\&&+120
   {g'}^2 \lambda _1 \tilde{g}^2+80 {g'}^2
   \lambda _2 \tilde{g}^2+48 {g'}^2 \tilde{g}^2
   Y_N^2-76 {g'}^2 \tilde{g}^2 Y_t^2-\frac{713}{3}
   {g'}^2 \tilde{g}^4\nonumber \\&&-45 g_2^2 {g'}^2
   \tilde{g}^2-64 {g'}^4 Y_t^2-11 \lambda _3^3
   \eea

\section{ Neutrino mass eigenstates in type-I seesaw}
To investigate the nature of the neutrino mass eigenstates we introduce a six dimensional vector in flavor space $\psi = (\nu_L,\nu_R^c)$ so that Eq.~(\ref{TypeISeesawEffective}) becomes  
\bea \label{TypeISeesawEffective2}
- \mathcal L_{Y}^\nu =  \frac{1}{2}\, \psi^T \, \mathcal M \,  \psi + h.c. \, ,
\eea
with a $6 \times 6$ mass matrix $\mathcal M$ of the form 
\bea \label{TypeISeesawMassMatrix}
\mathcal M = \left( \begin{array}{cc} 0 & M^{T}_{d} \\  M_{d} & M_{m} \end{array} \right)\, = \frac{1}{\sqrt{2}}\left( \begin{array}{cc} 0 &  v\,Y^T_\nu  \\   v\,Y_\nu & 2\,v'\, Y_N \, \end{array} \right)\,.
\eea
For the sake of simplicity, all the flavor indices have been omitted.
The mass matrix obtained in Eq.~(\ref {TypeISeesawMassMatrix}) can be formally set in a block-diagonal form with the help of a unitary matrix $\mathcal U$, implicitly defined as
\bea \label{diagonalization}
\left( \begin{array}{cc}  M_{\nu_l}  & 0 \\  0 & M_{\nu_h} \end{array} \right) = \mathcal{U}^{T} \left( \begin{array}{cc} 0 & M^{T}_{d} \\  M_{d} & M_{m} \end{array} \right) \mathcal{U} \,,
\eea 
where $M_{\nu_l}$ and $M_{\nu_h}$ are (generally non diagonal) $3 \times 3$ blocks. When expressed in terms of the rotated states $\psi' = \left( \nu_l,\nu^{c}_h \right) = \mathcal{U}^{-1}\psi$ the following matrix structure arises
\bea
- \mathcal L_{Y}^\nu &=&  \frac{1}{2}\, M_{\nu_l}^{ij} \nu_l^i \nu_l^j + \frac{1}{2}\, M_{\nu_h}^{ij} \nu^i_h \nu^j_h + h.c. \, ,
\eea 
Notice that the Dirac mass matrix $M_{d}$ has entries  at most of the order of the electroweak scale, meanwhile the Majorana mass $M_{m}$ for the sterile neutrinos is usually allowed to rise up to
the GUT scale or even up to Planck scale. \\
Such hierarchy is necessary for the implementation of the seesaw mechanism and translates into a hierarchy between $M_{\nu_l}$ and $M_{\nu_h}$. The same separation in scales justifies the use of a particular Ansatz \cite{Ibarra:2010xw} for the unitary matrix $\mathcal U$ employed to block-diagonalize the mass matrix in Eq.~(\ref{TypeISeesawMassMatrix}) 
\bea
\label{Uexp}
\mathcal U = \exp \left( \begin{array}{cc} 0 & \mathcal V \\  - \mathcal V^{\dagger} & 0 \end{array} \right) \simeq \left( \begin{array}{cc} 1-\frac{1}{2}\,\mathcal V \mathcal V ^{\dagger} & \mathcal V \\  - \mathcal V^{\dagger} & 1-\frac{1}{2}\,\mathcal V ^{\dagger}\mathcal V  \end{array} \right) \, ,
\eea
where $\mathcal V$ is a non-diagonal $3\times 3$ matrix with small elements. Inserting the matrix expansion given in Eq.~(\ref{Uexp}) into Eq.~(\ref{diagonalization}) one obtains in the following constraints
\bea \label{constraints}
0 &\simeq & M_{d} - M _{m}\mathcal V^{\dagger}  - \frac{1}{2} \, \mathcal V^{T}  \mathcal V^{*}  M_d - \mathcal V^{T}  M_d^T \mathcal V^{\dagger} - \frac{1}{2} \, M_d \mathcal V \mathcal V^{\dagger}\, , \nn \\
M_{\nu_l} & \simeq & -\mathcal V^{*} M_d - M_d^T \mathcal V^{\dagger} + \mathcal   V^{*} M_m  \mathcal V^{\dagger} \, , \nn \\
M_{\nu_h} & \simeq & M_m +\mathcal V^T M_d^T + M_d \mathcal V - \frac{1}{2} \mathcal V^{T} \mathcal V^{*} M_m - \frac{1}{2} M_m \mathcal V^{\dagger} \mathcal V  \, .
\eea
The first equality in Eq.~(\ref{constraints}), evaluated at leading order in $\mathcal V$, is used to fix $\mathcal V^{\dagger} \simeq M_m^{-1} M_d $. From the remaining conditions we are now able to recover the usual type-I seesaw relations for the light and the heavy mass matrices given in Eq.~(\ref{TypeISeesawMasses1}).\\
In order to completely resolve the neutrino spectrum, the mass matrices $m_{\nu_l}$ and $m_{\nu_h}$ must be separately diagonalized, leading to
\bea
\label{diag2}
 L^{T} M_{\nu_l} L^{-1} &=&  \textrm{diag}\left(m_{\nu_{l1}}, m_{\nu_{l2}}, m_{\nu_{l3}} \right) \,, \nn \\
H^{T} M_{\nu_h} H^{-1} & =& \textrm{diag}\left(m_{\nu_{h1}}, m_{\nu_{h2}}, m_{\nu_{h3}} \right) \, ,
\eea
for the light and the heavy neutrino mass eigenstates. This further step is of fundamental importance to investigate the (non-unitary) nature of neutrino mixing in a type-I seesaw scenario, given its role in defining the corresponding Pontecorvo-Maki-Nakagawa-Sakata (PMNS) matrix \cite{Pontecorvo:1957qd,Maki:1962mu} as $\mathcal U_{PMNS} = L_{c}^{\dagger} (1+\eta) L$, where $L_{c}$ diagonalizes the mass matrix for the charged leptons and $\eta = - \mathcal V \mathcal V^\dag /2$. \\
Such fine-grained analysis, however, is an unnecessary complication in our study and, for the sake of simplicity, 
we consider diagonal Yukawa couplings.

\providecommand{\href}[2]{#2}\begingroup\raggedright\endgroup


\begin{thebibliography}{10}

\bibitem{Cabibbo:1979ay}
N.~Cabibbo, L.~Maiani, G.~Parisi, and R.~Petronzio, {\it {Bounds on the
  Fermions and Higgs Boson Masses in Grand Unified Theories}},  {\em
  Nucl.Phys.} {\bf B158} (1979) 295--305.

\bibitem{Hung:1979dn}
P.~Q. Hung, {\it {Vacuum Instability and New Constraints on Fermion Masses}},
  {\em Phys. Rev. Lett.} {\bf 42} (1979) 873.

\bibitem{Lindner:1985uk}
M.~Lindner, {\it {Implications of Triviality for the Standard Model}},  {\em
  Z.Phys.} {\bf C31} (1986) 295.

\bibitem{Lindner:1988ww}
M.~Lindner, M.~Sher, and H.~W. Zaglauer, {\it {Probing Vacuum Stability Bounds
  at the Fermilab Collider}},  {\em Phys.Lett.} {\bf B228} (1989) 139.

\bibitem{Ford:1992mv}
C.~Ford, D.~Jones, P.~Stephenson, and M.~Einhorn, {\it {The Effective potential
  and the renormalization group}},  {\em Nucl.Phys.} {\bf B395} (1993) 17--34,
  [\href{http://xxx.lanl.gov/abs/hep-lat/9210033}{{\tt hep-lat/9210033}}].

\bibitem{Bezrukov:2012sa}
F.~Bezrukov, M.~Y. Kalmykov, B.~A. Kniehl, and M.~Shaposhnikov, {\it {Higgs
  Boson Mass and New Physics}},  {\em JHEP} {\bf 1210} (2012) 140,
  [\href{http://xxx.lanl.gov/abs/1205.2893}{{\tt arXiv:1205.2893}}].

\bibitem{Branchina:2013jra}
V.~Branchina and E.~Messina, {\it {Stability, Higgs Boson Mass and New
  Physics}},  {\em Phys. Rev. Lett.} {\bf 111} (2013) 241801,
  [\href{http://xxx.lanl.gov/abs/1307.5193}{{\tt arXiv:1307.5193}}].

\bibitem{Buttazzo:2013uya}
D.~Buttazzo, G.~Degrassi, P.~P. Giardino, G.~F. Giudice, F.~Sala, {\em
  et.~al.}, {\it {Investigating the near-criticality of the Higgs boson}},
  {\em JHEP} {\bf 1312} (2013) 089,
  [\href{http://xxx.lanl.gov/abs/1307.3536}{{\tt arXiv:1307.3536}}].

\bibitem{Branchina:2014usa}
V.~Branchina, E.~Messina, and A.~Platania, {\it {Top mass determination, Higgs
  inflation, and vacuum stability}},  {\em JHEP} {\bf 09} (2014) 182,
  [\href{http://xxx.lanl.gov/abs/1407.4112}{{\tt arXiv:1407.4112}}].

\bibitem{Branchina:2014rva}
V.~Branchina, E.~Messina, and M.~Sher, {\it {Lifetime of the electroweak vacuum
  and sensitivity to Planck scale physics}},  {\em Phys. Rev.} {\bf D91} (2015)
  013003, [\href{http://xxx.lanl.gov/abs/1408.5302}{{\tt arXiv:1408.5302}}].

\bibitem{DiLuzio:2015iua}
L.~Di~Luzio, G.~Isidori, and G.~Ridolfi, {\it {Stability of the electroweak
  ground state in the Standard Model and its extensions}},
  \href{http://xxx.lanl.gov/abs/1509.05028}{{\tt arXiv:1509.05028}}.

\bibitem{Anderson:1990aa}
G.~W. Anderson, {\it {New Cosmological Constraints on the Higgs Boson and Top
  Quark Masses}},  {\em Phys. Lett.} {\bf B243} (1990) 265--270.

\bibitem{Arnold:1991cv}
P.~B. Arnold and S.~Vokos, {\it {Instability of hot electroweak theory: bounds
  on m(H) and M(t)}},  {\em Phys. Rev.} {\bf D44} (1991) 3620--3627.

\bibitem{Espinosa:1995se}
J.~R. Espinosa and M.~Quiros, {\it {Improved metastability bounds on the
  standard model Higgs mass}},  {\em Phys. Lett.} {\bf B353} (1995) 257--266,
  [\href{http://xxx.lanl.gov/abs/hep-ph/9504241}{{\tt hep-ph/9504241}}].

\bibitem{Espinosa:2007qp}
J.~R. Espinosa, G.~F. Giudice, and A.~Riotto, {\it {Cosmological implications
  of the Higgs mass measurement}},  {\em JCAP} {\bf 0805} (2008) 002,
  [\href{http://xxx.lanl.gov/abs/0710.2484}{{\tt arXiv:0710.2484}}].

\bibitem{Kobakhidze:2013tn}
A.~Kobakhidze and A.~Spencer-Smith, {\it {Electroweak Vacuum (In)Stability in
  an Inflationary Universe}},  {\em Phys. Lett.} {\bf B722} (2013) 130--134,
  [\href{http://xxx.lanl.gov/abs/1301.2846}{{\tt arXiv:1301.2846}}].

\bibitem{Enqvist:2013kaa}
K.~Enqvist, T.~Meriniemi, and S.~Nurmi, {\it {Generation of the Higgs
  Condensate and Its Decay after Inflation}},  {\em JCAP} {\bf 1310} (2013)
  057, [\href{http://xxx.lanl.gov/abs/1306.4511}{{\tt arXiv:1306.4511}}].

\bibitem{Bezrukov:2014ipa}
F.~Bezrukov, J.~Rubio, and M.~Shaposhnikov, {\it {Living beyond the edge: Higgs
  inflation and vacuum metastability}},
  \href{http://xxx.lanl.gov/abs/1412.3811}{{\tt arXiv:1412.3811}}.

\bibitem{Fairbairn:2014nxa}
M.~Fairbairn, P.~Grothaus, and R.~Hogan, {\it {The Problem with False Vacuum
  Higgs Inflation}},  {\em JCAP} {\bf 1406} (2014) 039,
  [\href{http://xxx.lanl.gov/abs/1403.7483}{{\tt arXiv:1403.7483}}].

\bibitem{Enqvist:2014bua}
K.~Enqvist, T.~Meriniemi, and S.~Nurmi, {\it {Higgs Dynamics during
  Inflation}},  {\em JCAP} {\bf 1407} (2014) 025,
  [\href{http://xxx.lanl.gov/abs/1404.3699}{{\tt arXiv:1404.3699}}].

\bibitem{Kobakhidze:2014xda}
A.~Kobakhidze and A.~Spencer-Smith, {\it {The Higgs vacuum is unstable}},
  \href{http://xxx.lanl.gov/abs/1404.4709}{{\tt arXiv:1404.4709}}.

\bibitem{Herranen:2014cua}
M.~Herranen, T.~Markkanen, S.~Nurmi, and A.~Rajantie, {\it {Spacetime curvature
  and the Higgs stability during inflation}},  {\em Phys. Rev. Lett.} {\bf 113}
  (2014), no.~21 211102, [\href{http://xxx.lanl.gov/abs/1407.3141}{{\tt
  arXiv:1407.3141}}].

\bibitem{Kamada:2014ufa}
K.~Kamada, {\it {Inflationary cosmology and the standard model Higgs with a
  small Hubble induced mass}},  {\em Phys. Lett.} {\bf B742} (2015) 126--135,
  [\href{http://xxx.lanl.gov/abs/1409.5078}{{\tt arXiv:1409.5078}}].

\bibitem{Shkerin:2015exa}
A.~Shkerin and S.~Sibiryakov, {\it {On stability of electroweak vacuum during
  inflation}},  {\em Phys. Lett.} {\bf B746} (2015) 257--260,
  [\href{http://xxx.lanl.gov/abs/1503.0258}{{\tt arXiv:1503.0258}}].

\bibitem{Espinosa:2015qea}
J.~R. Espinosa, G.~F. Giudice, E.~Morgante, A.~Riotto, L.~Senatore, A.~Strumia,
  and N.~Tetradis, {\it {The cosmological Higgstory of the vacuum
  instability}},  \href{http://xxx.lanl.gov/abs/1505.04825}{{\tt
  arXiv:1505.04825}}.

\bibitem{Rose:2015lna}
L.~Delle~Rose, C.~Marzo, and A.~Urbano, {\it {On the fate of the Standard Model
  at finite temperature}},  \href{http://xxx.lanl.gov/abs/1507.06912}{{\tt
  arXiv:1507.06912}}.

\bibitem{Hook:2014uia}
  A.~Hook, J.~Kearney, B.~Shakya and K.~M.~Zurek, {\it {Probable or Improbable Universe? Correlating Electroweak Vacuum Instability with the Scale of Inflation}},
   {\em JHEP} {\bf 1501} (2015) 061, [\href{http://xxx.lanl.gov/abs/1404.5953}{{\tt arXiv:1404.5953}}].

\bibitem{Kearney:2015vba}
  J.~Kearney, H.~Yoo and K.~M.~Zurek, {\it {Is a Higgs Vacuum Instability Fatal for High-Scale Inflation?}}, 
{\em Phys.Rev.} {\bf D91} (2015) no.~12 123537, [\href{http://xxx.lanl.gov/abs/1503.05193}{{\tt arXiv:1503.05193}}].

\bibitem{Masina:2012tz}
I.~Masina, {\it {Higgs boson and top quark masses as tests of electroweak
  vacuum stability}},  {\em Phys.Rev.} {\bf D87} (2013), no.~5 053001,
  [\href{http://xxx.lanl.gov/abs/1209.0393}{{\tt arXiv:1209.0393}}].

\bibitem{Alekhin:2012py}
S.~Alekhin, A.~Djouadi, and S.~Moch, {\it {The top quark and Higgs boson masses
  and the stability of the electroweak vacuum}},  {\em Phys. Lett.} {\bf B716}
  (2012) 214--219, [\href{http://xxx.lanl.gov/abs/1207.0980}{{\tt
  arXiv:1207.0980}}].

\bibitem{Basso:2010yz}
L.~Basso, S.~Moretti, and G.~M. Pruna, {\it {Phenomenology of the minimal B-L
  extension of the Standard Model: the Higgs sector}},  {\em Phys.Rev.} {\bf
  D83} (2011) 055014, [\href{http://xxx.lanl.gov/abs/1011.2612}{{\tt
  arXiv:1011.2612}}].

\bibitem{Basso:2010jm}
L.~Basso, S.~Moretti, and G.~M. Pruna, {\it {A Renormalisation Group Equation
  Study of the Scalar Sector of the Minimal B-L Extension of the Standard
  Model}},  {\em Phys.Rev.} {\bf D82} (2010) 055018,
  [\href{http://xxx.lanl.gov/abs/1004.3039}{{\tt arXiv:1004.3039}}].

\bibitem{Basso:2013vla}
L.~Basso, {\it {Minimal Z' models and the 125 GeV Higgs boson}},  {\em Phys.
  Lett.} {\bf B725} (2013) 322--326,
  [\href{http://xxx.lanl.gov/abs/1303.1084}{{\tt arXiv:1303.1084}}].

\bibitem{Datta:2013mta}
A.~Datta, A.~Elsayed, S.~Khalil, and A.~Moursy, {\it {Higgs vacuum stability in
  the B-L extended standard model}},  {\em Phys.Rev.} {\bf D88} (2013), no.~5
  053011, [\href{http://xxx.lanl.gov/abs/1308.0816}{{\tt arXiv:1308.0816}}].

\bibitem{Chakrabortty:2013zja}
J.~Chakrabortty, P.~Konar, and T.~Mondal, {\it {Constraining a class of B-L
  extended models from vacuum stability and perturbativity}},  {\em Phys. Rev.}
  {\bf D89} (2014), no.~5 056014,
  [\href{http://xxx.lanl.gov/abs/1308.1291}{{\tt arXiv:1308.1291}}].

\bibitem{DiChiara:2014wha}
S.~Di~Chiara, V.~Keus, and O.~Lebedev, {\it {Stabilizing the Higgs potential
  with a Z$'$}},  {\em Phys. Lett.} {\bf B744} (2015) 59--66,
  [\href{http://xxx.lanl.gov/abs/1412.7036}{{\tt arXiv:1412.7036}}].

\bibitem{Anchordoqui:2015fra}
L.~A. Anchordoqui, V.~Barger, H.~Goldberg, X.~Huang, D.~Marfatia, L.~H.~M.
  da~Silva, and T.~J. Weiler, {\it {Majorana dark matter through the Higgs
  portal under the vacuum stability lamppost}},  {\em Phys. Rev.} {\bf D92}
  (2015), no.~6 063504, [\href{http://xxx.lanl.gov/abs/1506.0470}{{\tt
  arXiv:1506.0470}}].

\bibitem{Oda:2015gna}
S.~Oda, N.~Okada, and D.-s. Takahashi, {\it {Classically conformal U(1)
  extended standard model and Higgs vacuum stability}},  {\em Phys. Rev.} {\bf
  D92} (2015), no.~1 015026, [\href{http://xxx.lanl.gov/abs/1504.0629}{{\tt
  arXiv:1504.0629}}].

\bibitem{Haba:2015rha}
N.~Haba and Y.~Yamaguchi, {\it {Vacuum stability in the $U(1)_\chi$ extended
  model with vanishing scalar potential at the Planck scale}},  {\em PTEP} {\bf
  2015} (2015), no.~9 093B05, [\href{http://xxx.lanl.gov/abs/1504.05669}{{\tt
  arXiv:1504.05669}}].

\bibitem{Coriano:2014mpa}
C.~Corian\`o, L.~Delle~Rose, and C.~Marzo, {\it {Vacuum Stability in U(1)-Prime
  Extensions of the Standard Model with TeV Scale Right Handed Neutrinos}},
  {\em Phys. Lett.} {\bf B738} (2014) 13--19,
  [\href{http://xxx.lanl.gov/abs/1407.8539}{{\tt arXiv:1407.8539}}].

\bibitem{Das:2015nwk}
A.~Das, N.~Okada, and N.~Papapietro, {\it {Electroweak vacuum stability in
  classically conformal B-L extension of the Standard Model}},
  \href{http://xxx.lanl.gov/abs/1509.01466}{{\tt arXiv:1509.01466}}.

\bibitem{Basso:2008iv}
L.~Basso, A.~Belyaev, S.~Moretti, and C.~H. Shepherd-Themistocleous, {\it
  {Phenomenology of the minimal B-L extension of the Standard model: Z' and
  neutrinos}},  {\em Phys. Rev.} {\bf D80} (2009) 055030,
  [\href{http://xxx.lanl.gov/abs/0812.4313}{{\tt arXiv:0812.4313}}].

\bibitem{Basso:2010hk}
L.~Basso, S.~Moretti, and G.~M. Pruna, {\it {Constraining the $g'_1$ coupling
  in the minimal B-L Model}},  {\em J.Phys.} {\bf G39} (2012) 025004,
  [\href{http://xxx.lanl.gov/abs/1009.4164}{{\tt arXiv:1009.4164}}].

\bibitem{Basso:2010pe}
L.~Basso, A.~Belyaev, S.~Moretti, G.~M. Pruna, and C.~H.
  Shepherd-Themistocleous, {\it {$Z'$ discovery potential at the LHC in the
  minimal B-L extension of the Standard Model}},  {\em Eur. Phys. J.} {\bf
  C71} (2011) 1613, [\href{http://xxx.lanl.gov/abs/1002.3586}{{\tt
  arXiv:1002.3586}}].

\bibitem{Basso:2011na}
L.~Basso, S.~Moretti, and G.~M. Pruna, {\it {Theoretical constraints on the
  couplings of non-exotic minimal $Z'$ bosons}},  {\em JHEP} {\bf 08} (2011)
  122, [\href{http://xxx.lanl.gov/abs/1106.4762}{{\tt arXiv:1106.4762}}].

\bibitem{Basso:2012sz}
L.~Basso, K.~Mimasu, and S.~Moretti, {\it {Z' signals in polarised top-antitop
  final states}},  {\em JHEP} {\bf 09} (2012) 024,
  [\href{http://xxx.lanl.gov/abs/1203.2542}{{\tt arXiv:1203.2542}}].

\bibitem{Basso:2012ux}
L.~Basso, K.~Mimasu, and S.~Moretti, {\it {Non-exotic $Z'$ signals in
  $\ell^+\ell^-$, $b\bar b$ and $t\bar t$ final states at the LHC}},  {\em
  JHEP} {\bf 11} (2012) 060, [\href{http://xxx.lanl.gov/abs/1208.0019}{{\tt
  arXiv:1208.0019}}].

\bibitem{Accomando:2015cfa}
E.~Accomando, A.~Belyaev, J.~Fiaschi, K.~Mimasu, S.~Moretti, and
  C.~Shepherd-Themistocleous, {\it {Forward-Backward Asymmetry as a Discovery
  Tool for Z' Bosons at the LHC}},
  \href{http://xxx.lanl.gov/abs/1503.02672}{{\tt arXiv:1503.02672}}.

\bibitem{Accomando:2015ava}
E.~Accomando, A.~Belyaev, J.~Fiaschi, K.~Mimasu, S.~Moretti, and
  C.~Shepherd-Themistocleous, {\it {$A_{FB}$ as a discovery tool for $Z'$
  bosons at the LHC}},  \href{http://xxx.lanl.gov/abs/1504.03168}{{\tt
  arXiv:1504.03168}}.

\bibitem{Appelquist:2002mw}
T.~Appelquist, B.~A. Dobrescu, and A.~R. Hopper, {\it {Nonexotic neutral gauge
  bosons}},  {\em Phys.Rev.} {\bf D68} (2003) 035012,
  [\href{http://xxx.lanl.gov/abs/hep-ph/0212073}{{\tt hep-ph/0212073}}].

\bibitem{Khalil:2007dr}
S.~Khalil and A.~Masiero, {\it {Radiative B-L symmetry breaking in
  supersymmetric models}},  {\em Phys.Lett.} {\bf B665} (2008) 374--377,
  [\href{http://xxx.lanl.gov/abs/0710.3525}{{\tt arXiv:0710.3525}}].

\bibitem{Abreu:1994ria}
{\bf DELPHI Collaboration} Collaboration, P.~Abreu {\em et.~al.}, {\it {A Study
  of radiative muon pair events at $Z^0$ energies and limits on an additional
  $Z'$ gauge boson}},  {\em Z.Phys.} {\bf C65} (1995) 603--618.

\bibitem{delAguila:1988jz}
F.~del Aguila, G.~Coughlan, and M.~Quiros, {\it {Gauge Coupling Renormalization
  With Several U(1) Factors}},  {\em Nucl.Phys.} {\bf B307} (1988) 633.

\bibitem{delAguila:1995rb}
F.~del Aguila, M.~Masip, and M.~Perez-Victoria, {\it {Physical parameters and
  renormalization of $U(1)^a \times U(1)^b$ models}},  {\em Nucl.Phys.} {\bf B456}
  (1995) 531--549, [\href{http://xxx.lanl.gov/abs/hep-ph/9507455}{{\tt
  hep-ph/9507455}}].

\bibitem{Chankowski:2006jk}
P.~H. Chankowski, S.~Pokorski, and J.~Wagner, {\it {Z-prime and the
  Appelquist-Carrazzone decoupling}},  {\em Eur.Phys.J.} {\bf C47} (2006)
  187--205, [\href{http://xxx.lanl.gov/abs/hep-ph/0601097}{{\tt
  hep-ph/0601097}}].

\bibitem{Cacciapaglia:2006pk}
G.~Cacciapaglia, C.~Csaki, G.~Marandella, and A.~Strumia, {\it {The Minimal Set
  of Electroweak Precision Parameters}},  {\em Phys.Rev.} {\bf D74} (2006)
  033011, [\href{http://xxx.lanl.gov/abs/hep-ph/0604111}{{\tt
  hep-ph/0604111}}].

\bibitem{Dawson:2009yx}
S.~Dawson and W.~Yan, {\it {Hiding the Higgs Boson with Multiple Scalars}},
  {\em Phys.Rev.} {\bf D79} (2009) 095002,
  [\href{http://xxx.lanl.gov/abs/0904.2005}{{\tt arXiv:0904.2005}}].

\bibitem{Lyonnet:2013dna}
F.~Lyonnet, I.~Schienbein, F.~Staub, and A.~Wingerter, {\it {PyR@TE:
  Renormalization Group Equations for General Gauge Theories}},  {\em Comput.
  Phys. Commun.} {\bf 185} (2014) 1130--1152,
  [\href{http://xxx.lanl.gov/abs/1309.7030}{{\tt arXiv:1309.7030}}].

\bibitem{Ibarra:2010xw}
A.~Ibarra, E.~Molinaro, and S.~Petcov, {\it {TeV Scale See-Saw Mechanisms of
  Neutrino Mass Generation, the Majorana Nature of the Heavy Singlet Neutrinos
  and $(\beta\beta)_{0\nu}$-Decay}},  {\em JHEP} {\bf 1009} (2010) 108,
  [\href{http://xxx.lanl.gov/abs/1007.2378}{{\tt arXiv:1007.2378}}].

\bibitem{Pontecorvo:1957qd}
B.~Pontecorvo, {\it {Inverse beta processes and nonconservation of lepton
  charge}},  {\em Sov.Phys.JETP} {\bf 7} (1958) 172--173.

\bibitem{Maki:1962mu}
Z.~Maki, M.~Nakagawa, and S.~Sakata, {\it {Remarks on the unified model of
  elementary particles}},  {\em Prog.Theor.Phys.} {\bf 28} (1962) 870--880.

\bibitem{Salvioni:2009jp}
  E.~Salvioni, A.~Strumia, G.~Villadoro and F.~Zwirner,
   {\it {Non-universal minimal $Z'$ models: present bounds and early LHC reach}},
  {\em JHEP} {\bf 1003} (2010) 010,
  [\href{http://arxiv.org/abs/arXiv:0911.1450}{{\tt arXiv:0911.1450}}]
  
\bibitem{Alon1}
A.~E.~Faraggi and D.~V.~Nanopoulos, {\it A superstring $Z'$ at O (1-TeV) ?}
{\em  Mod.\ Phys.\ Lett.\ A} {\bf 6} (1991) 61.

\bibitem{Faraggi:2014ica} 
  A.~E.~Faraggi and J.~Rizos,
  {\it A light $Z'$ heterotic-string derived model}
  {\em Nucl.\ Phys.\ B} {\bf 895} (2015) 233,
[\href{http://arxiv.org/abs/arXiv:1412.6432}{{\tt arXiv:1412.6432}}]
 
\bibitem{Abbas:2007ag}
M.~Abbas and S.~Khalil,
{\it Neutrino masses, mixing and leptogenesis in TeV scale B-L extension of the standard model},
                        {\em JHEP} {\bf 04}, (2008), 056.
                        
 \bibitem{Iso:2010mv}
S.~Iso, N.~Okada and Y.~Orikasa,
{\it Resonant Leptogenesis in the Minimal B-L Extended Standard Model at TeV},
 {\em Phys. Rev.}{\bf D83} (2011), 093011.
 
 \bibitem{Blanchet}
S.~Blanchet, Z.~Chacko, S.~S.~ Granor and R.~N.~Mohapatra,
{\it Probing Resonant Leptogenesis at the LHC}
{\em Phys. Rev.}{\bf D82} (2010), 0904.2174.





  \end{thebibliography}
\end{document}